\documentclass[iop]{emulateapj}

\usepackage[usenames,dvipsnames]{xcolor}
\usepackage{color}

\newcommand{\lya}{Ly$\alpha$}
\newcommand{\halpha}{H$\alpha$}
\newcommand{\ha}{H$\alpha$}
\newcommand{\hbeta}{H$\beta$}
\newcommand{\hb}{H$\beta$}
\newcommand{\hg}{H$\gamma$}

\newcommand{\wha}{{\it W}(H$\alpha$)}
\newcommand{\wlya}{{\it W}(Ly$\alpha$)}

\newcommand{\lsol}{$L_\odot$}
\newcommand{\hi}{H{\sc i}}
\newcommand{\hii}{H{\sc ii}}
\newcommand{\nii}{[N{\sc ii}]}
\newcommand{\sii}{[S{\sc ii}]}
\newcommand{\oii}{[O{\sc ii}]}
\newcommand{\oiii}{[O{\sc iii}]}

\newcommand{\flya}{$f_{\rm Ly\alpha}$}

\newcommand{\llya}{$L_{\rm Ly\alpha}$}
\newcommand{\lha}{$L_{\rm H\alpha}$}
\newcommand{\ergsec}{erg/s}

\newcommand{\egs}{erg/s/cm$^{2}$}
\newcommand{\egsa}{erg/s/cm$^{2}$/\AA}
\newcommand{\msun}{$M_\odot$}
\newcommand{\fesclya}{$f_\mathrm{esc}^{\mathrm{Ly}\alpha}$}
\newcommand{\lfuv}{$L_\mathrm{FUV}$}
\newcommand{\ebv}{$E_{B-V}$}
\newcommand{\kms}{km/s}

\newcommand{\lsun}{$L_\odot$}
\newcommand{\zsun}{$Z_\odot$}

\newcommand{\mufuv}{$\mu_{\rm FUV}$}
\newcommand{\mulya}{$\mu_{\rm Ly\alpha}$}

\newcommand{\msqa}{mag$/\square\arcsec$}
\newcommand{\EWo}{{\it W}(Ly$\alpha$)$_0$}

\newcommand{\NHI}{$N_\mathrm{HI}$}
\newcommand{\Folya}{{\it FWHM}(Ly$\alpha$)$_0$}
\newcommand{\mrk}{Mrk\,259}

\shorttitle{LARS01}
\shortauthors{\"Ostlin, Hayes, Duval et al.}

\begin{document}

\title{The Lyman-alpha Reference Sample:  \\
I. Survey outline and first results for Markarian\,259}


\author{
G\"oran \"Ostlin\altaffilmark{2}, 
Matthew Hayes\altaffilmark{2},
Florent Duval\altaffilmark{2}, 
Andreas Sandberg\altaffilmark{2}, 
Th\o ger Rivera-Thorsen\altaffilmark{2}, 
Thomas Marquart\altaffilmark{2,3}, 
Ivana Orlitov\'a\altaffilmark{4,14}
Angela Adamo\altaffilmark{2}, 
Jens Melinder\altaffilmark{2}, 
Lucia Guaita\altaffilmark{2}, 
Hakim Atek\altaffilmark{5}, 
John M. Cannon\altaffilmark{6}, 
Pieter Gruyters\altaffilmark{3}, 
Edmund Christian Herenz\altaffilmark{7}, 
Daniel Kunth\altaffilmark{8}, 
Peter Laursen\altaffilmark{2,9}, 
J.Miguel Mas-Hesse\altaffilmark{10}, 
Genoveva Micheva\altaffilmark{2,11}, 
H\'ector Ot\'i-Floranes\altaffilmark{12,15}, 
Stephen A. Pardy\altaffilmark{6,13}, 
Martin M. Roth\altaffilmark{7}, 
Daniel Schaerer\altaffilmark{4},  and 
Anne Verhamme\altaffilmark{4}
}



\altaffiltext{1}{Based on observations made with the NASA/ESA Hubble Space Telescope, obtained at the Space Telescope Science Institute, which is operated by the Association of Universities for Research in Astronomy, Inc., under NASA contract NAS 5-26555. These observations are associated with program \#12310,\#11522.}
\altaffiltext{2}{Department of Astronomy, Stockholm University, Oscar Klein Centre, AlbaNova, Stockholm SE-106 91, Sweden}
\altaffiltext{3}{Department of Physics and Astronomy, Division of Astronomy and Space Physics, Uppsala University, Box 516, 75120 Uppsala, Sweden}
\altaffiltext{4}{Observatoire de Gen\`eve, Universit\'e de Gen\`eve, Chemin des Maillettes 51, 1290
      Versoix, Switzerland }
\altaffiltext{5}{Laboratoire dÕAstrophysique, Ecole Polytechnique F\'ed\'erale de Lausanne, Observatoire de Sauverny, CH-1290 Versoix, Switzerland}
\altaffiltext{6}{Department of Physics \& Astronomy, Macalester College, 1600 Grand Avenue, Saint Paul, MN 55105}
\altaffiltext{7}{Leibniz-Institute for Astrophysics Potsdam (AIP),
  innoFSPEC, An der Sternwarte 16, 14482 Potsdam, Germany }
\altaffiltext{8}{Institut d'Astrophysique Paris, 98bis Bd Arago, 75014 Paris}
\altaffiltext{9}{Dark Cosmology Centre, Niels Bohr Institute, University of Copenhagen, DK-2100, Denmark}
\altaffiltext{10}{Centro de Astrobiolog'a (CSIC-INTA), Departamento de Astrof'sica, POB 78, 28691, Villanueva de la Ca\~nada, Spain}
\altaffiltext{11}{Subaru Telescope,
National Astronomical Observatory of Japan,
650 North A'ohoku Place,
Hilo, HI 96720, U.S.A.}
\altaffiltext{12}{Instituto de Astronom'a, Universidad Nacional Aut—noma de MŽxico,
Apdo. Postal 106, Ensenada B. C. 22800, Mexico}
\altaffiltext{13}{University of Wisconsin - Madison, 475 N. Charter Street, Madison, WI 53706, USA}
\altaffiltext{14}{Astronomical Institute, Academy of Sciences of the Czech Republic, Bo\v n{\'\i} II 1401,
      141 00 Prague, Czech Republic.}
\altaffiltext{15}{Centro de Radioastronom\''a y Astrof\''sica, UNAM, Campus Morelia, Mexico}


\begin{abstract}
The Lyman-alpha reference sample (LARS) is a substantial program with the Hubble Space
Telescope (HST) that  provides a sample of local universe laboratory galaxies in which to 
study the detailed astrophysics of the visibility and strength of the Lyman-alpha
(\lya) line of neutral hydrogen. \lya\ is the dominant spectral line in use for characterizing
high redshift ($z$) galaxies. This article presents an overview of the survey, its selection function and 
HST imaging observations. The sample was selected from the combined GALEX+SDSS catalogue at 
 $z=0.028-0.19$, in order to allow \lya\ to be captured with combinations of long pass
filters in the Solar Blind Channel (SBC) of the Advanced Camera for Surveys (ACS) on board HST. 
In addition, LARS utilises \ha\ and \hb\ narrow, and $u, b, i$ broad-band imaging 
with ACS and the Wide Field Camera 3 (WFC3). 
In order to study galaxies in which large numbers of \lya\ photons are 
produced (whether or not they escape) we demanded an \ha\ equivalent width \wha$\ge100$\AA .
The final sample of 14 galaxies  covers  far UV (FUV, $\lambda\sim1500$\,\AA) luminosities 
that overlaps with those of high-$z$ \lya\ emitters and Lyman Break Galaxies (LBGs), making LARS
a valid comparison sample. 
We present the reduction steps used to obtain the \lya\ images, including 
our  LARS eXtraction software (LaXs) which utilises pixel-by-pixel spectral 
synthesis fitting of the energy distribution to determine and subtract the continuum at \lya .
We demonstrate that the use of SBC long pass filter combinations increase the signal to noise with
an order of magnitude compared to the nominal \lya\ filter available in SBC. 
To exemplify the science potential of LARS, we also present some first results for a single galaxy, Mrk\,259 (LARS\,\#1). 
This irregular galaxy shows bright extended (indicative
of resonance scattering), but strongly asymmetric  \lya\ emission. 
%
Spectroscopy from HST/COS centered on the brightest UV knot show a moderate 
outflow in the neutral interstellar medium (probed by low ionization stage absorption features), 
and \lya\ emission with an asymmetric profile. Radiative transfer modeling  is able to reproduce
 the essential  features of the \lya\ line profile, and confirms the presence of an outflow.
From the  integrated photometry we measure a \lya\ luminosity of \llya$=1.3\times 10^{42}$~\ergsec\ 
an equivalent width \wlya =45\,\AA\ and a far UV absolute magnitude  $M_{\rm FUV}=-19.2$ (AB). 
\mrk\ would hence be detectable in high-$z$ \lya\ and LBG surveys. 
The   total \lya\ escape 
fraction is  12\,\%. This number is higher than the low-$z$
average, but similar to that at $z>4$ demonstrating that LARS provides a valid comparison sample for high-$z$
galaxy studies.

\end{abstract}

\section{Introduction}
\label{intro}

The Lyman-alpha emission line (\lya) fulfills several extremely important roles
in observations of the high-redshift ($z$) universe. Firstly, as recombination
nebulae re-process $\sim 1/3$ of the raw ionizing power of hot stars into a
single narrow ($\sim 1$\AA) spectral emission feature, it acts as a luminous
spectral beacon by which to identify galaxies at the highest redshifts
\citep{1996Natur.382..231H, 1998AJ....115.1319C,
2002ApJ...565L..71M,2005MNRAS.359..895V,2006ApJ...642L..13G,2006PASJ...58..313S,
2007A&A...471...71N,2008ApJS..176..301O,Hayes10,2011ApJS..192....5A}.
Secondly, at the highest redshift it provides at least the potential for the
all-important spectroscopic confirmation of galaxies selected by other methods,
transforming the status of candidates to objects with secure redshifts
\citep{Steidel1996,2005ApJ...622..772C,Vanzella2009,2010MNRAS.408.1628S,
2010Natur.467..940L,Curtis-Lake2012}. 

\lya\ is also a resonance line and photons may scatter wherever they encounter
neutral hydrogen atoms. This scattering may occur in the interstellar medium
(ISM) of the galaxies themselves, or after emission from galaxies, in the
intergalactic medium (IGM) that immediately surrounds them. Inside the ISM,
this resonance scattering leads to a complicated radiative transport of \lya,
in which the visibility of the line may be influenced by large number of
factors, including raw dust content \citep{1993ApJ...415..580C,
2008A&A...491...89V,2009A&A...506L...1A,Hayes10,Atek2014}, dust geometry
\citep{2009ApJ...704L..98S}, the neutral gas content and kinematics
\citep{1998A&A...334...11K,2003ApJ...598..858M}, and gas geometry
\citep{1991ApJ...370L..85N,1996ApJ...466..831G,2006MNRAS.367..979H,
Laursen13,Duval2014}.  These radiative transport issues, and the fact that to
date no clear order of precedence has been established between them (although
see \citealt{Mallery2012}), implies that the total escape fraction of \lya\
photons (\fesclya; defined as the ratio of observed to intrinsic \lya\
luminosity; \citealt{2005A&A...438...71H}), cannot be predicted from knowledge
of any single of the above quantities.  The direct consequence of this is that
photometric measurements of \lya\ will reflect the underlying properties of
galaxies only in the very broadest statistical sense, and for individual
galaxies, the observed  \lya\ flux would not reveal much about the underlying
physical quantities.  Moreover, if it is not understood what conditions that
regulate  galaxies' ability to emit the \lya\ photons that are produced within
them, studies of the galaxy population and cosmology by means of \lya\ will be
subject to unknown biases.

\begin{figure*}[t!]
\centering
\includegraphics[angle=0,scale=0.9]{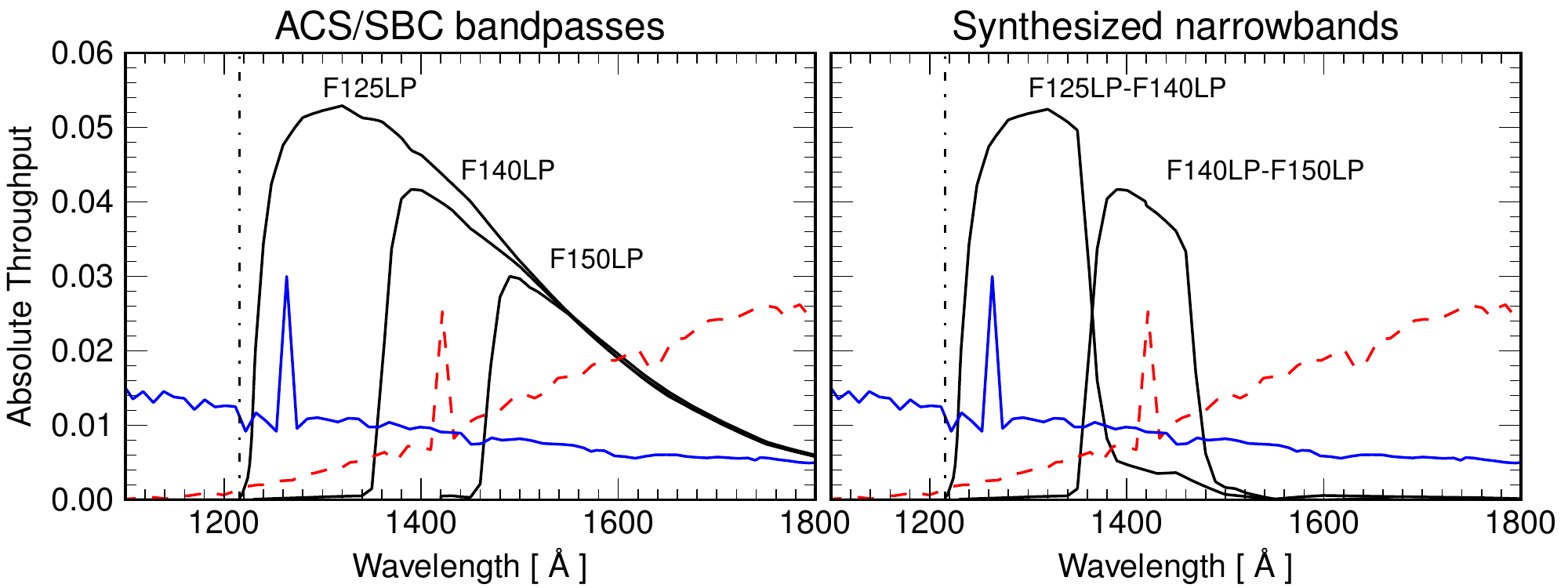}
\caption{ACS/SBC bandpass combinations (left) that yield effective \lya\
bandpasses (right).  Bandpasses are labeled, as are the synthesized narrowbands
that result from the subtraction of adjacent long-pass filters.  The subtracted
bandpasses are both well-defined, and both cut on and off sharply; the extended
wing of the F125LP--F140LP combination is just 5\% of the integrated filter
surface. The combinations F125LP--F140LP and F140LP-F150LP sample \lya\ at
redshifts $z=0.028-0.109$ and $z=0.134-0.190$, respectively, which hence
defines the redshift ranges permitted by our sample selection (see
Sections~\ref{sect:strat} and \ref{sect:sel} for details).  The maximal
throughput of the effective bandpasses are 5 and 4\% for the low- and high-$z$
setups; for comparison the dedicated \lya\ filter in SBC (F122M) has an
integrated peak system throughput of 0.9\%.  Overlaid in blue solid and red
dashed lines are \textsc{starburst99} models of blue and red continuum spectra
with \lya\ lines added, and redshifted to $z=0.04$ and 0.17, the typical values
for low- and high-$z$ LARS galaxies (the choice made in this figure to put the
blue spectrum at low-$z$ and the red one at high-$z$ is of course completely
arbitrary). The geocoronal \lya\ line, shown by the dot-dashed vertical line at
1216 \AA\ is not transmitted by these long-pass filters.} 
\label{fig:sbc}
\end{figure*}

However, regarding the IGM the situation of \lya-emission vis-a-vis
radiative transport is not so pessimistic. Again due to the high absorption
cross section in \hi, \lya\ observations provide an independent test of the
intermediate to late stages of reionization \citep{2002ApJ...576L...1H,
2004MNRAS.349.1137S, 2004ApJ...617L...5M, 2008A&A...491...89V,
2010MNRAS.408..352D}, where it is sensitive to intermediate volume-ionized
fractions of a few 10 per-cent \citep{2008ApJS..176..301O}. In contrast the
Gunn-Peterson trough observed in quasar absorption spectra
\citep[e.g.][]{2006ARA&A..44..415F}  only probes the very final stages ($<1\%$
neutral). Obviously these \lya\ tests are reliant upon the condition that, at
the highest redshifts, a substantial fraction of the \lya\ photons that are
produced are able to avoid being trapped by their own ISM. While this condition seems
not to be fulfilled at low and intermediate redshift
\citep{2010ApJ...711..928C,2011ApJ...730....8H}  \lya\ emission seems to
undergo a strong cosmic evolution as evidenced by three primary results. With
increasing $z$: (a) the volumetric \lya\ escape fraction increases
\citep{2011ApJ...730....8H}; (b) the fraction of UV-selected galaxies with
strong \lya\ increases  \citep{2010MNRAS.408.1628S}; and (3) the UV luminosity
functions of UV- and \lya-selected galaxies appear to converge
\citep{2008ApJS..176..301O}. Thus the future of \lya\ as a probe of
reionization and of galaxy evolution in general may be an optimistic one if we
can understand on what conditions a galaxy transmits its \lya\ photons. 

Recent years have seen considerable effort go into the modeling of \lya\
transport using intensive computer simulations, working in idealized galaxy
geometries \citep{2006ApJ...649...14D, Verhamme06, 2011MNRAS.411.1678C}, and
quasi-realistic galaxies drawn from computer simulations
\citep{2006ApJ...645..792T, Laursen09, 2012A&A...546A.111V,
2012MNRAS.424..884Y}. Results have been as insightful as they have been
CPU-intensive but are still, taken as a whole, unsatisfactory. Regarding
transport studies in simulated galaxies, it is not clear that galaxy formation
models probe the necessary scales, that photons are injected in the correct
places, that clumping of the ISM is realistic, or that dust production,
destruction or coupling with metallicity necessarily conform with nature.
Perhaps most important, simulated galaxies lack a realistic treatment of
galactic winds. Regarding idealized geometries, it is not clear that sufficient
parameter spaces are covered, if vital quantities are omitted, or whether distributions
are realistic (particularly regarding the multiphase ISM). What is badly needed
in this field is stronger empirical constraints on  the \lya\ transport
process. 

\subsection{The case for low redshift \lya\ imaging}
\lya\ is used intensively to study the galaxy population at high-$z$, and
luminosity functions (LF; see compilation in \citealt{2011ApJ...730....8H}) and
equivalent width (EW) distributions are now becoming well populated
\citep{2007ApJ...667...79G, 2008ApJS..176..301O, 2010MNRAS.408.1628S}.  Yet
while it is exactly these properties that are used to describe the high-$z$
galaxy population, these galaxies all lie at distances too great for us to
determine the physical processes that actually shape the aforementioned
distributions.  Notably standard nebular diagnostics of dust attenuation,
metallicity, and ionization state are redshifted to wavelengths where they are
difficult or impossible to observe, and that forces the community to turn to
sub-optimal and biased observables/indices.  Particularly for \lya-selected
galaxies, which typically exhibit faint stellar continua, measuring features
like the 4000\AA\ break also becomes increasingly troublesome. Far UV
absorption features (either interstellar or photospheric) may be visible but
again, only for the brightest galaxies. These problems are typically solved
through stacking analyses, which destroys almost all information relating to
the diversity of the populations. Moreover, spatial sampling scales at high-$z$
are around 6--8 kpc/arcsec, meaning most high-$z$ galaxies are unresolved in
ground-based data and surface brightness, which decreases proportionally to
$(1+z)^{3(4)}$ in an unresolved (resolved) emission line implies
order-of-magnitude losses.  To measure (or robustly constrain) all the relevant
quantities complicit in \lya\ transport in individual galaxies, and
to simultaneously probe more detailed scales in the ISM, we need to turn to the
low-$z$ universe.

Naturally this requires large investments of time on UV-capable space-based
platforms, in which field the IUE, HST, and GALEX have been pivotal. By
following up various blue-compact and emission line galaxies IUE first
demonstrated that radiative transport effects were most likely at play, as
dust-corrected \lya/Balmer line ratios dramatically fail to reconcile with the
predictions of recombination theory \citep{1996ApJ...466..831G}, and
demonstrating the importance of gas geometry as well as direct dust
attenuation. Using HST high resolution spectroscopy,
\citet{1994A&A...282..709K,1998A&A...334...11K} and \citet{1995A&A...301...18L}
first highlighted the importance of neutral gas kinematics (see also, e.g.
\citealt{2003ApJ...588...65S} and \citealt{2007A&A...467...63T} for
observational results at high-z) which later enabled
\citet{1999MNRAS.309..332T} and \citet{2003ApJ...598..858M} to cast the \lya\
emission/absorption as an evolutionary sequence following the age of the
star-formation episode. However, the nature of spectroscopic observations meant
that the global emission properties of the galaxies studied remained unknown,
and called for the need for photometric \lya\ imaging studies of local star
forming galaxies.

While earlier HST cameras (WFPC, FOC, WFPC2, STIS) included \lya\ imaging
filters, the low integrated system throughputs ($<0.3$\%) prevented their use
for studying faint targets such as external galaxies. The arrival of ACS and
its solar blind channel (SBC) resulted in a big step in far UV sensitivity and
the first \lya\ imaging study of six star forming galaxies in the local
universe was initiated \citep{2003ApJ...597..263K}.  The spatial scattering of
\lya\ photons could be observed for the first  time \citep{2005A&A...438...71H,
2007MNRAS.382.1465H, Atek08, 2009AJ....138..923O}.  In these studies we found
that when \lya\ is found in emission, its luminosity is dominated by large
scale, largely  diffuse components.  The large scale diffuse nature of the
emission also means that UV spectroscopy centered of the UV-brightest parts of
galaxies would miss most of the \lya\ flux. This first pilot study aimed at
exploring \lya\ imaging of local galaxies, and while it found the first
evidence for the importance of resonant scattering  the small sample size and
its hand picked nature prevented general conclusions and statistically
significant trends to be drawn. 

\subsection{The Lyman Alpha Reference Sample} 
To remedy this situation we
started, in HST cycle 18, \emph{LARS -- the Lyman-alpha Reference Sample} (GO program 12310).
\emph{LARS} is a sample of 14 star forming galaxies in the nearby universe
selected from the cross-correlated GALEX General Release 2 and SDSS DR6
catalogues.  In addition to \lya, LARS includes HST imaging in \ha, \hb,
optical and UV continuum, and UV spectroscopy with COS sampling \lya\ and
metallic lines arising in the interstellar medium of the galaxies.  The 
UV continuum luminosity range spanned by LARS is similar to \lya-emitters and
LBGs at $z\sim 3$, but works at a physical resolution up to two orders of
magnitude smaller than is possible at high-$z$. Furthermore we have obtained
direct \hi\ masses for the first time in a \lya\ sample (Pardy et al., 2014
submitted to ApJ) and other crucial datasets including: far infrared
spectroscopy and photometry to study dust and photodissociated gas, 3D
spectroscopy for ionized gas kinematics, ground-based narrowband images in
nebular lines and the continuum, and deep infrared imaging to examine the host
galaxies in detail. Thus in the long term, LARS will provide what is easily the
most extensive data set in which to study detailed \lya\ astrophysics.  With
LARS we will, in this and subsequent papers, investigate how \lya\ is
transported out of galaxies, what fraction of \lya\ that survives, and the 
\lya\ morphologies. We will do this as a function of physical parameters like
dust and gas content, metallicity, kinematics, and stellar population properties.
In order to rapidly disseminate results from the program presented here,
we presented a letter \citep{Hayes2013} showing that \lya\ emission profiles in
LARS galaxies typically extend over scales twice as large as those measured in
the UV continuum and \halpha. The current paper (Paper I) presents the survey 
outline, sample selection and data analysis methods, plus a science demonstration
 for the first galaxy in the sample. In paper II  \citep{Hayes2014} 
we present resolved photometry of \lya\ and other quantities for the full LARS sample.
Subsequent paper in the series discuss integrated \hi\ properties (Pardy et al. 2014),
COS spectroscopy (Rivera-Thorsen et al. 2014 in prep), radiative transfer modeling (Orlitova 
et al. 2014 in prep), a detailed study of an individual target (Duval et al. 2014, in prep), 
comparative studies of morphologies (Guaita et al. 2014 in prep), with several more 
papers in preparation.
 
The current  paper is
organised as follows: In section 2 we review some observational/technical
considerations that impacts the designs of the project.  In section 3 we
describe the sample. In section 4 we describe the HST imaging data, and in
section 5 how we use these to arrive at high fidelity \lya\ images. Section 6
exemplifies what can be deducted from the HST imaging and spectroscopic data by
showing preliminary results for the first galaxy in the sample. Section 7
contains discussion and summary.  Throughout we assume a cosmology of $(H_0,
\Omega_\mathrm{M}, \Omega_\Lambda) = (70~\mathrm{km~s}^{-1}~\mathrm{Mpc}^{-1},
0.3, 0.7)$. All magnitudes given in the paper are in the AB system

\section{LARS survey strategy and continuum subtraction}\label{sect:strat}
While several possible methods exist by which  \lya\ can be imaged at $z\sim 0$
with HST, there are a number of important considerations regarding the
available bandpasses that restrict our methods, survey strategy, and ultimately
sample selection.  \lya\ imaging in the local universe requires access to space
platforms, which in practice means HST.  Whereas several HST cameras (WFPC,
WFPC2, STIS, ACS) have had  \lya\ filters for $z\sim0$, these have suffered
from imperfect band pass shapes and low throughput. The sample of six local
($z=0.007-0.029$) galaxies presented in \citet{2009AJ....138..923O} used the
'narrow' ($\Delta\lambda\sim120 $\AA\ or 10\%) \lya\ filter (F122M) available in the 
solar blind channel (SBC) of
ACS. This bandpass has an extended red wing, low throughput and  high
background since the geocoronal \lya\ emission is included in the bandpass.
Moreover, in order to sample the continuum, only filters at significantly
(20\%) longer wavelengths are available, which makes the continuum subtraction
very sensitive to the spectral shape  on the red side of \lya. 
 The \lya\ EW for star forming populations has a maximum of $\sim 200$\AA\ at zero 
age (Schaerer 2003). While diffuse regions dominated by resonantly scattered \lya\ may
have very high EW, the central UV bright regions have much lower values and the integrated 
\wlya\  of local starbursts ranges between negative (absorption) and a few 10's of \AA\  
 \citep{1996ApJ...466..831G, 2009AJ....138..923O}. Hence the \lya\ line may 
contribute only a few percent to the flux of the F122M filter, requiring a very 
accurate continuum subtraction, and any attempt to just subtract the continuum
using a broad filter on the red side would fail dramatically (Kunth et al. 2003; Hayes and \"Ostlin 2006). 

 The near UV continuum 
of starbursts can in general be well characterised by a power law with index $\beta$ 
($f_\lambda \propto \lambda^\beta$). Hence it may seem plausible that it should be 
possible to continuum subtract \lya\  by determining $\beta$ and extrapolate the 
continuum to the wavelength of  \lya\ .  However, near \lya, the continuum evolve rapidly in 
wavelength with the age and dust reddening of the stellar population, and unfortunately
a single UV colour is insufficient to estimate the continuum
at \lya\ \citep{2005A&A...438...71H}.

The solution, as shown in \citet{2005A&A...438...71H,2009AJ....138..911H} is to
estimate the continuum under \lya\ through modeling the galaxy spectrum as a
composite population of young stars, old  stars and nebular gas. By measuring
the UV spectral slope ($\beta$), the 4000 \AA\  break plus its fractional contributions from an
old underlying stellar population and nebular gas emission, the continuum under
\lya\ can be accurately estimated. In practice this approach requires imaging
in several band passes in addition to those capturing \lya\ and its nearest
continuum point.

As an alternative setup to using the dedicated \lya\ filter (F122M),
combinations of the ACS long-pass filters can be employed.  All the long-pass
filters have nearly identical long wavelength performance and the subtraction
of two adjacent filters hence forms a well defined bandpass. In  Fig.
\ref{fig:sbc} we show the effective bandpass formed by the combination of
F125LP--F140LP and F140LP--F150LP. The long pass filters have very high
transmission redwards of the cut, and consequently the throughput is
enhanced by a factor $\sim5$ compared to the nominal \lya\ filter in SBC
(F122M). Furthermore, the geocoronal \lya\ background, that provides the
dominant source of noise in the F122M filter, is completely suppressed by using
these passbands.  The F125LP--F140LP bandpass combination is effective in
capturing \lya\ for redshifts $z=0.028$ to $0.109$, and F140LP--F150LP for
$z=0.134-0.190$. 
 While the bandpass shapes of these effective filters are much better than for F122M,
the widths of these combinations are still $\sim 120$ \AA\ and
since we are interested in measuring the \lya\ line also in the case of low
equivalent width or even absorption, the same considerations as for F122M still hold:
the effective bandpasses will in many regions be strongly dominated by continuum emission.
Hence, the continuum has to be modeled in the same manner but will yield more accurate 
results \citep{2009AJ....138..911H}.

For  the starburst galaxy NGC\,6090, at $z=0.029$ \citep[included in][]{2009AJ....138..923O}, data was available for F122M,
F125LP, F140LP and F150LP prior to the submission of the LARS proposal in Cycle 18, 
and we could thus compare the two methods using real data. Despite the
shorter exposure time in F125LP vs F122M, the signal to noise in the resulting
\lya\ image is increased with a factor of 7 (see appendix A.1 for further details).
Given the very favorable outcome of this comparison, the combination of F125LP
with F140LP was the baseline setup chosen for LARS.  The  higher redshift
window defined by the combination of F140LP with F150LP was expected to perform
similarly well as the throughput is slightly lower, which, however, is
compensated by a yet lower background.  Its main drawback is the loss of
physical resolution and an increased cosmological surface brightness dimming.

\begin{figure*}[t!]  
\centering
\includegraphics[angle=0,scale=0.75]{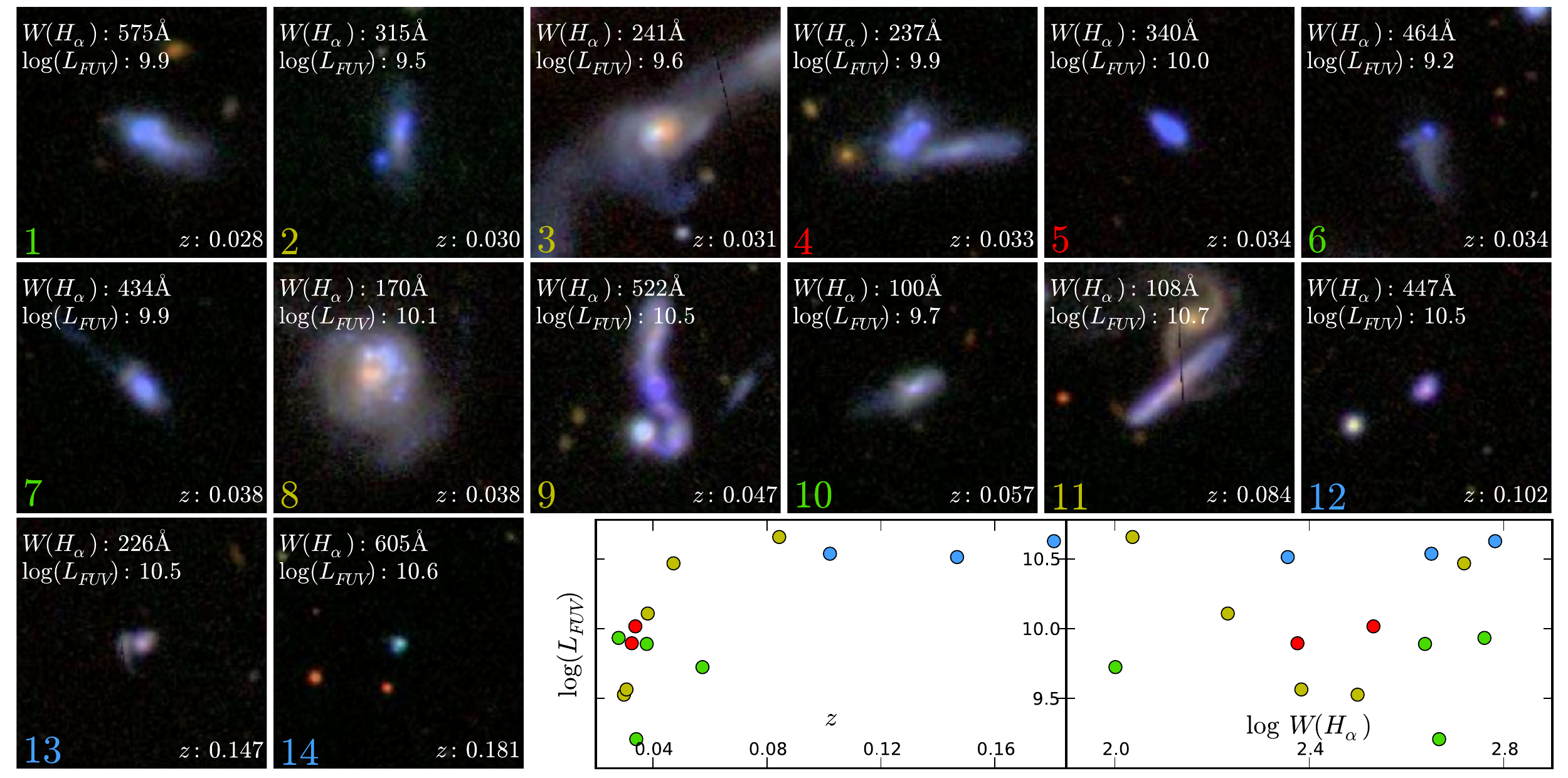}
\caption{LARS sample as seen by SDSS, in $40\arcsec \times 40\arcsec$
thumbnails. Each panel is numbered with its LARS id number (increasing with
redshift) in the lower left corner, its redshift in the lower right corner, and
the \wha\ and \lfuv\ values upper left.  The LARS id numbers are color coded
according to their selection: green (COS GTO targets), red (from McCandliss
sample) and blue (Lyman Break Analogs) numbers means that the target was originally
included in another sample (see text for details) while numbers are yellow for
genuine LARS targets.   Lower right panels show the \lfuv\ vs $z$ and \wha
distributions, where the color coding has the same meaning. } 
\label{fig:selimages}
\end{figure*}

\begin{deluxetable*}{ccccccccccc}[t!] 
\tabletypesize{\scriptsize}
\tablecaption{LARS Selection\label{tab:galprop}}
\tablehead{
\colhead{LARS} & 
		\colhead{SDSS} & 
		\colhead{redshift} & 
		\colhead{\wha} & 
		\colhead{$m_\mathrm{FUV}$} & 
		\colhead{$m_\mathrm{NUV}$} & 
		\colhead{UV slope} & 
		\colhead{$E_{B-V}^\mathrm{MW}$} & 
		\colhead{$\log (L_{\rm FUV})$} &
		\colhead{common name} \\
\colhead{ID \#} & 
		\colhead{object ID} & 
		\colhead{z}  &
		\colhead{\AA} & 
		\colhead{AB} &  
		\colhead{AB} & 
		\colhead{$\beta$} &
		\colhead{mag} & 
		\colhead{\lsun} \\
\colhead{(1)} & 
		\colhead{(2)} & 
		\colhead{(3)} & 
		\colhead{(4)} & 
		\colhead{(5)} &  
		\colhead{(6)} &  
		\colhead{(7)} & 
		\colhead{(8)} &  
		\colhead{(9)} &  
		\colhead{(10)}}
\startdata
  1 & J132844.05+435550.5 & 0.028 &  575 &   $16.65 \pm 0.03$ & $16.45 \pm 0.02$ & $-1.55$ & $0.017$ & $ 9.92$ & Mrk\,259\\
  2 & J090704.88+532656.6 & 0.030 &  315 &   $17.91 \pm 0.02$ & $17.62 \pm 0.01$ & $-1.37 $ & $0.018$ & $ 9.48 $ \\
  3 & J131535.06+620728.6 & 0.031 &  241 &   $17.89 \pm 0.08$ & $16.92 \pm 0.01$ & $ ~~0.17 $ & $0.023$ & $ 9.52 $ & Arp\,238\\
  4 & J130728.45+542652.3 & 0.033 &  237 &   $16.97 \pm 0.04$ & $16.68 \pm 0.02$ & $-1.36 $ & $0.019$ & $ 9.93 $ \\
  5 & J135950.91+572622.9 & 0.034 &  340 &   $16.81 \pm 0.03$ & $16.81 \pm 0.02$ & $-2.01 $ & $0.011$ & $10.01 $ &  Mrk\,1486\\
  6 & J154544.52+441551.8 & 0.034 &  464 &   $18.91 \pm 0.11$ & $18.58 \pm 0.06$ & $-1.25 $ & $0.020$ & $ 9.20 $ & KISS\,2019\\
  7 & J131603.91+292254.0 & 0.038 & 434 &   $17.68 \pm 0.03$ & $17.03 \pm 0.01$ & $-0.53 $ & $0.010$ & $ 9.75 $ & IRAS\,1313+2938\\
  8 & J125013.50+073441.5 & 0.038 &  170 &   $16.91 \pm 0.01$ & $16.37 \pm 0.01$ & $-0.80 $ & $0.034$ & $10.15 $ \\
  9 & J082354.96+280621.6 & 0.047 &  522 &   $16.59 \pm 0.03$ & $16.25 \pm 0.02$ & $-1.23 $ & $0.032$ & $10.46 $ & IRAS0820+2816 \\
10 & J130141.52+292252.8 & 0.057 &  100 &   $18.66 \pm 0.03$ & $18.29 \pm 0.01$ & $-1.17 $ & $0.012$ & $ 9.74 $ & Mrk\,061\\
11 & J140347.22+062812.1 & 0.084 &  108 &   $17.24 \pm 0.02$ & $16.80 \pm 0.01$ & $-1.01 $ & $0.025$ & $10.70 $ \\
12 & J093813.49+542825.1 & 0.102 &  447 &   $18.06 \pm 0.02$ & $17.95 \pm 0.01$ & $-1.75 $ & $0.018$ & $10.53 $ & SBS\,0934+547 \\
13 & J015028.39+130858.4 & 0.147 &  226 &   $19.18 \pm 0.03$ & $19.01 \pm 0.02$ & $-1.61 $ & $0.072$ & $10.60 $ & IRAS\,0147+1254 \\
14 & J092600.40+442736.1 & 0.181 &  605 &   $19.00 \pm 0.10$ & $19.10 \pm 0.07$ & $-2.22 $ & $0.019$ & $10.69 $ 
\enddata
\tablecomments{Column (1) LARS ID number (2) SDSS name (3) SDSS redshift (4) SDSS equivalent width in \ha\ 
(5-6) GALEX FUV and NUV magnitudes.  GALEX bandpasses centers use the pivot wavelengths: 
1524~\AA\ and 2297~\AA\ for the FUV and NUV channels, respectively. 
(7) UV continuum slope assuming standard parameterization: $\beta \propto \lambda^\beta$. Accuracy ranges from 0.01 to 0.05. (8) 
Foreground extinction in \ebv, as reported in the SDSS database and derived from 100~\micron\ dust emission 
maps. (10) FUV luminosity in logarithmic monochromatic solar 
luminosities, $L_{\rm FUV}=\lambda\times L_{\lambda}$. The precise restframe wavelength varies from target to 
target and is evaluated at 1524~\AA$/(1+z)$ }
\end{deluxetable*}

\begin{figure*}[t!]  
\centering
\includegraphics[angle=0,scale=0.293]{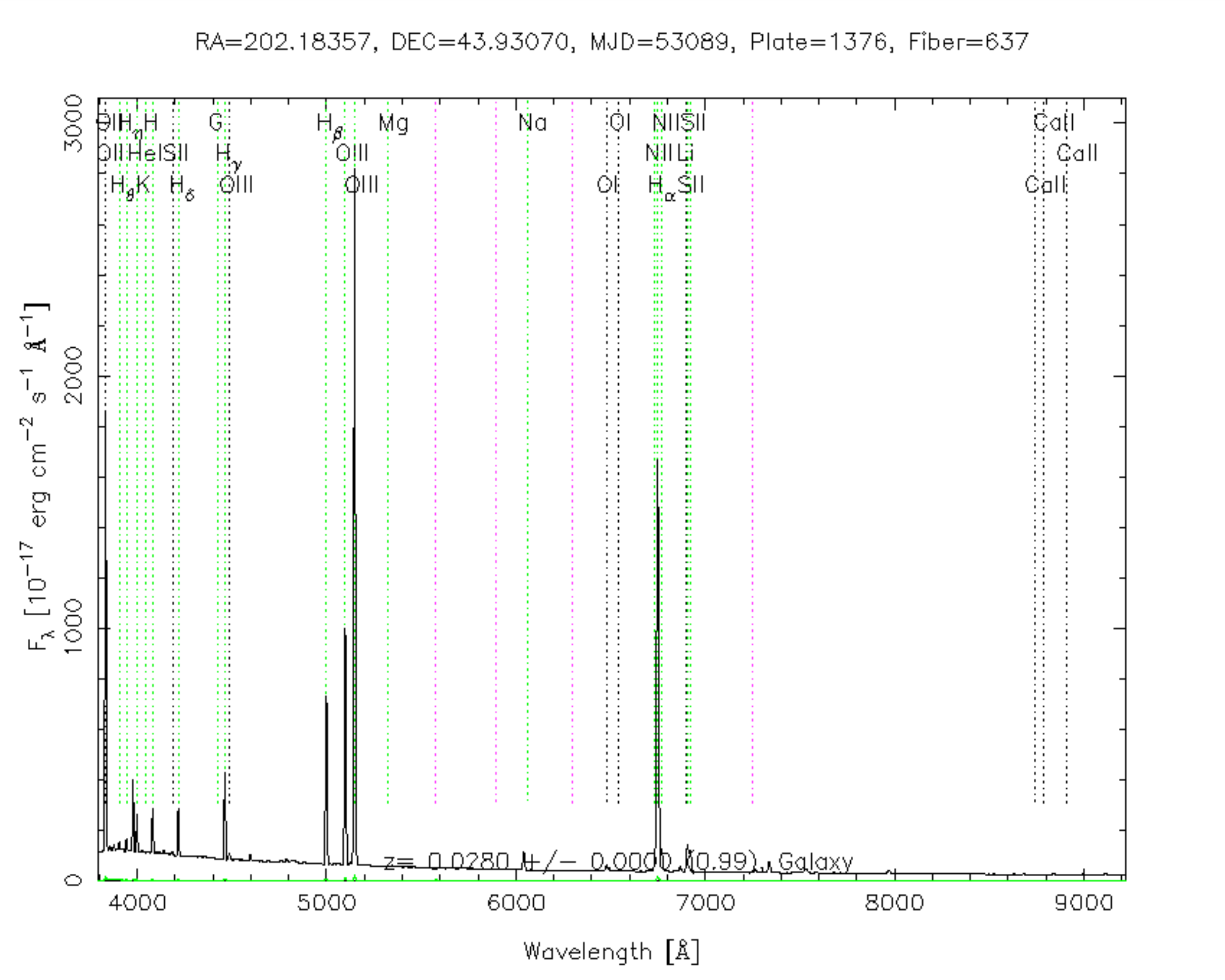}
\includegraphics[angle=0,scale=0.293]{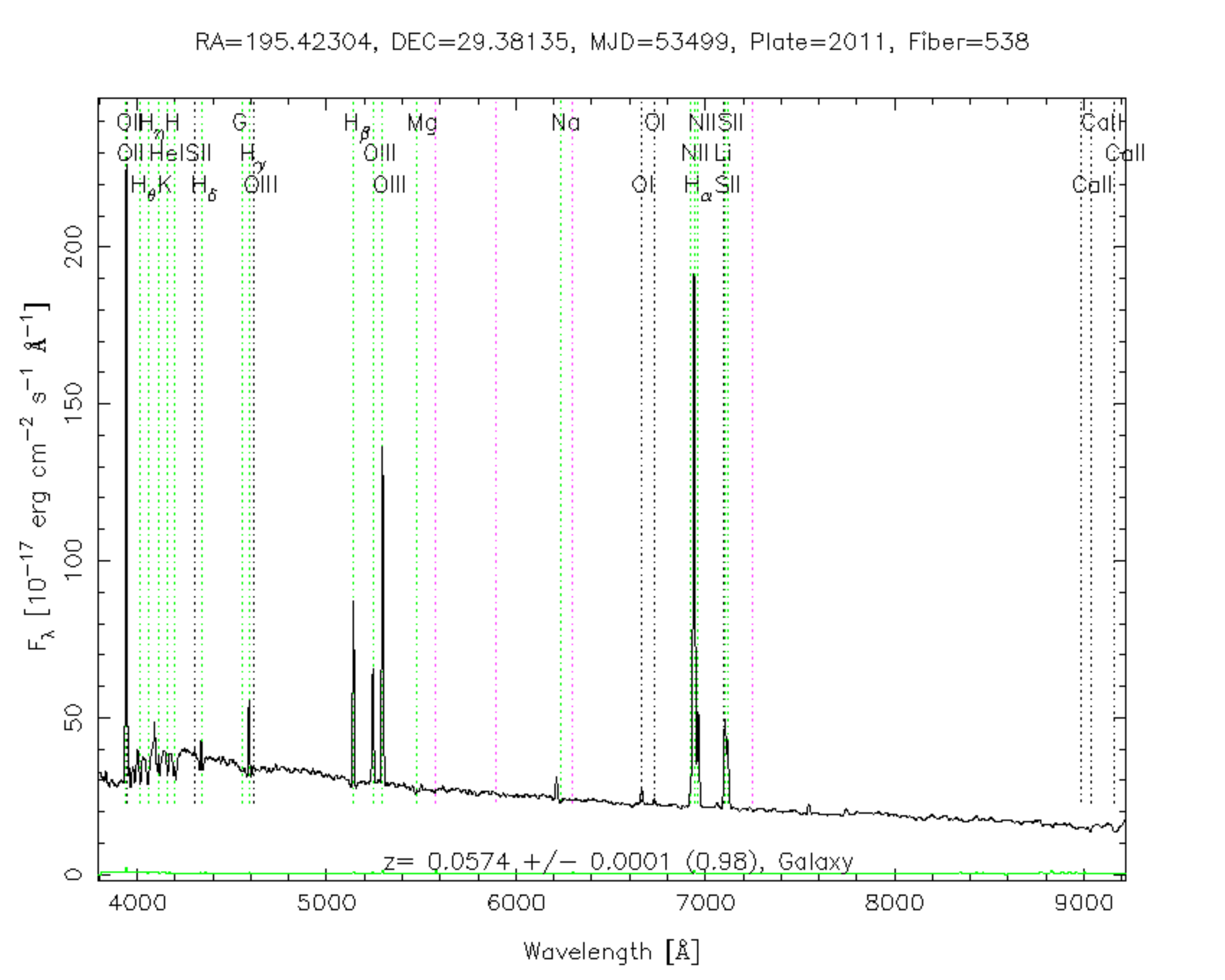}
\caption{SDSS spectra of lars\#1 and 10. LARS\,\#1 belongs to the objects with the highest \wha\ according
to SDSS while Lars\#10 has the lowest, and therefore the most apparent contribution of an underlying stellar
population.} 
\label{fig:sdssspec}
\end{figure*} 

\section{LARS sample selection}\label{sect:sel}
Our sample selection starts out from the combined SDSS (DR6) and GALEX (DR3)
data releases, which allow us to draw a UV selected sample of galaxies with
known redshift.  From the combined catalogs we use the redshift and Galactic
reddening information from SDSS  and the GALEX  far UV (FUV) flux to calculate
the FUV luminosity for all cross matched sources in the two possible redshift
ranges (Section~\ref{sect:strat}). 

To ensure that our sample only contains actively star forming galaxies, i.e.
galaxies that produce \lya\ (whether or not it escapes), we have required a
rest frame \ha\ equivalent  width of: \wha$ \ge 100 $ \AA .  A galaxy with
constant star formation rate still has a \wha$ >100$ \AA\ after 1 Gyr
\citep{1999ApJS..123....3L}. A lower \wha\ cut could hence lead to disk
galaxies with slowly declining star formation dominating the sample. Indeed, we
find that \wha$ \ge 100$ \AA\ leads to a sample dominated by compact ($I_{\rm
FUV} > 10^8$\lsol/kpc$^2$) systems \citep{Heckman2005}, mainly of irregular
morphology,  whereas for $0 < $\wha$ <  50$ \AA\ ordinarily looking disks seem
to dominate.  For an instantaneous burst with a Salpeter IMF and $Z=0.004$
\wha\ drops to 100 \AA\ after 7 Myr.  If radiative transfer effects are ignored
the same population, at age 7~Myr, has \wlya $\approx 25$ \AA, which is close
to the rest frame limits (usually 20 \AA ) for \lya\ candidate selection in
current high$-z$ surveys, implying that our  100 \AA\ cut mostly rejects
objects that could anyway not show strong \lya.  It is worth noting though that
for extended star formation histories it is in principle possible to have both
\wha $<100$ and  \wlya$> 20$ \AA\ since the continuum dominating near \ha\
would not contribute much near \lya .  Having a sample unbiased in \wha\ could
in a sense have been preferable but would risk being contaminated by post starburst galaxies
and given the limited sample size we have
focused on galaxies with very active star formation that are likely to have
intrinsically strong \lya\ production\footnote{A follow up project, eLARS,  to
image 28 more galaxies was approved in Cycle 21. Here the cut is \wha$>40$\AA .
}

Since LARS aims to investigate the \lya\ properties of star forming galaxies we
rejected objects with a strong AGN component by requiring the line width in
\ha\ to be smaller than FWHM 300 km/s (line of sight $\sigma < 130$ km/s) and
that the galaxies are classified as H{\sc ii}-like based on their \oiii/\hb,
\nii/\ha\ and \sii/\ha\ ratios.

In selecting targets, we aimed to cover a range of FUV luminosities, from
$\sim10^9$ in monochromatic solar luminosities ($\nu L_\nu$ / \lsun), to as
high as possible without including AGN dominated objects. Priority was given to
low $z$ objects since this maximizes the physical resolution, but in order to
include the more FUV luminous objects we needed to sample larger volumes. In
practice this is what led to the inclusion of a high-$z$ (F140LP-F150LP) setup.
In order to maximize the efficiency of observing at FUV wavelengths, priority
was furthermore given to objects with low foreground extinction in the SDSS
database (see Table 1).

Because LARS needs many different observing modes (cameras, filters, spectra),
observing a substantial sample can quickly become observationally expensive. In
principle, the expense can be reduced by using archival data, when available,
under the proviso that the selection function is not compromised. Hence we
investigated which of the objects that fulfilled the selection criteria that
also have archival data valuable for LARS. If the archival data was deemed
sufficiently deep, such targets were then prioritized. SDSS cutout images, and
the \lfuv, redshift, and \wha\ distributions of the final LARS sample are shown
in Figure~\ref{fig:selimages}.  Regarding relevant archival HST observations LARS overlaps
with:
\begin{itemize}
\item Two galaxies from GO\,11110 (PI McCandliss), the sample of which was
originally selected on FUV flux and declination and have 3 orbits of SBC far UV
imaging in the required filters each.  These galaxies are represented in
Figure~\ref{fig:selimages} with red ID numbers and points.  The remaining
twelve galaxies from GO\,11110 fail either the redshift, \wha\ or AGN criteria.

\item Three galaxies (shown in blue in Fig. \ref{fig:selimages}) from the Lyman
break analog sample by Heckman (programs 10920, 11107) including COS
spectroscopy and partial SBC imaging.  This sample was selected on FUV
luminosity and surface brightness. The remaining objects fail either the $z$,
\wha\ and/or AGN criteria.

\item Four galaxies (shown in green in Fig. \ref{fig:selimages}) from the COS
GTO programs (11522, 12027) of star forming galaxies (PI Green), having spectra
covering \lya\ and ISM absorption features.  These galaxies have originally
been selected on \ha\ brightness and compactness (due to the 2.5\arcsec
aperture of COS) from the KISS objective prism survey
\citep{2001AJ....121...66S}.  The remainder of the sample fail the \wha\ and/or
$z$ criteria.

\end{itemize}

Thus in total, LARS could be directly optimized by the inclusion of seven
objects, fulfilling our selection criteria, that have relevant data from other
programs.  In order to build a sufficient sample, 7 more targets were selected
that do not have prior SBC or COS observations. An aspect that we considered
for the final seven targets was that the total sample should a have a
reasonable distribution in FUV luminosity. 

The total sample of 14 objects populate a UV luminosity  range between
$\log(L_\mathrm{FUV}/L_\odot)=9.2$ (corresponding to 
$SFR_\mathrm{FUV}\approx 0.5M_\odot$/yr, similar to the objects that expected
to dominate re-ionization and that match future JWST goals) and 10.7 (roughly
the characteristic luminosity, $L_\star$,  for LBGs, \citealt{Reddy2008}).  The
$L_\mathrm{FUV}$ range also encompasses that of  the 'double blind'  \lya\ +
\ha\ sources at $z=2$ \citep{Hayes10}, high-$z$ LAEs
\citep[e.g.][]{2009A&A...498...13N}, GALEX selected LAEs \citep{Deharveng2008}
and the  $z\sim 0.2$ GALEX selected 'LBG analogs' \citep{Hoopes2007,
2008ApJ...677...37O}. This implies that the results from LARS can be
meaningfully compared to numerous other studies. 

Five of the LARS galaxies have available IRAS fluxes, and follow the IRX-beta
relation by \citet{1999ApJ...521...64M}.  Indicative stellar masses and star
formation rates from the SDSS added value catalogs
\footnote{http://www.mpa-garching.mpg.de/SDSS/DR7/\#derived}
\citep{2003MNRAS.341...33K, 2007ApJS..173..267S, 2004MNRAS.351.1151B} shows
that these LARS galaxies lies above the SFR-Mass main sequence
\citet{2011A&A...533A.119E}, indicating that they are ``true" starburst
galaxies, which is likely to be a consequence of the \wha\ selection of LARS. 

On a cautionary note, the targets with available SBC/COS data were originally
selected by others, which means that we do not have full control of the
selection function. For the Green program potential targets with large
angular extent are not included. The size criterium becomes less of an issue
as we go out in redshift, and the Green targets do not stand out as being
systematically smaller than the other LARS targets. 

The resulting sample of 14 galaxies and its distribution in $z$, \lfuv\ and
\wha\ are presented in Fig. \ref{fig:selimages}.  In Table \ref{tab:galprop} we
give the redshift and \wha\ as obtained from SDSS and ultraviolet photometric
properties as measured by GALEX, and the foreground Galactic reddening. The
median $E_{B-V}^\mathrm{MW} =0.019$  corresponds to 15\% of the far UV photons
being absorbed in the Milky Way, and only LARS \#13 has a significantly higher
value.  Consequently, our targets also have high galactic latitudes.

We remeasured the fluxes of the most prominent emission lines in the SDSS
spectra, which we present in Table \ref{tab:sdssspec1}. We use these to derive
the internal $E(B-V)$, i.e. the average value within the SDSS aperture, and  to
estimate the Oxygen abundance of our targets. When the `auroral'
\oiii$\lambda$4363~\AA\ line is detected with sufficient accuracy (9 galaxies)
we use this line to calculate O/H using the direct temperature sensitive
method. When this line was not detected with sufficient S/N we used two
empirical calibrations from the compilation by \citet{2007A&A...462..535Y}. The
4363 \AA\ line becomes weaker with increasing metallicity and the 5 LARS
targets without a well detected auroral line are consequently the most metal
rich in the sample, according to both of the empirical calibrations used.  The
values found for the Oxygen abundance range from $\sim 0.10-1$~\zsun. 

The \ha\ equivalent widths for the selection were taken from SDSS, obtained for
a 3$\arcsec$ aperture. Most targets fulfilling our selection criteria are
rather compact and aperture effects will be moderate, and in any case our
criteria will guarantee that any target will at least host a central active
starburst.  Nevertheless, with the HST images now in hand, we can recalculate
the \wha\ (corrected for \nii\ assuming the values from the SDSS spectrum to
hold globally). We find that  \wha\ derived from SDSS spectra and HST images
are tightly correlated. Two galaxies, LARS \#10 and 11, which have the lowest
\wha\ according to SDSS, fall below the threshold of 100 \AA\ if measured from
the reduced HST \ha\ images.  In  Table  \ref{tab:sdssspec2}  we provide the
integrated \ha\ equivalent widths for the full sample based on the
a posteriori obtained HST \ha\ images.

In Fig. \ref{fig:sdssspec} we show the SDSS spectrum for LARS \#1 which has one
of the highest \wha\ and LARS \#10 which has the lowest \wha\ as measured in
SDSS spectra. The high \wha\ objects show strong Balmer and \oiii\ emission
lines and featureless blue continua. In the lower \wha\ sources we see in
addition underlying absorption in the higher order Balmer lines and a 4000~\AA\
break.  In Fig. \ref{fig:bpt} we show the  BPT \citep{1981PASP...93....5B}
classification for the LARS sample, indicating that they are primarily powered
by star formation.

\begin{deluxetable*}{ccccccccc}  
\tabletypesize{\scriptsize}
\tablecaption{SDSS spectroscopic remeasurements for LARS sample. \label{tab:sdssspec1}}
\tablehead{
\colhead{ LARS ID} &      \colhead{ \ha} &            \colhead{ \hb }&           \colhead{ \hg} &           
      \colhead{\oii$_{3727}$} &        \colhead{ \oiii$_{4363}$} &      \colhead{ \oiii$_{5007}$} &           \colhead{ \nii $_{6584}$} &         \colhead{ \sii$_{6717}$  }\\
}
\startdata
1       &  2070.6 $\pm$ 23.9 &  668.2 $\pm$ 7.9 &  308.8  $\pm$ 4.6 
&  1512.9 $\pm$ 41.9 &  30.91 $\pm$ 0.06 &  2766.0 $\pm$ 28.3 &    117.5 $\pm$ 2.1 &  144.1 $\pm$ 2.4 \\
2       &    396.4 $\pm$ 4.5 &   129.0 $\pm$ 1.7 &   60.0 $\pm$ 1.2 
&    294.4 $\pm$ 7.9 &   7.33 $\pm$ 0.26 &    580.1 $\pm$ 5.3 &     23.0 $\pm$ 0.6 &   30.4 $\pm$ 0.7 \\
3       &  2650.2 $\pm$ 30.7 &  444.1 $\pm$ 5.2 &  150.4 $\pm$ 2.7 
&   678.0 $\pm$ 10.8 &             --- &    458.6 $\pm$ 5.7 &  1011.6 $\pm$ 10.5 &  319.3 $\pm$ 4.4 \\
4       &  1146.3 $\pm$ 11.0 &  330.0 $\pm$ 4.4 &  147.4 $\pm$ 2.5 
&   817.2 $\pm$ 22.5 &  15.06 $\pm$ 0.11 &  1579.4 $\pm$ 15.6 &     62.1 $\pm$ 1.3 &   80.9 $\pm$ 1.4 \\
5       &  1173.4 $\pm$ 14.5 &  382.6 $\pm$ 4.9 &  167.5 $\pm$ 2.9 
&   652.1 $\pm$ 23.7 &  39.65 $\pm$ 0.05 &  1879.4 $\pm$ 19.0 &     51.9 $\pm$ 1.3 &   60.7 $\pm$ 1.5 \\
6       &    175.6 $\pm$ 2.2 &   ~59.4 $\pm$ 1.2 &   27.7 $\pm$ 0.9 
&    138.4 $\pm$ 5.8 &   5.32 $\pm$ 0.23 &    274.3 $\pm$ 3.5 &      6.4 $\pm$ 0.4 &   14.3 $\pm$ 0.6 \\
7       &  1280.3 $\pm$ 13.9 &  377.5 $\pm$ 4.7 &  158.4 $\pm$ 2.8 
&   729.3 $\pm$ 22.5 &  22.65 $\pm$ 0.11 &  1554.8 $\pm$ 16.8 &    118.1 $\pm$ 2.0 &  105.2 $\pm$ 1.9 \\
8       &    219.1 $\pm$ 2.9 &   ~53.6 $\pm$ 1.1 &   22.0 $\pm$ 0.9 
&    140.1 $\pm$ 6.3 &             --- &    109.2 $\pm$ 1.8 &     57.4 $\pm$ 1.1 &   39.3 $\pm$ 0.9 \\
9       &  1827.2 $\pm$ 17.8 &  493.4 $\pm$ 5.4 &  207.5 $\pm$ 2.9 
&   798.5 $\pm$ 18.3 &  22.03 $\pm$ 0.15 &  2242.6 $\pm$ 20.4 &    199.6 $\pm$ 2.6 &  120.4 $\pm$ 1.8 \\
10      &    225.8 $\pm$ 2.7 &   ~57.0 $\pm$ 1.2 &   21.0 $\pm$ 0.8 
&    140.3 $\pm$ 7.7 &             --- &    107.6 $\pm$ 1.9 &     39.5 $\pm$ 1.0 &   37.6 $\pm$ 0.9 \\
11      &    250.4 $\pm$ 2.7 &   55.3 $\pm$ 1.1 &   19.0 $\pm$ 0.8 
&    116.3 $\pm$ 4.6 &             --- &     51.9 $\pm$ 1.1 &     74.6 $\pm$ 1.2 &   43.0 $\pm$ 1.0 \\
12      &    645.0 $\pm$ 6.6 &  199.6 $\pm$ 2.5 &   84.6 $\pm$ 1.5 
&   358.2 $\pm$ 23.2 &  10.26 $\pm$ 0.41 &    844.6 $\pm$ 8.4 &     58.1 $\pm$ 1.1 &   50.2 $\pm$ 1.0 \\
13      &    277.5 $\pm$ 2.3 &   ~68.4 $\pm$ 1.5 &   28.5 $\pm$ 0.7 
&   151.1 $\pm$ 11.8 &             --- &    170.5 $\pm$ 1.8 &     60.1 $\pm$ 0.8 &   32.0 $\pm$ 0.7 \\
14      &    263.2 $\pm$ 9.1 &   76.5 $\pm$ 1.5 &   33.0 $\pm$ 1.0 
&    119.4 $\pm$ 9.9 &   8.53 $\pm$ 0.78 &    418.7 $\pm$ 4.8 &     10.5 $\pm$ 0.7 &   13.2 $\pm$ 0.7 \\
\enddata
\tablecomments{Line fluxes in units of $10^{-16}$ \egs . Only \oiii$_{4363}$ measurements which were considered reliable are included. }
\end{deluxetable*}

\begin{deluxetable*}{ccccccccccc}  
\tabletypesize{\scriptsize}
\tablecaption{SDSS derived spectral properties for LARS sample  \label{tab:sdssspec2}}
\tablehead{ 
\colhead{ LARS ID} &  \colhead{ z}  &       \colhead{ \wha}  &  \colhead{ \ha/\hb} &  \colhead{ \nii/\ha}  &           \colhead{ E(B-V) } &  
     \colhead{ P}  &  \colhead{ O/H$_{T_{\rm e}}$}  &  \colhead{ O/H$_ {O3N2}$}  &  \colhead{ O/H$_ P$}  & \colhead{ \wha$_{\rm HST}$} \\
& & \colhead{ (\AA )} & & & & & & & & \colhead{ (\AA ) } 
}

\startdata
1       &     0.028 &   560 $\pm$ 6 &  3.099 &   0.057 &  0.089 $\pm$ 0.015 &   
0.709 &   8.070 &       8.242 &    8.241 &  400\\
2       &     0.030 &   326 $\pm$ 4 &  3.073 &   0.058 &  0.082 $\pm$ 0.016 &   
0.725 &   8.041 &       8.226 &    8.251 & 252\\
3       &     0.031 &   277 $\pm$ 3 &  5.968 &   0.382 &  0.688 $\pm$ 0.015 &   
0.410 &     --- &       8.414 &    8.915 & 208\\
4       &     0.033 &   580 $\pm$ 6 &  3.474 &   0.054 &  0.194 $\pm$ 0.015 &   
0.721 &   8.191 &       8.191 &    8.221 & 247\\
5       &     0.034 &   336 $\pm$ 4 &  3.067 &   0.044 &   0.080 $\pm$ 0.016 &   
0.794 &   7.800 &       8.124 &    8.135 & 442\\
6       &     0.034 &   529 $\pm$ 7 &  2.958 &   0.036 &  0.047 $\pm$ 0.022 &   
0.726 &   7.864 &       8.082 &    8.053 & 138\\
7       &     0.038 &   428 $\pm$ 6 &  3.392 &   0.092 &  0.172 $\pm$ 0.015 &   
0.740 &   7.911 &       8.352 &    8.447 & 366\\
8       &     0.038 &   186 $\pm$ 3 &  4.091 &   0.262 &  0.343 $\pm$ 0.023 &   
0.511 &     --- &       8.505 &    8.751 & 111\\
9       &     0.047 &   544 $\pm$ 5 &  3.703 &   0.109 &  0.252 $\pm$ 0.013 &   
0.786 &   8.051 &       8.366 &    8.519 & 265\\
10      &     0.057 &   105 $\pm$ 1 &  3.962 &   0.175 &  0.314 $\pm$ 0.023 &   
0.506 &     --- &       8.505 &    8.577 & ~84\\
11      &     0.084 &   115 $\pm$ 1 &  4.529 &   0.298 &  0.436 $\pm$ 0.021 &   
0.374 &     --- &       8.436 &    8.807 & ~77\\
12      &     0.102 &   445 $\pm$ 5 &  3.232 &   0.090 &  0.128 $\pm$ 0.015 &   
0.760 &   8.008 &       8.342 &    8.437 & 489\\
13      &     0.147 &   264 $\pm$ 2 &  4.056 &   0.216 &  0.335 $\pm$ 0.022 &   
0.600 &     --- &       8.503 &    8.669 & 204\\
14      &     0.181 &  ~700 $\pm$ 31 &  3.442 &   0.040 &  0.185 $\pm$ 0.037 &   
0.824 &   7.823 &       8.055 &    8.231 & 927\\
\enddata
\tablecomments{Derived quantities from the SDSS spectra. \wha\ is the
equivalent width of \ha\ remeasured from the SDSS spetra.  P=\oiii / (\oii +
\oiii ) is similar to the excitation parameter.   \nii/\ha is the \nii
$_{6584}$/ha\ value, used to correct the HST narrowband \ha\ images for \nii\
contamination, and to estimate the metallicity.  O/H$_{T_{\rm e}}$ is the
oxygen abundance in units of 12+log(O/H), determined with the direct method
utilising the temperature sensitive \oiii$_{4363}$ line.  O/H$_ {O3N2}$  is the
oxygen abundance derived by the empirical O3N2 relation, and  O/H$_ P$ the
R$_{23}-P$ relation, both from \citet{2007A&A...462..535Y}.  The last column
give the integrated \wha\ derived from the continuum subtracted and \nii\
corrected HST \ha\ images (see Paper\,II, Hayes et al. 2014, for details). Note
that this integrated value is in two cases smaller than the selection threshold,
which was based on the value within the SDSS aperture.}
\end{deluxetable*}

\begin{figure}  
\centering
\includegraphics[angle=0,scale=0.56]{Lars_sdss_bpt_n2.pdf}
\caption{BPT type classification of the LARS target galaxies compared to the
SDSS DR7 spectroscopic database from MPA-JHU\footnote{URL:
http://www.mpa-garching.mpg.de/SDSS/DR7/} and the starburst-AGN dividing lines
according to Kewley (2001) and Kaufmann (2003). All targets appear dominated by
star formation, with LARS\#3 being rather close to the 'Liner' region. } 
\label{fig:bpt}
\end{figure}

\begin{deluxetable*}{ccccccccc}[t!]
\tabletypesize{\scriptsize}
\tablecaption{LARS imaging exposure times (in seconds).\label{tab:imgobs}}
\tablehead{
\colhead{{\em bandpass alias}} & 
		\colhead{FUV-1} & 
		\colhead{FUV-2} & 
		\colhead{FUV-3} & 
		\colhead{$U$} & 
		\colhead{$B$} & 
		\colhead{$i$} & 
		\colhead{\hb } & 
		\colhead{\ha }  \\
\colhead{{\em instrument/camera}} & 
		\colhead{ACS/SBC} & 
		\colhead{ACS/SBC} & 
		\colhead{ACS/SBC} & 
		\colhead{WFC3/UVIS} & 
		\colhead{WFC3/UVIS} & 
		\colhead{WFC3/UVIS} & 
		\colhead{WFC3/UVIS} & 
		\colhead{WFC3/UVIS}  \\
\colhead{{\em spectral element}} & 
		\colhead{F125LP} & 
		\colhead{F140LP} & 
		\colhead{F150LP} & 
		\colhead{F336W} & 
		\colhead{F438W} & 
		\colhead{F775W} & 
		\colhead{F502N} & 
		\colhead{F656N}  \\ \\
\colhead{{\bf LARS ID \#}} & 
		\colhead{} & 
		\colhead{} & 
		\colhead{} & 
		\colhead{} & 
		\colhead{} &  
		\colhead{} & 
		\colhead{} &
		\colhead{}  
		}
\startdata
01 & 1900 & 1807  & 1610  & 920 & 800  & 600  & 1900 & 800  \\
02 & 2012 & 1840  & 1740  & 2000 & 1230  & 1208  & 3920 & 1265  \\
03 & 2002 & 1984  & 1750  & 1318 & ~1440$^{ a}$  & ~840$^a$  & 3030 & 1200  \\
04 & ~2648$^b$ & ~2648$^b$   &  ~2648$^b$ & 1004 & 800  & 620  & 2000 & 805  \\
05 & ~2669$^b$ & ~2669$^b$  & ~2669$^b$  & 1020 & 850  & 620  & 2000 & 871  \\
06 & 1886 & 1850  & 1650  & 1840 & 1600  & 840  & 2852 & 1600  \\
\hline \\
  &  &  & &  &  &  &  ACS/WFC &  ACS/WFC \\
&  &  & &  &  &  & F502N &  FR656N \\
 \\
07 & 1872 & 1806  & 1540  & 900 & 800  & 534  & 1467 & 720  \\
08 & 1870 & 1768  & 1550  & 900 & 800  & 519  & 1452 & 720  \\
\hline \\
 &  &  & &  &  &  &  FR505N &  FR716N \\
 \\
09 & 1900 & 1780  & 1538  & 900 & 800  & 534  & 1440 & 708  \\
10 & 1872 & 1803  & 1543  & 900 & 800  & 534  & 1440 & 708  \\
\hline \\
 &  &  & &  &  &  &  FR551N &  FR716N \\
\\
11 & 1900 & 1688  & 1490  & 900 & 800  & 519  & 1457 & 720  \\
12 & 1545 & 1300  & ~2640$^c$  & 1020 & 800  & 604  & 1640 & 740  \\
\hline \\
&  &  & & F390W & F475W & F850LP & FR551N  &   FR782N \\
 \\
13 & -- & 2711  & ~2580$^d$  & 900 & 800  & 519  & 1350 & 720  \\
14 & -- & 2805  & ~2610$^c$  & 1239 &1098  & ~2274$^d$  & 2472 & 2340$^d$  \\
\enddata
\tablecomments{This table specifies the HST imaging observations that are used
in LARS to produce \lya\ images and other data products. Left column gives the
LARS ID number and the other columns the exposure times (in seconds) for each
filter in question. Each column is labelled with the bandpass shorthand
abbreviation, the instrument/camera and spectral element used. The LARS ID
number increases with redshift and hence the filters used to capture \ha\ and
\hb\ differs as we go to higher ID numbers, as indicated. The other filters
stay the same except for  LARS \#13 and \#14 where also the $U,B,I$ filters
have changed.  Each exposure was split in 2-4 sub exposures. Entries with
superscripts come from the HST archive: {\em a}: Instead of WFC3 we use F435W
and F814W taken with ACS/WFC under program 10592 (PI Evans); {\em b}: Data from
program 11110 (PI McCandliss), {\em c}: Data from program 11107 (Heckman), {\em
d}: Data from program 10920 (Heckman).}
\end{deluxetable*}

\section{LARS HST imaging data}\label{sect:imdata}

Twelve of the LARS galaxies (LARS \#1 to \#12) fall in the low-$z$ window. As
mentioned above, \lya\  is here captured by the F125LP and F140LP filter
combination. We also include F150LP which combined with F140LP gives
information on the FUV continuum slope although the spectral baseline is small.
The F125LP filter omits the bright geocoronal \lya\ background, but includes
O{\sc i}$\lambda 1302$\AA\ and thus still has the highest background of the SBC
filters. Therefore the observations in this filter were placed in the middle of
the orbits when HST is in earth shadow. This dramatically  reduces the airglow,
and indeed, for those COS spectra of LARS targets that were obtained when HST 
was in earth shadow, the O{\sc i}$\lambda 1302$\AA\  line is not visible.

The additional filters required for modeling the continuum near \lya\ should
sample both sides of the 4000~\AA\ break as well as longer wavelengths that are
sensitive to any older underlying population.  For young stellar populations,
even the broad band fluxes may be dominated by nebular line and continuum
emission \citep{2002A&A...390..891B}. While this can be included in the
modeling, it relies on the assumption of the nebular and stellar emission being
co-spatial. For a well resolved galaxy, this is often not the case as the
nebular emission is often seen to be concentrated to shells and filaments.
Therefore, for the current application where spatially resolved \lya\
photometry is the main objective, it is clearly preferable to estimate the nebular
contribution from an \ha\ image. The Hydrogen emission lines and the nebular
continuum can then be readily modeled in a way that is only (sensitively)
dependent on the reddening.  Other strong emission lines, notably those of
\oiii , depend more on the physical parameters of the nebulae, many of which
are unknown. Hence we chose the complementary broad band filters in such a way
that the \oiii\ lines at 4959 and 5007 \AA\ are not transmitted, and if
possible, we have also avoided the \oii\ doublet at 3727 \AA . The optimum
filter set for the low-$z$ window is WFC3/UVIS F336W, F438W and F775W.  

\begin{figure}[t!] 
\centering
\includegraphics[angle=0,scale=0.475]{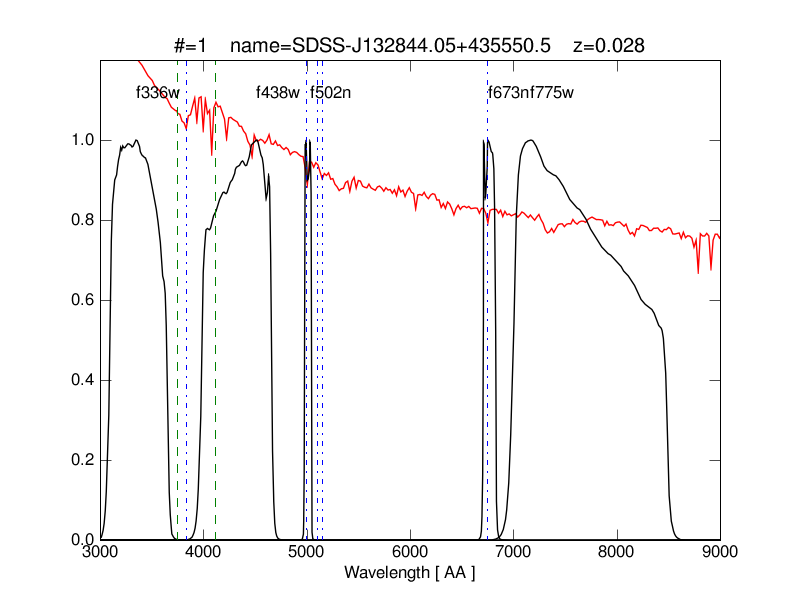}
\caption{WFC3/UVIS filter setup used for LARS target \#1, together with an
arbitrary chosen SB99 model spectrum.  Vertical blue dash-dot lines show
redshifted wavelengths of the \oii , \hb , \oiii, and \ha\  emission lines.
Vertical green dashed lines show the location of the Balmer and  4000\AA\
breaks. } 
\label{fig:uvisfilters}
\end{figure}

By also including \hb\ imaging we can perform a \lya\ continuum subtraction
that is not dependent on assumptions on how the nebular reddening relates to
that of the stellar population -- instead, we can test how \lya\ emission
depends on this vital parameter.  The \ha\ and \hb\ filters were obtained
through WFC3/UVIS narrow band filters whenever one that matched the wavelength
of the redshifted Hydrogen lines (while not including the \oiii\ lines near \hb
) existed. When there was no matching conventional narrow band filter in WFC3
or ACS, we used instead the  2\% wide tunable linear ramp filters (LRF) in
ACS/WFC. The filters used to observe \ha\ generally partly transmits also the
adjacent \nii\ lines (which we use the SDSS spectra to correct for assuming the
values in Table \ref{tab:sdssspec2} to hold globally\footnote{We are in the
process of acquiring ground based spatially resolved \nii/\ha\ maps to study
the validity of this assumption and which will be used in future LARS data
releases.}).  In Fig. \ref{fig:uvisfilters} we show the near-UV/optical band
pass selection for LARS\#1 together with the placement of strong emission
lines. 

For the higher redshift targets, we used F140LP and F150LP with SBC, and the
near-UV/optical filters chosen were WFC3/UVIS/F390W, F475W, F850LP.

The LARS imaging (except for archival data) was executed between October 2010 and February 2012. All
imaging observations are presented in Table~\ref{tab:imgobs}
 
\begin{figure*}[t!] 
\centering
\includegraphics[angle=0,scale=0.9]{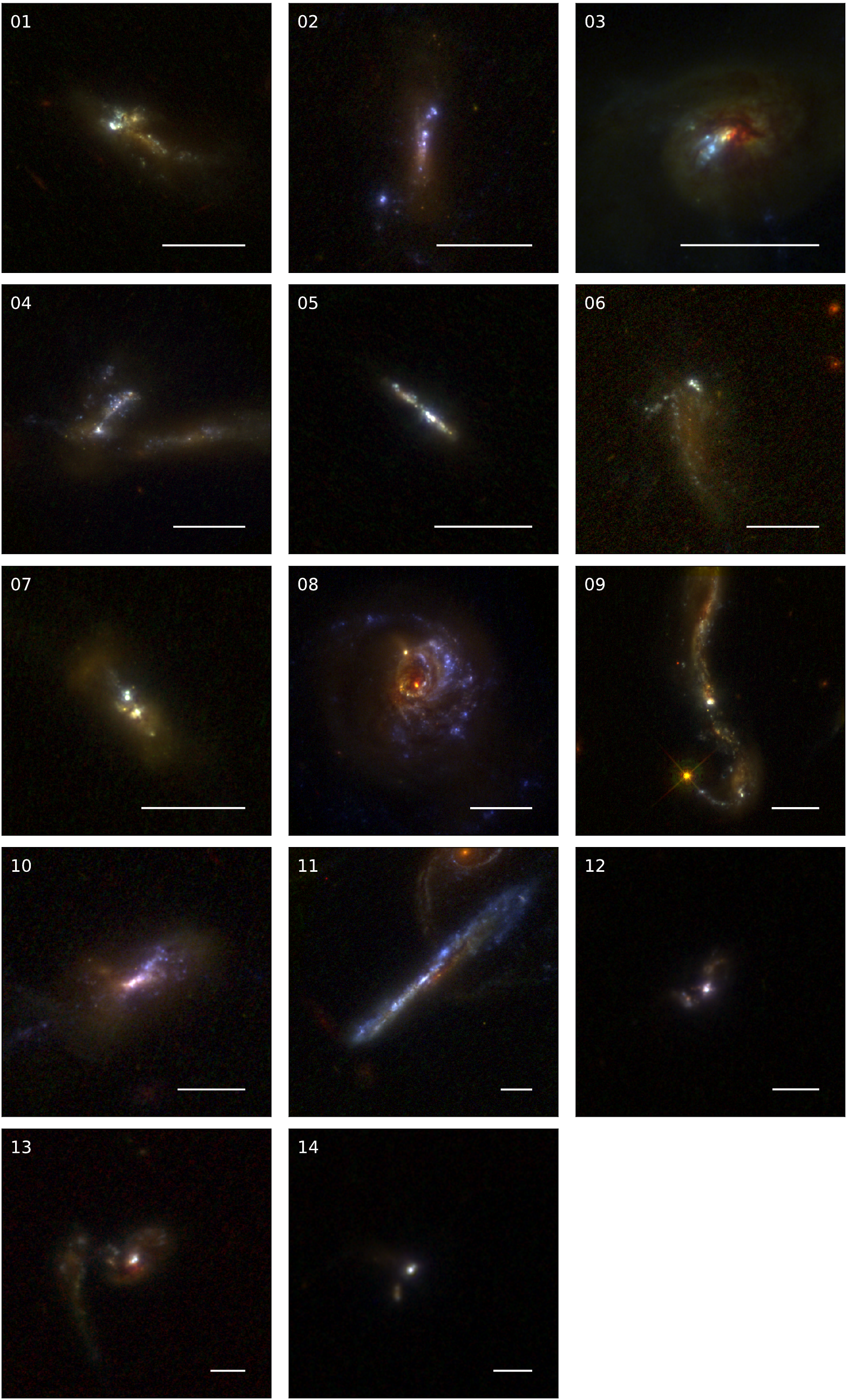}
\caption{The full LARS sample as seen by HST. RGB composite made from FUV, $B$ and $i$-band.
North is up and East to the left.  White bars in the lower right corner shows a constant physical size of 5\,kpc.} 
\label{fig:sample_hst}
\end{figure*}

\section{Image  processing and  \lya\ image generation}\label{sect:obs}
\subsection{Reductions}\label{sect:reduc}
Images were drizzled onto a common pixel scale and orientation with the
MultiDrizzle tools in STScI/Pyraf.  For this we adopt the 0.04 arcsec/pixel
scale of WFC3/UVIS,  with which most optical data was obtained and which is
intermediate between ACS/SBC and ACS/WFC. Additional tweaking of the alignment
between images obtained in different visits was then performed using well
centred star clusters and the {\sc geomap/geotran} tasks in {\sc Pyraf}.

Corrections for Charge Transfer Efficiency (CTE) losses in ACS were performed
using the tools developed by Anderson and Bedin (ISR ACS
2010-03)\footnote{http://www.stsci.edu/hst/acs/software/CTE}. The WFC3 camera
is newer and had not suffered much radiation damage at the time when the LARS
imaging was being carried out (LARS imaging begun 19 months after WFC3 was
installed during SM4), hence CTE performance is better.  When the first
LARS imaging data processing was carried out, no CTE correction tool had been
released for WFC3, but it will be included in our future LARS public data
release. 

On analyzing the HST imaging results and comparing with spectroscopy we found a
small systematic difference in the flux calibration of SBC images and COS
spectra which on further investigation of archival data is found to arise from
the SBC zero points.  We recalibrated these data by using combined SBC imaging
and prism spectroscopy of the globular cluster  NGC\,6681, see Appendix A.2 for
details. 

The SBC images have very low background, especially since the F125LP
observations where done in SHADOW mode when the telescope is in earth shadow.
Remaining background were removed by estimating its level near the corners of
the camera, far from the sources. The background in the filters obtained with ACS/WFC 
or WFC3/UVIS were removed in the same manner.

{\bf PSF matching:}
Matching of the point-spread function (PSF) was also performed. If the fraction
of energy received from a point source in the central pixel varies from filter
to filter we would get spurious results on small spatial scales. PSF matching
is thus necessary because the \lya\ extraction software uses multiple
bandpasses to model the continuum near \lya\ on a pixel to pixel (or spaxel)
level in order to obtain a spatially resolved \lya\ image. It is particularly
important for matching the FUV SBC data to the optical data, since the SBC PSF
shape is very different from that of  ACS/WFC and WFC3/UVIS. Ordinarily this would be
done by building empirical PSFs from point-sources located within the field,
but the  small field of view and far UV imaging observations preclude this
method. As such we rely upon synthetic PSFs generated for each of our
observational setups with TinyTim. We measure the FWHM of the synthetic PSF
generated for each bandpass and use the broadest PSF image as the reference. We
use the {\sc Pyraf/psfmatch} task to generate the convolution kernel that will
smear the PSF of each image to the best possible representation of that of the
reference image. 

This process is sub-optimal, and from simulations we note a
relative flux residual error in the core of objects at the 20\% level. 
We note also that this experiment has been performed on synthetic data. 
For future LARS data
releases we are in the process of developing an improved scheme for PSF
matching such that the results can be trusted down to the individual pixel
level (Melinder et al. in prep).

{\bf Voronoi tessellation:}
As shown by \citet{2009AJ....138..911H}, reliable continuum subtraction
requires a local S/N of 5 or greater. The S/N per pixel in the centre of LARS
targets easily exceeds this allowing data analysis at full HST resolution
(modulo the PSF matching effects). In the faint outer regions, on the other
hand, the S/N per pixel may be well below one.  The procedure of Voronoi
tessellation -- adaptive binning -- enables us to maintain the spatial
resolution when S/N is high, and where S/N is low aggregate pixels together to
improve local S/N. For this we used the WVT binning algorithm by
\citet{Diehl2006}, which is a generalization of the Voronoi binning algorithm
of \citet{Cappellari2003}. We produce a binning pattern mask from the F140LP
image requiring a local S/N of 10 and a maximum bin size of 625 pixels (1
square arcsec). The minimum bin size is artificially set to 4 pixels  (i.e.
larger than the HST PSF FWHM in any of the filters used) in order
to mitigate effects of imperfect PSF matching. The binning pattern defined by
the F140LP image is then applied to the other filters. Since massive star
formation tends to be concentrated to the knots, while the underlying stellar
distribution is distributed on larger scale, the longer wavelength broadband
filters tend to have S/N larger than F140LP except in the smallest spaxels.

\begin{figure*}[t!] 
\centering
\includegraphics[angle=0,scale=0.85]{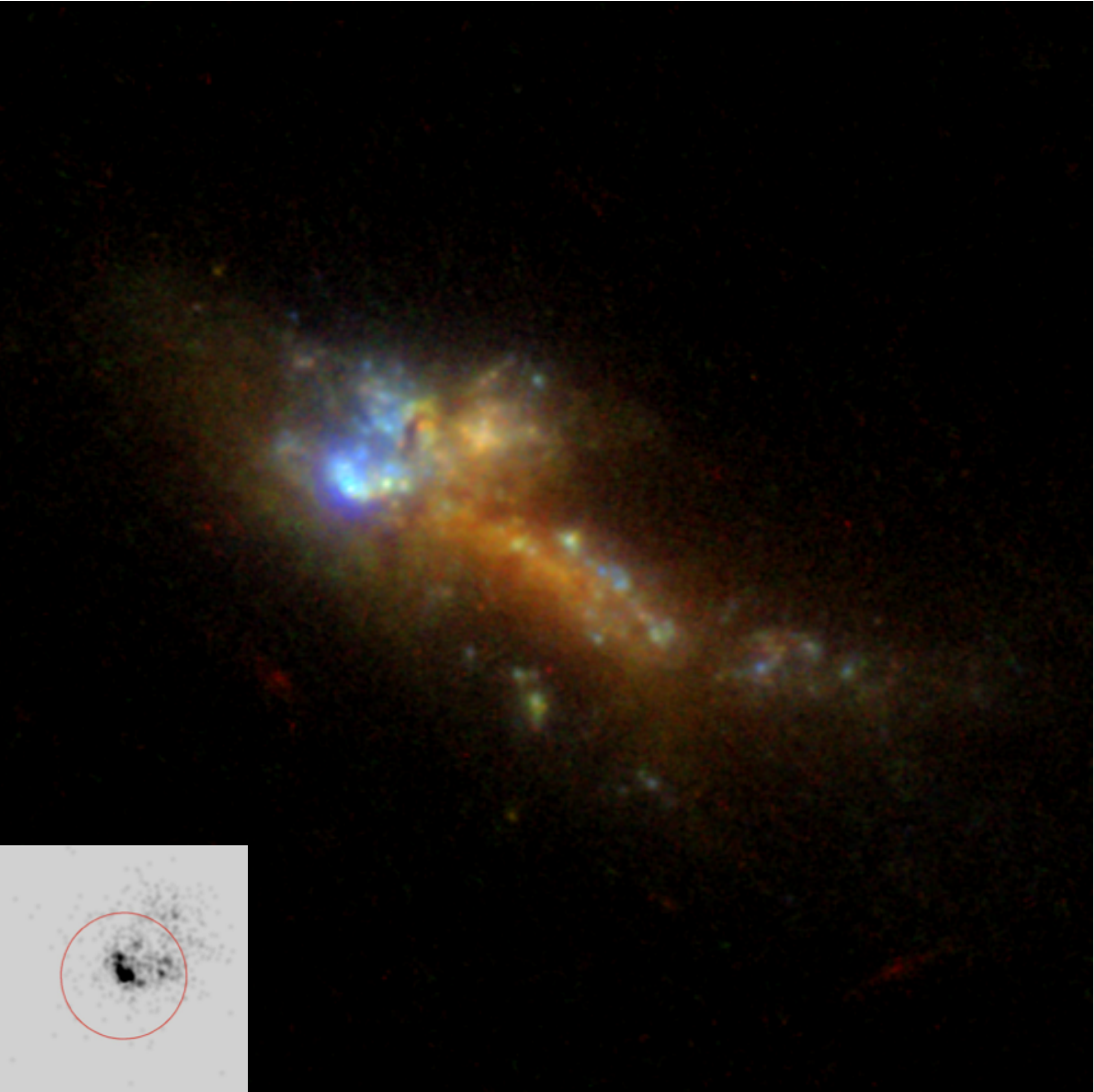}
\caption{Mrk\,259 in false color, showing the F140LP filter (FUV) as blue,
F438W ($B$) as green, and F775W ($i$) as red.
FOV$=21\arcsec.5\times21\arcsec.5$ ($12\times12$kpc), north is up, east is
left.  The inset on the lower left shows the COS NUV acquisition image with the
COS spectroscopy aperture (diameter 2.5\arcsec) indicated with the red circle. 
The COS aperture was placed on the brightest UV (blue) knot. } 
\label{fig:rgb}
\end{figure*}

\begin{figure*}[t!] 
\centering
\includegraphics[angle=0,scale=0.75]{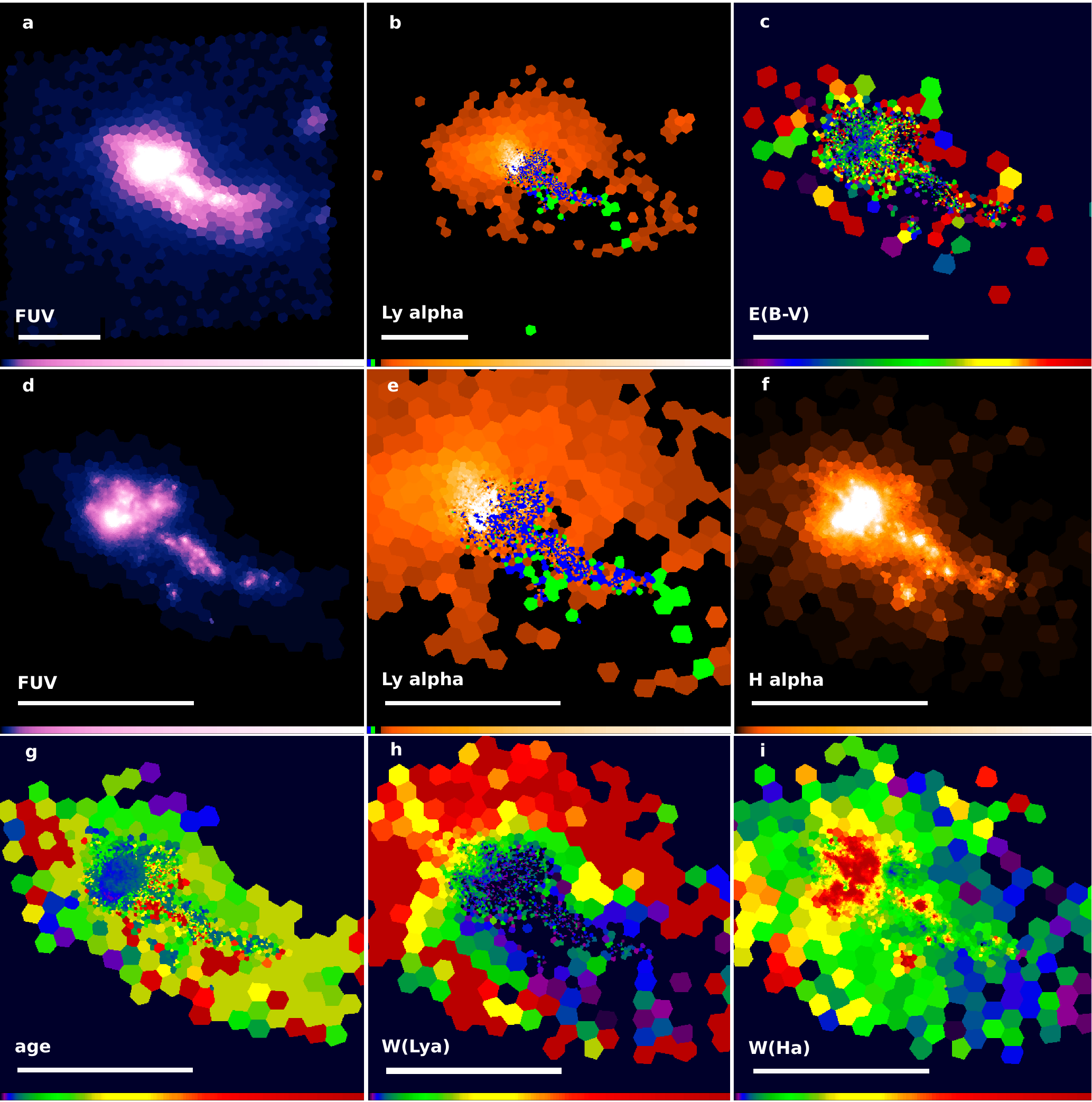}
\caption{ Voronoi tessellated images of Mk\,259. For all panels the scale bar
has a length of 10\arcsec\ (or 5.6 kpc) and north is up, east is left. The zero and
maximum intensity limits for each panel is indicted in square brackets. Panels
{\bf a} and {\bf b} have a size of $40\arcsec\times 40\arcsec$ while the other panels
are zoomed in to $20\arcsec\times 20\arcsec$.
 {\bf Upper left (a):}  Far UV (F140LP) continuum in logarithmic intensity scaling  [0,$1.25]\times10^{-16}$  \egs/arcsec$^2$.  
{\bf Upper middle (b):}  \lya\ line intensity in log scaling [-0.2, 6.25]$\times10^{-14}$ \egs/arcsec$^2$. Black regions corresponds to zero,
while green show (slight) and blue (more) absorption.  
{\bf Upper right (c):} E(B-V) based on \ha / \hb\ in linear scale from [0,0.5]. 
{\bf Middle left (d):}   Zoom in on the far UV continuum in logarithmic scaling  [0,3] $\times10^{-16}$  \egs/arcsec$^2$. 
{\bf Center (e):}   \lya\ as above but zoomed in (same intensity scale).
{\bf Middle right (f):} \ha \ intensity in log scaling [0,2]$\times10^{-14}$ \egs/arcsec$^2$.  
{\bf Lower left (g):}   Age of young population in log scale, [1,60] Myr. 
{\bf Lower middle (h):} \wlya\  in log scale [0,900]\AA . Black regions have zero or negative \wlya .  
 {\bf Lower right (i):} \wha\ in log scale [0,900] \AA .
For panels c and g regions fainter than \mulya$=25.5$ \msqa\ have been masked out, while for panels h and i the mask level is  $26.2$ \msqa .} 
\label{fig:3x3}
\end{figure*}
 
\subsection{Continuum subtraction -- LaXs, generation of \lya\ images}\label{sect:contsub}
As discussed in \citep{2005A&A...438...71H,2009AJ....138..911H}, the production
of continuum subtracted \lya\ images from raw ACS/SBC data is a non-trivial
process because of the broad nature of the bandpass that captures \lya. As
such, it becomes necessary to pay close attention to the behaviour of the
continuum at all wavelengths sampled by F125LP filter (or F140LP for the high-z
window).  In order to estimate this continuum behaviour we need an accurate
picture of the age of the stellar population that dominates the light output at
$\lambda < 1500$\AA\ and the attenuation it suffers due to dust.  Thus we need
a spatially resolved spectral energy distribution (SED) fitting tool, and a
data set that has enough wavelength sampling points across the UV continuum
(sensitive to both age and dust) and the 4000\AA\ break (particularly age
sensitive) to recover these two quantities in a non-degenerate fashion. Unlike
in most applications of SED fitting, our galaxies are well spatially resolved
and we are able to resolve composite populations including young stars, their
associated nebulae, and underlying stellar populations. These additional
populations of gas and old stars will not contribute significantly to the
continuum in F125LP, but they may be substantial and even dominate our
observational determination of the 4000\AA\ break (restframe $U-B$), biasing
our fit. To mitigate the influence of an underlying population and nebular gas
we include a longer wavelength sampling point ($i$ band) and the direct
measurement of the nebular contribution through an \ha\ image.

The continuum subtraction methodology, including composite stellar components
and nebular gas is presented in \citet{2009AJ....138..911H}.  In short we convolve
theoretical/model SEDs with the HST filter response function given by SYNPHOT 
 (which includes any red leaks). First we generate the best possible nebular gas SED,
compute its contribution to every filter and  subtract it from the observed
data-points. Then we perform a two component stellar SED (using one young and one 
old population) fit to the broadband data
points. We then reconstruct the contributions of the various components to the
F125LP bandpass and subtract it from the observation, leaving us with net \lya\
line fluxes, either in emission or interstellar absorption against the UV
continuum. 

The continuum subtraction is implemented by the \emph{LARS eXtraction software (LaXs)}, which
performs a full stellar population analysis in each pixel and reconstructs the
predicted flux due to continuum processes in F125LP, and also the narrowbands
that sample \halpha\ and \hbeta.  For the template libraries we use the
\emph{Starburst99} population synthesis models
\citep{1999ApJS..123....3L,2005ApJ...621..695V}. In \citet{2009AJ....138..911H}
we showed assumptions about the IMF to have a negligible impact upon our
ability to recover \lya\ quantities, but that stellar metallicity needed to be
known to within a factor of about 2. For the first generation of LARS images we
assume a Salpeter IMF between the mass limits of 0.1 and 100\msun\ and stellar
metallicity of $Z=0.008$.  We assume that each spaxel can be approximated by
an instantaneous burst (simple stellar population, i.e. characterized by just
its age) together with an underlying old component.  We first take the optical
broadband data and perform a rapid 1-component SED fit which we use to
continuum-subtract the \halpha\ (correcting for the contribution of \nii\ using
the SDSS values and the filter transmission at their redshifted wavelengths)
and \hbeta\ narrowband observations. Since at low nebular equivalent widths
these can be strongly affected by stellar absorption, we correct for this by
performing a multi-component SED fit, and measure the intrinsic stellar
absorption in \halpha\ and \hbeta\ from the high-resolution templates
\citep{2005A&A...436.1049M}. With the (stellar absorption corrected) \halpha\
and \hbeta\ fluxes we calculate the nebular dust attenuation (parameterized by
\ebv\ assuming the SMC law \citealt{Prevot84}).  As shown by
\citet{Mas-Hesse1999} and discussed in \citep{2005A&A...438...71H}, we expect
the strong UV radiation field in sources such as these to destroy the bulk of
the graphite molecules, leaving a dust composition without the absorption bump
at $\approx 2200$\AA.  

From dust corrected \halpha\ we generate the intrinsic nebular continuum
spectrum from points presented in \citet{1984ASSL..112.....A}, and add nebular
lines according to the prescription of \citet{2003A&A...401.1063A}. The whole
nebular spectrum is then reddened by the inferred \ebv. Note that this is the
same prescription for adding nebular emission as adopted by
\citet{2009A&A...502..423S,2010A&A...515A..73S}, and allows us to first order
to account for the effect of weak emission lines falling within our broadbands. After
convolving this scaled nebular spectrum with our broadband filter profiles we
subtract the gas spectrum from all our images leaving us with ``true" stellar
emission images. We then perform a two component stellar SED fit to the data
and by the same methods, produce a map of the estimated flux due to purely
continuous processes in F125LP.  This is subtracted from the observation
leaving us with the \lya-only image. 

The fitting procedure is based upon a brute-force minimization of the $\chi^2$
statistic, with four free parameters;  $\chi^2$ is defined as: \begin{equation}
\chi^2 = \sum _i  \frac{d_{\mathrm{stel},i} - [ n_{1,i}\cdot
m_{1,i}(\mathrm{age},E_{B-V}) + n_{2,i}\cdot m_{2,i} ]} {\sigma_i^2}
\end{equation} where subscript $i$ refers to the $i^{th}$ data-point
(corresponding, in the low redshift case to filters F140LP, F150LP, F336W,
F439W and F775W) $d_\mathrm{stel}$ refers to the observational data-points that
have had had their nebular component subtracted, $\sigma_i$ refers to the
observational error on $d_\mathrm{tot}$. Furthermore, $n$ refers  to the
normalization and $m$ the modeled fluxes, and subscripts 1 and 2 refer to the
young and old stellar components respectively. Synthetic fluxes in the young
population are a function of age, and stellar attenuation. Thus the free
parameters in the fit are the normalization of the two stellar populations, and
the the age and dust attenuation in the young population.  Since we have five
datapoints included in the $\chi^2$ fit, we have limited the number of free
parameters to four in order not to overfit the problem. Therefore, the age of
the old population was fixed at age of 4 Gyr and is assumed to have the same
reddening as the young population. While these last two constraints are not
very well motivated on scientific grounds, they have quite limited impact on
the derived \lya\ properties -- the important aspect is that one red population
should be included to account for possible underlying populations which could
otherwise affect the 4000-break.


\begin{figure*}[t!] 
\centering
\includegraphics[angle=0,scale=0.91]{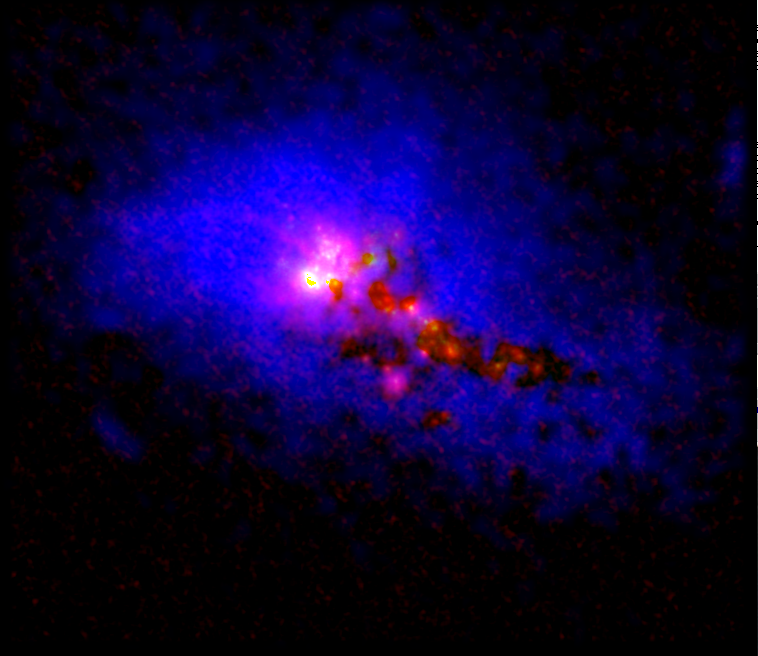}
\caption{Mrk\,259 in false color with arcsinh intensity scaling and cuts levels
set to show diffuse \lya .  \ha\ is shown in red, FUV continuum in green and
\lya\ in blue.  The \lya\  image is based on the non-binned output from LaXs
and has been adaptively smoothed to suppress noise on large scales while
preserving small scale detail.  FOV$=30\arcsec.4\times26\arcsec.8$
($17\times15$ kpc)} 
\label{fig:rgb_fa}
\end{figure*}


\section{First results for LARS\#1 (Markarian\,259)}\label{sect:obj}

To demonstrate the potential of the LARS program, in this section we present 
some early results for the first and nearest galaxy in the sample: LARS\#1,
also known under the common name Markarian\,259.  Its optical appearance as
seen by SDSS is shown in Figure \ref{fig:selimages} together with its SDSS
spectrum in Fig. \ref{fig:sdssspec}. Measured spectral information is
summarised in Tables \ref{tab:sdssspec1} and \ref{tab:sdssspec2}. For the
assumed cosmology the scale is 573 pc/arcsec.

In the the SDSS spectrum we are able to identify and measure a number of
optical emission lines relevant for studying the fundamental nebular properties
of Mrk\,259, which are listed in Table~\ref{tab:sdssspec2}.  From these lines
we are able to determine a slight nebular reddening, measuring \ebv=0.089
magnitudes based upon  \halpha/\hbeta .
We de-redden the  observed line fluxes and derive an electron temperature of
12,400~K, and a nebular oxygen abundance of $12 + \log_{10} (\mathrm{O/H}) =
8.07$. 
At 4817~\AA, we identify Wolf-Rayet features, redshifted from 4686~\AA\ implying
the presence of very young stars, 2-5 Myr, see \citet{Wofford2013}. The lack of a strong
4000\AA\ break also points to a very young dominating cluster without any
significant contribution from an evolved underlying population within the SDSS 
fiber aperture.

\begin{figure*}[t!] 
\centering
\includegraphics[angle=0,scale=0.451]{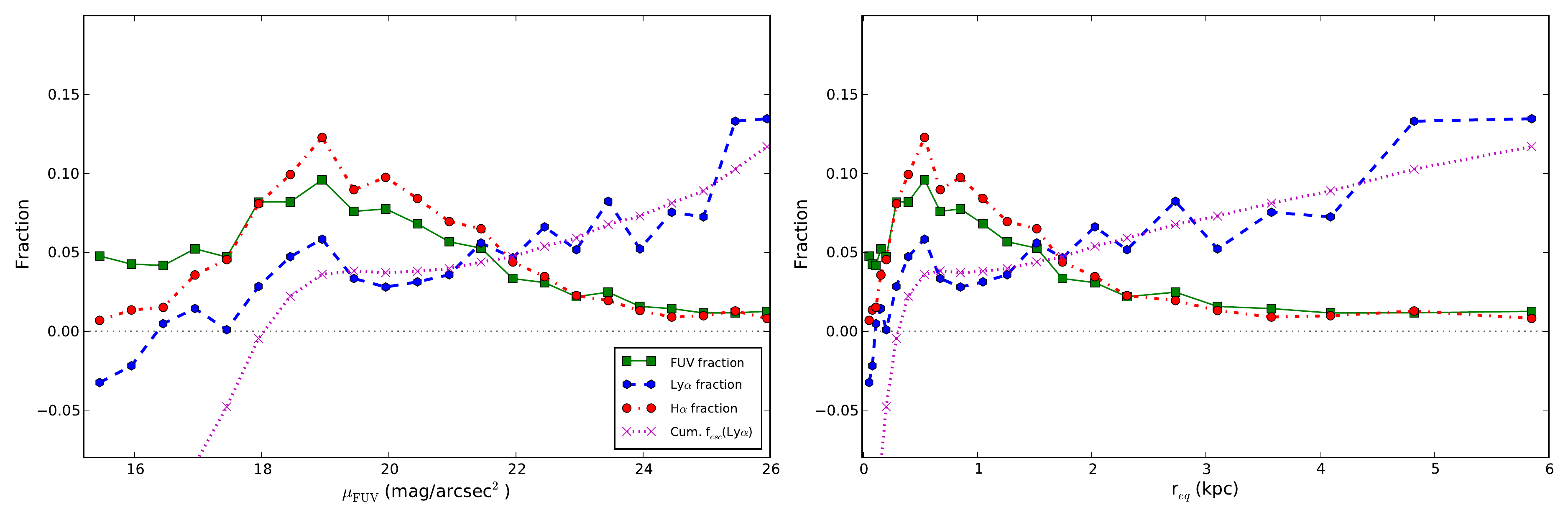} 		
\caption{Escape fraction and spatial distribution of \lya , \ha , and FUV
continuum.  The \emph{left} panel shows the relative flux distribution of \lya
, \ha , and FUV continuum in \mrk\ as a function of FUV surface brightness (AB magnitude
per arcsec$^2$) as in \"Ostlin et al. (2009). The green squares show the fraction of the total UV flux contained in each half
magnitude bin ranging from $\mu_{\rm FUV}= 15.2$  to 26.2 AB mag/arcsec$^2$.
The red circles similarly show how the \ha\ flux is distributed in regions with
different $\mu_{\rm FUV}$ and the blue circles show the same for \lya. The
magenta crosses show the cumulative escape fraction obtained when integrating
summing over all brighter UV resolution elements.  In the \emph{right} plot we
show the same quantities but as a function of equivalent radius
($r_{eq}=\sqrt{area/\pi}$ of the isophotal areas used in the \emph{left} panel).  } 
\label{fig:fesc}
\end{figure*} 

\begin{figure*}[t!] 
\centering
\includegraphics[angle=0,scale=0.5]{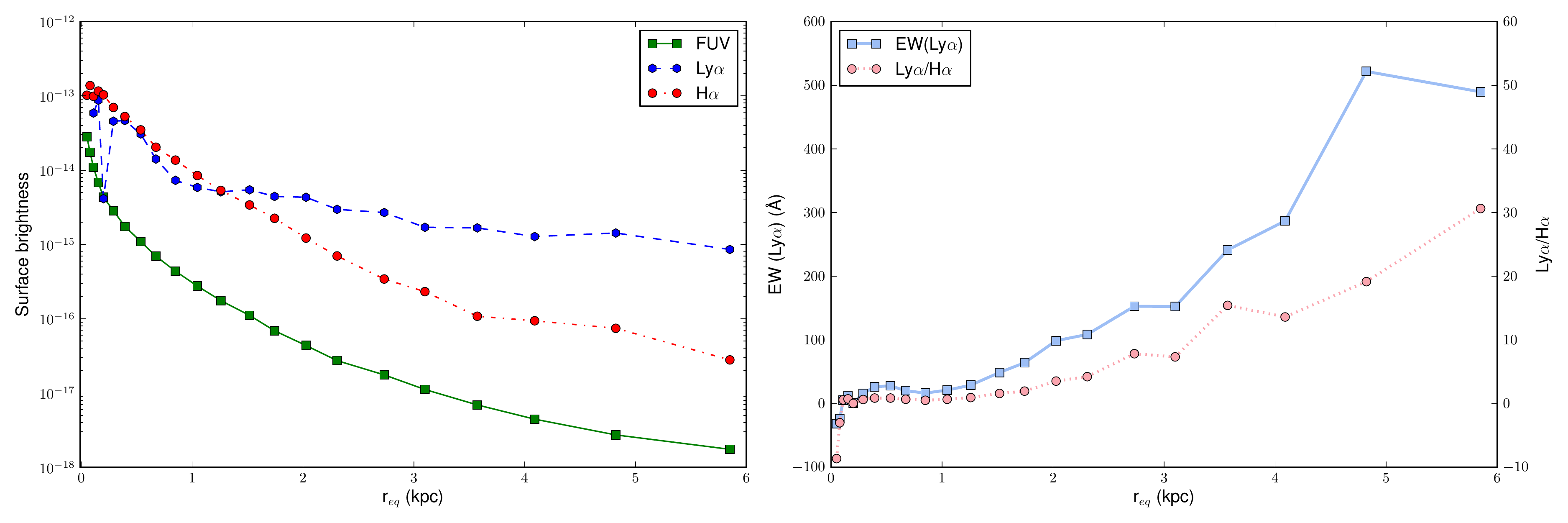}
\caption{The \emph{left} panel shows the surface brightness distribution of \mrk\ obtained by isophotal
integration. Green shows logarithm of the FUV continuum (in units of \egsa ),
while red and blue shows the logarithm of \ha\ and \lya\ respectively (both in
\egs).  The entire FUV profile is very well fitted by Sersic profile with
$n=4$ and \ha\ with $n=2$. Except for the inner parts \lya\ has an exponential
behavior ($n=1$). In the \emph{right} panel, the blue squares shows \wlya\ as a function of
radius, while the red circles shows \lya/\ha\ as a function of radius. } 
\label{fig:surf}
\end{figure*}

\begin{figure*}[t!] 
\centering
\includegraphics[angle=0,scale=0.753]{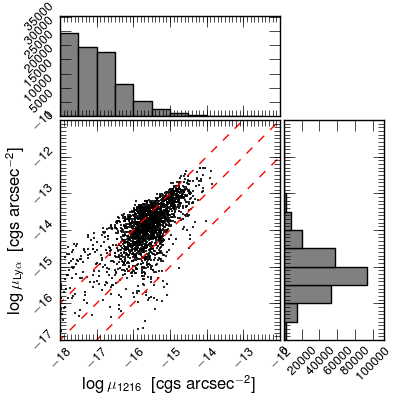}
\includegraphics[angle=0,scale=0.753]{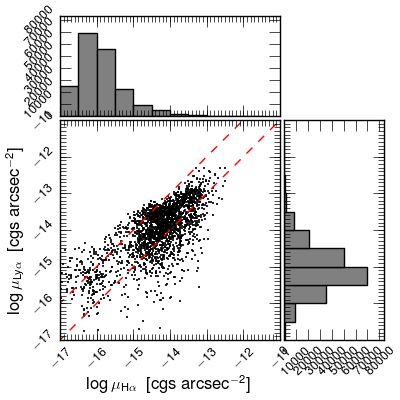}
\caption{Scatterplots showing the surface brightness in  \lya , FUV and \ha\ of
individual spaxels. The units are erg/s/cm$^2$/$\square\arcsec$ for \lya\ and
\ha, and  erg/s/cm$^2$/\AA $\square\arcsec$ for the FUV continuum. The \emph{left}
plot shows \lya\ vs FUV and the dashed lines correspond to constant values of
\wlya : 1, 10 a,d 100 \AA . The \emph{right} plot shows \lya\ vs \ha\ where the lower
line correspond to \lya = \ha\ and the upper line to \lya=9\ha , i.e. the
theoretical prediction from case B recombination and no reddening.  The spaxels
represent different areas, and in FUV faint areas they are made from summing up to
625 pixels - the histograms show the number of pixels at each brightness level.
Note that since these plots show logarithmic quantities, only spaxels with
positive flux are included.  } \label{fig:scat}
\end{figure*}

\subsection{LARS images of Mkn\,259}
In Fig. \ref{fig:rgb} we show the appearance of LARS\#1 in HST images with
F336W in blue, F438W in green and F775W in red.  The galaxy has an integrated
absolute magnitude of $M_B=-19.2$ and possesses a highly irregular morphology,
suggesting that it may have been produced in a small galaxy merger. Lots of
compact star clusters are seen like in the nearer targets studied in
\citet{2009AJ....138..923O} and come in different colors suggesting a spread in
ages. The brightness peaks NE of the center where we see a very blue UV bright
region. The geometrical centre is redder containing many dust absorption
features.

In Fig. \ref{fig:3x3} we show some of the LaXs output results for LARS\#1. Each
sub panel shows the scale with a white 10\arcsec\ long bar. In the upper left
panel (a)  we show the FUV image, where the Voronoi tessellation pattern is
obvious in the outskirts of the galaxy. In the upper central panel (b) we show
the resulting \lya\ image in the same spatial scale: emission is coded in red
to white, black shows close to zero emission, while spaxels where \lya\ is in
absorption is shown in green and blue.  Obviously, the asymmetry seen in the
UV/optical is even more pronounced in \lya . The \lya\ brightness peak roughly
coincides with the UV brightest region (although this region also presents
significant \lya\ absorption).  However, most  of  the \lya\ emerges
from NE of this, whereas the SW region is much fainter and contains lots of
\lya\ absorption regions. 

The remaining panels have all been zoomed with a factor of two.  The mid-left
(d)  and mid-central (e) panels show the same images as above, but now with
more detail. In the remaining panels, we show $E(B-V)$ ©, \ha\ intensity (f),
the fitted age of the young population (g), and the equivalent widths of \lya\
(h) and \ha\ (i).  The intensity scales have been set to enhance detail. 

The UV bright complex is the very youngest in the galaxy, consistent with the
detection of WR features as discussed above and also coincides with the peak
\ha\ brightness.  Towards the SW, the
stellar population is older, except for some young star clusters (which appear
in blue  in Fig. \ref{fig:rgb}),  and we see the \lya\ map  varying between
emission and absorption on small scales, with the equivalent width ranging
between approximately --40 and +40 \AA ( see panel {\bf h}).  We interpret this
as some sight-lines having a smaller covering fraction of static \hi\ which
could be due to the ISM being porous, and/or small scale velocity structure.
Some regions which have a high \wha , e.g. star clusters in the SW tail, show
\lya\ in absorption or weak emission.

Even if the central UV-bright knot also contains the pixels with the highest
\lya\ brightness, this region has moderate \wlya  (10--30 \AA ) or \lya\ in
absorption. The highest \wlya\ is found in an asymmetric halo to the NW, with
values ($\gtrsim 500$\AA ), much higher than model predictions for young stellar
populations. In the same regions, \wha\ is moderate ($\sim 10$\AA ). This
indicates that the NE \lya\ halo originates in resonantly scattering photons
produced in the UV and \ha\ high surface brightness regions.  $E(B-V)$ appears
patchy with no obvious correlation with the strength of \lya . The same can be
said about the age map which, while anti-correlated with \wha\ as expected,
gives no or little guidance about \wlya .


In Fig. \ref{fig:rgb_fa} we show an RGB composite with the \ha\ line image in
red, the F140LP image in green, and \lya\ in blue.  Here we present, for
aesthetic reasons, results from running LaXs with non tessellated images;
instead, the \lya\ image has been adaptively  smoothed (using the {\sc midas}
task {\sc filter/adaptiv}) to suppress noise in low surface brightness regions.
This image now clearly shows the relative distribution of \lya\ with respect to FUV and
\ha\ emission and demonstrates the spatial decoupling of \lya\ from the far UV
sources that ionized the gas in the first place and the original recombination
sites probed by the \ha\ image.
It is furthermore obvious that the young star forming complexes that stretch
out from the center towards SW (see Fig.  \ref{fig:rgb}) and which are bright
in \ha\ do not result in strong \lya\ emission.

\subsection{Surface photometry} 
Further insights  can be obtained by comparing the radial distribution of
\lya\ compared to \ha\ and FUV, which we do in Fig. \ref{fig:fesc}.  The left
panel shows the fraction of \lya , \ha , and FUV  luminosity emitted as a
function of UV surface brightness integrated in 0.5 magnitude bins down to
$\mu_{\rm FUV}=26.2$ AB magnitudes per square arcsecond, which is the limit at
which the \lya\ image is judged reliable.  All quantities add up to unity when
summed over all surface brightness bins.  It can be seen that \ha\ follows the
FUV continuum relatively closely whereas the bulk of the \lya\ emission comes
out at lower UV surface brightness.  Most of the FUV and \ha\ luminosity comes
from regions with $\mu_{\rm FUV}=18$ to $21$ mag/$\square\arcsec$. The lower
fraction of \ha\ emission (compared to FUV) from the highest surface brightness
bins may be  a natural consequence of Str\"omgren spheres being larger than the
ionising sources.  For \lya\ the situation is again different: the brightest UV
regions
show net \lya\  absorption, and more than half of the cumulative \lya\ emission
emerges from regions with $\mu_{\rm FUV}\ge22$ mag/$\square\arcsec$. 

The right panel of Fig. \ref{fig:fesc} shows the same quantities as a function
of FUV isophotal equivalent radius ($r_{eq}=\sqrt{{\rm Area}/\pi}$).  Most of the FUV and \ha\
emission emerges from inside $r_{eq} = 1.5$ kpc and most of the \lya\ comes out
outside of this radius.  For FUV and \ha\ the contribution to the total flux
decreases monotonically outside $0.5$ kpc, while for \lya\ the flux increases
monotonically outside $0.7$ kpc. This result can most easily be understood in
terms of resonant scattering of \lya\ and that the photons we detect have
travelled far from their production sites. The latter should be best
represented by \ha . The \ha\ half light radius is 1.2 kpc whereas that for
\lya\ is 4.8 kpc., and neglecting extinction effects this indicates an average
total transverse scattering length of 3.6 kpc.  While we conservatively only integrated
the profiles out to $\mu_{\rm FUV}=26.2$mag/$\square\arcsec$ (after which the
\lya\ map breaks up into individual spaxels and noncontiguous areas,
indicating that we have reached the background noise limit for the chosen
maximum spaxel size of 1 $\square\arcsec$) the \lya\ flux contribution is still
increasing at the corresponding radius indicating that the \lya\ halo may be
considerably larger. At the smallest radii, the average \lya\ flux is negative,
suggesting a combination of dust absorption of \lya\ photons and that some
photons may have scattered out of the aperture. 

In both panels, the cumulative \lya\ escape fraction is sown by a magenta
dotted line. The escape fraction was calculated as:
\begin{equation}
f_\mathrm{esc}^{\mathrm{Ly}\alpha} = \frac{ F_{\mathrm{Ly}\alpha}^\mathrm{obs} }{F_{\mathrm{Ly}\alpha}^\mathrm{int}} = 
\frac{ F_{\mathrm{Ly}\alpha}^\mathrm{obs} }{8.7 \times F_{\mathrm{H}\alpha}^\mathrm{int}} = 
\frac{ F_{\mathrm{Ly}\alpha}^\mathrm{obs} }{8.7 \times F_{\mathrm{H}\alpha}^\mathrm{obs} \times 10^{0.4 \cdot E_{B-V} \cdot k_{6563} } }
\end{equation}
where the factor 8.7 derives from the intrinsic \lya/\halpha\ line ratio for
case B \citep{1987MNRAS.224..801H} and the superscripts 'obs' and 'int' refer to
observed and intrinsic quantities, respectively.  At high UV surface
brightness/small radius, the net \lya\ output is negative, but  the escape
fraction increases monotonically and reaches above 10\% at the limit to which
we can trace \lya\ at $\mu_{\rm FUV}= 26.2$  or $r\approx 6$kpc.

Figure \ref{fig:surf} shows the surface brightness profiles of FUV, \ha\ and
\lya , which has again been calculated as a function of the FUV equivalent
isophotal radius ($\sqrt{{\rm Area}/\pi}$). The FUV luminosity profile closely
follows a Sersic profile with index $n=4$. The morphology of Lars\#1 is
strongly irregular, suggestive of a merger origin, and the finding that its
young stars follow an $r^{1/4}$ law is consistent with observations of merger
remnants like NGC\,7252 (Schweizer 1982). The \ha\ profile is more extended and
best fit by $n\approx2 $  and for  \lya\  the profile outside $r>1$kpc is
approximately exponential, i.e. $n\approx 1 $.  The neutral gas distribution of
late type galaxies is usually exponential \citep{2012ApJ...756..183B} and if
resonant scattering by \hi\ is the dominant mode by which \lya\ is transported
through a galaxy, we expect the \lya\ brightness distribution to depend on the
\hi\ distribution. As \ha\ traces both young stars and gas we may consequently
expect an intermediate behaviour. The right panel of Fig. \ref{fig:surf} shows
the radial behavior of \wlya\ and the \lya/\ha\ ratio: for $r>3$kpc \wlya\
exceeds the maximum predictions for young stellar populations and  \lya/\ha\
exceeds the recombination ratio, again indicative of resonant scattering
\citep[see also][]{Atek08,2009AJ....138..923O}

We can also study the surface photometry of FUV, \ha\ and \lya\ on in a
spatially resolved manner, i.e. on the individual spaxel level (the area of a
spaxel ranges from 4 to 625 pixels, or from $0.0016\square\arcsec =560$ pc$^2$
to $1\square\arcsec=0.36$ kpc$^2$), which is what we show in \ref{fig:scat}. Each
Voronoi-binned spaxel is represented by a point, while the histograms take the
total number of contributing {\em pixels} into account. Since these are
logarithmic  plots, we only show spaxels that show  \lya\ in emission.  The
left panel shows \lya\ vs FUV surface brightness together with lines of constant \wlya shown in  red.
In the UV bright regions, pixels which have net \lya\ emission shows \wlya\
ranging between a few and 100 \AA . As we approach fainter  continuum levels
the scatter increases and notably a tail of spaxels with \wlya $\approx
1000$\AA\ develops. Such high \wlya\ values cannot be produced by normal
stellar populations. In the right hand panel we show \lya\ vs \ha. Similarly,
at bright levels, \lya/\ha -ratios for most spaxels range between 1 and 10 in
accordance with the simplest theoretical picture. In the faint tail however,
line ratios up to 100 are observed. 

All results presented in this section indicates that the \lya\ morphology of
LARS\#1 is formed by resonant scattering.  Moreover, the \lya\ emission is
markedly asymmetric with most of the emission coming out in the NE half of the
galaxy.  This could possibly be caused by an outflow of which the \lya\ image
is reminiscent. In the central region we see \lya\ changing from emission to
absorption on small scales. The 'tail' has many young star clusters/\hii
-regions from which we detect no direct \lya\ indicating a high covering
fraction of static \hi .

\begin{figure*}[t!]  
\centering
\includegraphics[angle=0,scale=0.35]{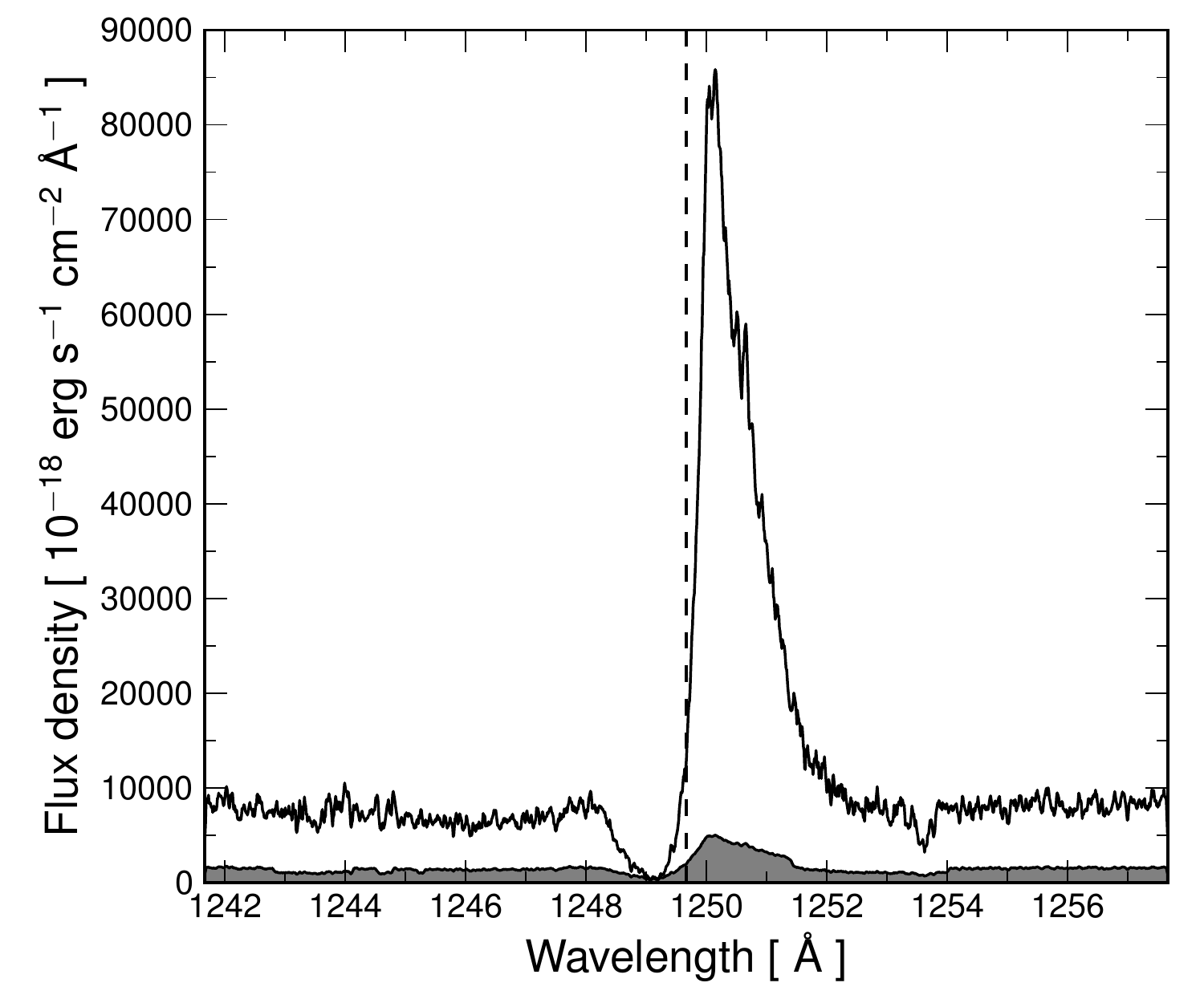}
\includegraphics[angle=0,scale=0.35]{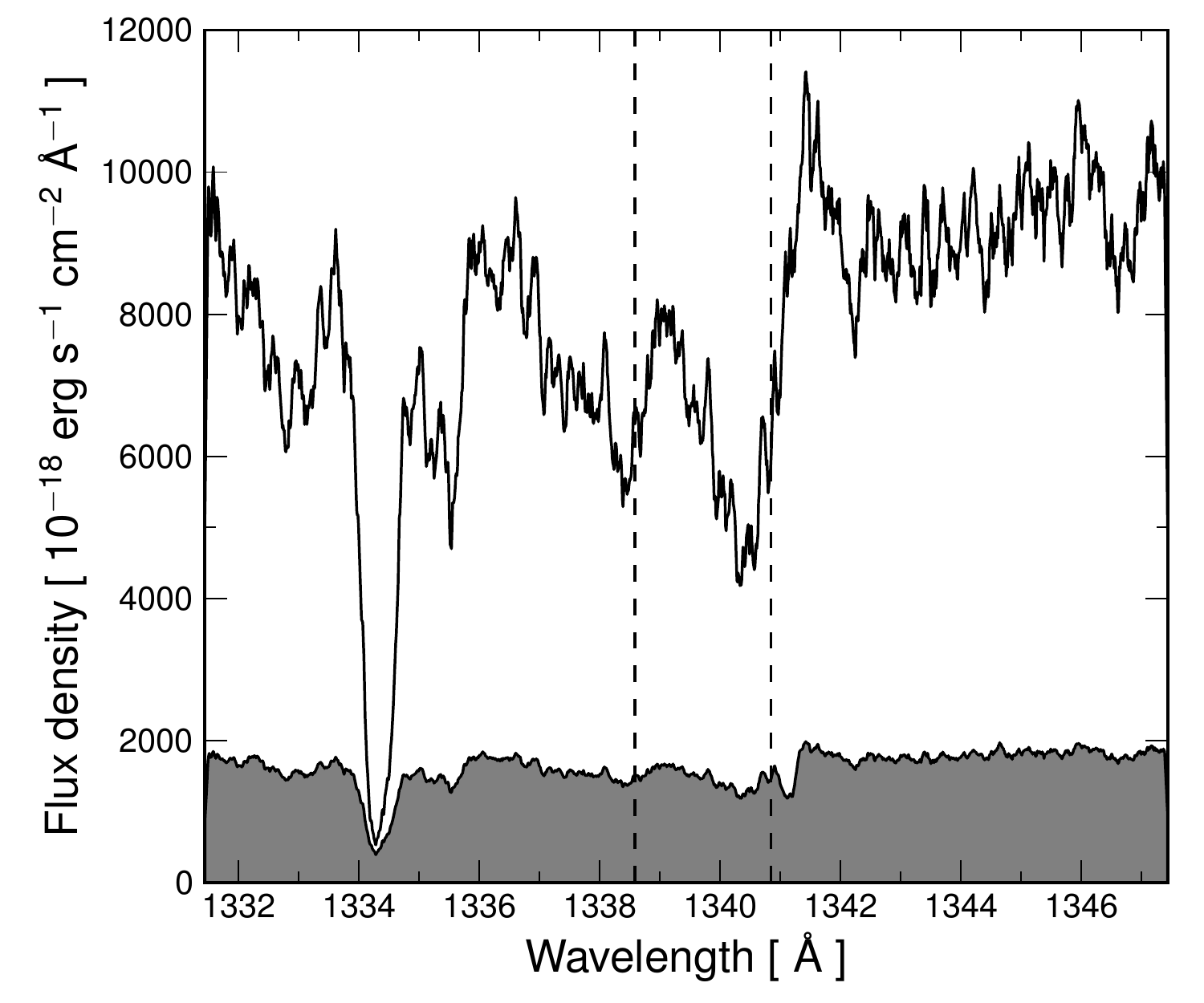}
\includegraphics[angle=0,scale=0.35]{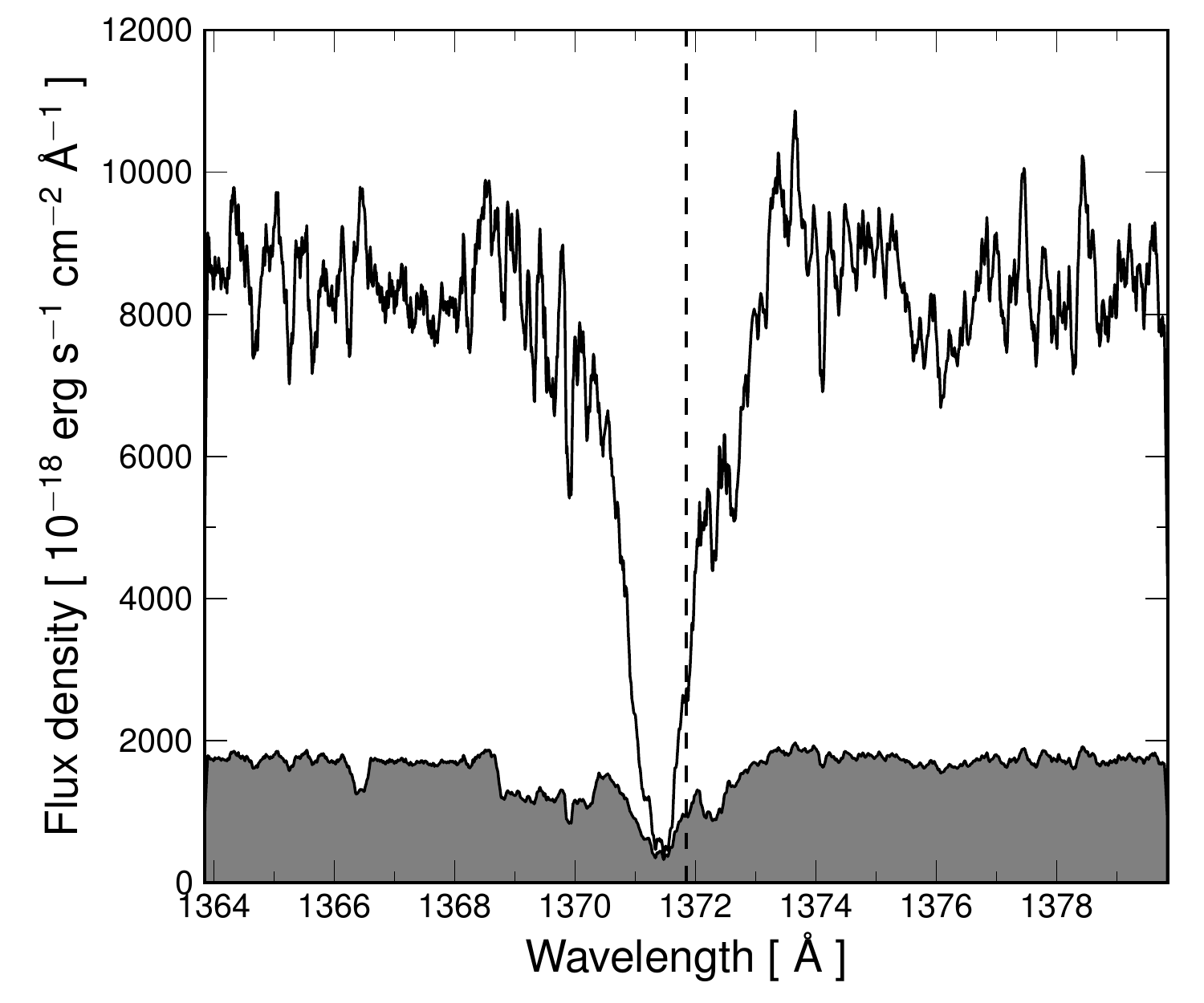}
\caption{COS spectra of Mrk\,259. On the {\bf left} the region around \lya\ is shown: the line shows a characteristic P-Cygni shape.
The {\bf middle} plot shows the spectrum around the redshifted interstellar O{\sc i}$\lambda\,1302$ and the {\bf right} panel shows
C{\sc ii}$\lambda1334$. In all panels, the vertical dashed line shows the  central wavelength each line would have if they had the
same redshift as \ha . The grey shaded area shows the uncertainty. 
 The \lya\ line peak is redshifted with $\sim 100$
km/s. The ISM lines are on the contrary systematically blue shifted and indicates the presence of outflows along the line of sight.   } 
\label{fig:cos}
\end{figure*}

\subsubsection{Global quantities}

Here we summarize the total photometric quantities for \mrk ,  derived in an aperture as
large as possible while still being able to trust the ingredient photometry as
well as the LaXs output. For \mufuv$\ge 26.5$\msqa\ Mkn\,259 starts to blend
with the background noise, i.e. the outer isophotes start to break up and we
find isolated spaxels outside the outermost connected isophote. We therefore
conservatively take as our grand total the quantities achieved when integrating
over areas with  \mufuv$\le 26.2$\msqa .

The integrated \lya\ flux we find is then \flya $= 6.95\times10^{-13}$ \egs\
which translates into \llya $= 1.3 \times 10^{42}$ \ergsec .  For the same
area, the integrated \wlya\ is 44.7 \AA . For \ha\ 
we find  \lha $ = 6.7\times10^{41}$ erg/s, and  hence \lya/\ha = 1.86.
Correcting \ha\ for the extinction map that LaXs produce from \ha/\hb\ we
derive \fesclya = 12.3\%. If instead \ha\ is corrected for the extinction
derived from LaXs fit to the young stellar SED we find 19.3\% (since \hb\
measurements are rare at high $z$ the escape fraction is normally computed from
SED based reddening estimates, if considered at all). Such a high a escape
fraction is rarely found at low redshift \citep[although LARS contains 3
galaxies with even higher values, ][paper\,II]{Hayes2014}   but is close to the
best estimate of the cosmic average at $z=4.5$ \citep{2011ApJ...730....8H}.
The star formation rate inferred for \ha\ before dust correction is 5.3
\msun/yr, and 9.2   \msun/yr after dust correction using the SFR calibration by \citet{Kennicutt1998}.
The integrated FUV luminosity is \lfuv $ = 2.7\times10^{40}$ erg/s,
corresponding to $M_{\rm FUV}=-19.2$ (AB) and log(\lfuv/\lsun)$=10.03$. Hence,
\mrk\ is detectable in deep Lyman Beak Galaxy surveys up to $z=10$
\citep{bouwens2014} and deep high-$z$ \lya \ surveys \citep{Hayes10}

Homogeneously derived global quantities for all LARS galaxies, based on the
version 1 data release, are given in paper\,II \citep{Hayes2014}. Our values
presented for \mrk\ above are similar to those presented in paper\,II.  The
difference comes from the current paper using isophotal integration, guided by
the FUV isophotes, whereas paper\,II used circular apertures and a smaller
equivalent radius (4 kpc compared to 6 kpc in the present paper).

\subsection{Hubble Space Telescope Spectroscopy}\label{mk259spec}
Spectroscopic observations of \mrk\ from the COS guaranteed time program (GO\,11522; P.I.
Green), have been obtained from the Mikulski Archive for Space Telescopes
(MAST).  The G130M grating was used, providing spectral coverage between 1150
and 1450~\AA, with nominal spectral resolution of around 15~km/s near \lya.
This spectrum has also been  discussed in \citet{Wofford2013}.  The 2.5\arcsec \
diameter entrance aperture of the COS spectrograph is shown together with the
COS NUV acquisition image as an inset in Fig.~\ref{fig:rgb}.

In order to estimate the true spectral resolution we take the F140LP image and
collapse and measure the one dimensional half-light radius, in the dispersion
direction. We obtain a value of $0.48$\arcsec , which can be expected
to reduce the spectral resolution by a factor of about 2.5, giving us a
continuum resolution of about 37~km/s.  In the \lya\ line the degrading of spectral 
resolution may be more pronounced due to the diffuse nature of \lya\ emission
(see Figures~\ref{fig:3x3} and \ref{fig:rgb}, as
well as \citealt{Hayes2013} and Paper\,II).  In the case that the \lya\ surface
brightness profile is completely flat and fills the COS aperture, we would
expect $R\approx 1\,500$.  From the analysis presented in the previous section, 
we find the \lya\ half light radius within the COS  aperture to be $\approx 1\arcsec$ 
indicating an effective resolution for \lya\ emission of $R\approx 2\,250$ or $\sim 130$ km/s.

The spectra obtained from MAST were pipeline-processed using \textsc{calcos},
in which the main potential source of error is the adoption of correct physical
coordinates for wavelength calibration. We first examine the two dimensional
COS/NUV  acquisition  image. The inset in Fig. \ref{fig:rgb} displays the COS 
acquisition image, with the 2.5\arcsec\ primary science aperture (PSA) overlaid.  
	The adopted centroid coordinates obtained
during the acquisition procedure differ from the peak of the light distribution
by just 2 pixels, corresponding to a negligible error in wavelength.  Similar
conclusions are reached by measuring the wavelength of the wings of the
geocoronal \lya\ emission.

The \textsc{calcos} output include a large number of potential error sources, 
which are binary-encoded in the data tables for every extracted pixel.
We re-sample extracted one dimensional spectra onto a common wavelength grid 
and combine them, using a conservative rejection 
scheme that rejects every element of questionable quality, regardless of the 
reason; this does result in a small number of un-sampled regions in the final
spectrum, but these do not fall in any of features of interest. 
The error spectrum is added in quadrature using the same rejection scheme. 

The resulting spectrum in the
regions of \lya\ and selected ISM absorption features are shown in Fig.
\ref{fig:cos}. The \lya\ line presents a redshift of $\approx100$ km/s as
measured by the difference between the \lya\ peak flux and the \ha\ velocity
measured in the SDSS spectrum. The low ionisation stage (LIS) ISM absorption 
lines, mainly from Si{\sc ii} are blueshifted by a similar amount. The  blueshifted
ISM lines indicates the presence of outflowing neutral gas with
a mean velocity of 100 km/s (see Rivera-Thorsen et al. 2014 in prep for a
full analysis). This value is similar to what has been found in high-$z$ \lya\
emitters \citep[e.g.][]{Hashimoto2013}.

A $\chi^2$ fit to  the COS UV spectrum, using the high resolution library of
SB99 for the same input parameters as those adopted for the imaging results produced 
by LaXs give an age of 4.04 Myr, and a stellar reddening of $E(B-V)=0.02$. Adopting a
lower metallicity ($Z=0.001$ or $0.004$) give comparably good fits, and nearly
identical results, while those for higher metallicity (Z=0.02) are
worse. If instead assuming a constant star formation rate, we find an age of 24
Myr and a reddening of $E(B-V)=0.04$. From the imaging data we derive a young
population age of 4.34 Myr, and zero reddening of the stellar component, from a
matched circular aperture. Hence the imaging SED fit produced by LaXs and the
SED-fit to the COS spectrum produce consistent results.



\subsection{Lyman-alpha radiative transfer modeling}
The \lya\ line of \mrk\ has a P-Cygni profile, with the observed equivalent
width \wlya$=9\pm2$\,\AA . A difficulty in deriving the spectroscopic equivalent
width is the continuum which possesses lots of features around \lya .   From the HST imaging we 
derived \wlya$\approx14$\,\AA \ for the area corresponding to  the COS aperture.  Given
the very different spectral baselines of imaging and spectroscopy these numbers are
quite consistent. 
The \lya\ absorption minimum is located at $\sim-130$\,\kms, approximately at
the mean outflow velocity as measured by the ISM absorption lines
(Rivera-Thorsen et al. 2014 in prep).  
The main \lya\ peak is located at $\sim +110$\,\kms. However, 
   as discussed in \cite{2003ApJ...598..858M}, this value is highly dependent 
   on the spectral resolution. This is an effect of the convolution of the 
   sharp, steep blue edge profile produced by scattering, and the 
    instrumental profile. Our measurement differs from that of \cite{Wofford2013}, due to 
       the revised systemic redshift, which we derived from our custom-made
       fits to the SDSS emission lines. A weak blue peak is present at $\sim
-400$\,\kms.  The red \lya\ peak has has a visible substructure at $\sim +230$\,\kms, and weaker ones 
       at $\sim +300$ and $\sim +390$\,\kms . In the homogeneous shell
model, these should ideally be multiples of the outflow velocity.

The observed \lya\ line profile thus encodes
      the information about the neutral gas and dust in which the
      \lya\ photons diffuse. 
      Radiative transfer models provide keys to decoding the ISM
      parameters, which is essential for the high-redshift
      galaxies where other data are scarce. LARS as a low-redshift
      sample, on the other hand, provides an opportunity for 
      detailed checking of the \lya\ model predictions and independent,
      multiwavelength data.      
             


We use the enhanced version of the {\em MCLya} 3D Monte Carlo radiation
transfer code of \cite{Verhamme06}, which computes the detailed physics of the
\lya\ line and the adjacent UV continuum, for an arbitrary 3D geometry
         and velocity field.

We here assume the geometry of an expanding, homogeneous, spherical shell of
neutral ISM (\hi\ and dust, uniformly mixed), which surrounds the starburst
producing UV continuum and \lya\ emission in \hii\ regions. This geometry is an
approximation of the outflows which seem ubiquitous in starburst galaxies
\citep{Shapley03}, and it has been successful at reproducing the \lya\ line
profiles of  low- and high-$z$ galaxies \citep{Verhamme08, Schaerer08, Dessauges10, Vanzella10,
Lidman12, Leitherer13}. 
 The shell is characterised by four parameters: the
radial expansion velocity $v_\mathrm{exp}$, the \hi\ column density \NHI, the
\hi\ Doppler parameter $b$, and the dust absorption optical depth
$\tau_\mathrm{a}.$ For each parameter set, a full Monte Carlo simulation was
run, constructing a library of $>6000$ synthetic models \citep{Schaerer11}.  We
assume the intrinsic spectrum to be composed of a flat stellar continuum and a
Gaussian \lya\ line, characterised by a variable full width at half maximum
\Folya, and equivalent width \EWo.  The theoretical spectra were smeared to the
observed spectral resolution derived from the \lya\ source extent (Sect. 6.3).

We used an automated line profile fitting tool 
exploring the full parameter space of the model grid, as described in
\cite{Schaerer11}: $0 < v_\mathrm{exp} < 700$\,\kms; $10^{16} <$\NHI$<
10^{22}$\,cm$^{-2};$ $0<\tau_\mathrm{a}<4;$ $10<b<160$\,\kms. The intrinsic
\lya\ equivalent width was varied between 0 and 300\,\AA, and the FWHM between
0 and 300\,\kms.

\subsubsection{Results of the \lya\ model fitting }

\begin{figure}[hbt]
 \includegraphics[width=0.5\textwidth]{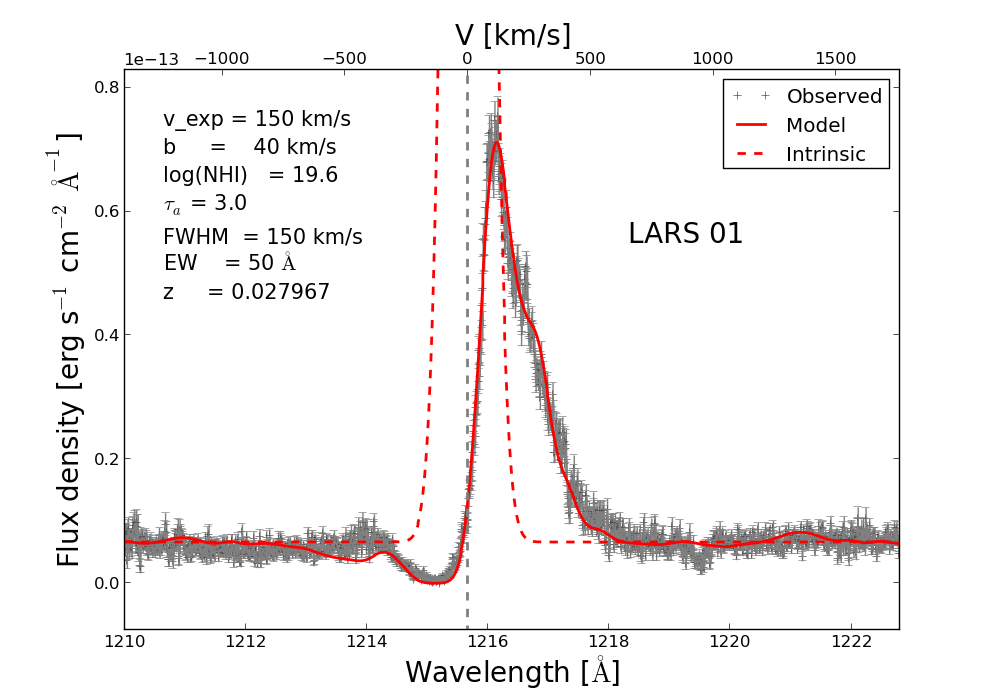}
 \caption{\label{fig:RT} Best-fitting model of the \lya\ emission-line
  profile, using the radiative transfer code. The observational data are
  plotted in black, the final model in solid red, and the intrinsic Gaussian
  \lya\
  profile (i.e., input in the radiative transfer code) in dashed red.
  The intrinsic profile has been artificially shifted down, to appear in the
  plotted scale. Therefore, the difference in the levels of the intrinsic and
  observed continuum does not correspond to the dust attenuation derived.
  A Milky-Way absorption line is visible in the red wing, masked for the 
  \lya\ profile fitting.   
}
\end{figure}

The best-fitting model of the \lya\ profile, from the simple shell model  is
presented in Fig.\,\ref{fig:RT}.  It reproduces the overall spectral shape, the
position and shape of the main peak (including some of the red-wing
substructure), the position, depth and width of the absorption trough, and the
presence of a weak blue peak (whose amplitude is underestimated by the model, though).  
The best-fit solution is characterised by:
$v_\mathrm{exp} = 150$\,\kms, \NHI$=4\times10^{19}$\,cm$^{-2},$
$\tau_\mathrm{a} = 3,$ $b=40$\,\kms, \EWo$=50$\,\AA, \Folya$=150$\,\kms.  
We now compare these numbers to what other constraints are available.

The inferred expansion speed is close to the mean velocity                      
derived from the LIS UV absorption lines 
           $v_\mathrm{LIS} = -100$\,\kms (Rivera-Thorsen et al. 2014 in prep).             
           Therefore, even though the LIS lines show a broad profile, 
           extending over $> 300$\,\kms, the idealized single shell model 
           with the expansion speed equal to the mean $v_\mathrm{LIS}$ appears 
           to be a reasonable approximation. 
            The \ha\ FWHM from SDSS (corrected for instrumental
dispersion) is $\sim~170$\kms, and \Folya\ is expected to be identical.  The
dust reddening inferred from the SDSS spectroscopy  is \ebv\,$=0.09$ 
(Table~\ref{tab:sdssspec2}) while the imaging analysis suggests a nebular
reddening of \ebv\,$=0.14$.  In general spectroscopic measurements should be
more accurate, but this has not been corrected for underlying Balmer absorption
(which the imaging data has within LaXs). Converted to the optical depth in the
vicinity of \lya, the SDSS value represents $0.5<\tau_\mathrm{a}<2$, depending
on the extinction law applied
\citep{1994ApJ...429..582C,1989ApJ...345..245C,Prevot84}, while the imaging
data suggest $0.8<\tau_\mathrm{a}<2.8$.  Our automated fit of the \lya\ profile
provided $\tau_\mathrm{a}=3,$ which suggests that the dust composition of
LARS\#1 is more akin to the SMC bar \citep{Prevot84}, and in particular has
significantly more UV extinction than predicted by the
\citet{1994ApJ...429..582C} attenuation law.  The excess of UV extinction may
also be an indication that the amounts of dust in \hii\ and \hi\ regions,
inferred from the Balmer emission decrement and the \lya\ line profile,
respectively, are not identical.
  
  No constraints are available for the Doppler parameter $b$ 
or the \hi\ column density \NHI. 
The gas-to-dust ratio derived from the \NHI\ obtained by the \lya\ fitting is
lower than the Galactic value \NHI/\ebv~$=5.8\times10^{21}$\,cm$^{-2}$
\citep{Bohlin78}. 
However,  as discussed earlier
\citep{Schaerer08,Verhamme08}, the \lya\ profile, sensitive to the \hi\
outflow, may trace only a fraction of the total neutral hydrogen content, and
thus \NHI\ derived from the \lya\ emission line profile may not be
representative of the entire \hi\ content.           

To estimate  \EWo\, we used the SDSS  \ha\ flux (Tab.\,\ref{tab:sdssspec2}) and
the UV continuum flux in the vicinity of \lya, measured from the COS data
(corrected for the size difference of the apertures).  We corrected the \ha\
flux for reddening using the SDSS value, and the UV flux with $\tau_\mathrm{a}$ derived from the
           optical SDSS value converted to UV by the extinction law.      
Scaling the extinction corrected \ha\ flux with 8.7
we arrive at the predicted \lya\ flux within the aperture, which divided by the
corrected UV flux gives  \EWo $=30-100 $\,\AA\ depending on the extinction law applied. 
          

We conclude that the expanding homogeneous shell models capture the essential
features of the \mrk\ \lya\ line profile, with no a-priori constraints applied.
In addition, the ISM parameters characterising the best-fitting model are
mostly consistent with those derived from independent data.  Further detailed
modelling exploring other possible solutions and their degeneracies will be
presented in Orlitov\'a et al. (2014, in preparation) for the entire LARS
sample.


\section{Conclusions and summary}

In this paper we have presented the Lyman Alpha Reference Sample (LARS), its
survey outline and the powerful source of information that the combination of
multiwavelength imaging and spectroscopy produce for understanding the Lya
production in the test-bench galaxy Mrk259. LARS comprises 14 galaxies and is
the first comprehensive \lya\ imaging study performed at low redshifts with a
well defined selection function. 

The study has been performed with the Hubble Space Telescope and with the Solar
Blind Channel (SBC)  of the Advanced Camera for Surveys (ACS), which is capable
of imaging at the rest wavelength of \lya . With ancillary imaging we also collected $u,
b, i$, \ha\ and \hb\ with the UVIS channel of the Wide Field Camera 3
(WFC3) and the Wide Field Camera (WFC) of ACS. 

The targets were selected from the combined SDSS DR6 + GALEX DR3 catalogue.  In
order to study systems with significant production of \lya\ photons, whether or
not these escape, we applied a cut in the \ha\ emission equivalent width of
\wha$\ge100$\AA .  Potential targets classified as AGN based on the width and ratios 
of emission lines  were removed in order to focus on systems dominated by star
formation. Objects with high foreground extinction were discarded.

\lya\ imaging necessitates continuum subtraction which is non-trivial due to
the strong evolution of the continuum near \lya\ as a function of dust
absorption and stellar population age. For low redshifts there is an additional
complication in that the available continuum filters of HST are broad in comparison to
the typical \lya\ emission line equivalent width, and moreover sample continuum
at significantly longer wavelengths than \lya . Therefore we have developed the
LARS  eXtraction software (LaXs) that utilises UV and optical continuum filters
and \ha\ and \hb\ images to perform a pixel-based SED fit in order to estimate
the continuum at \lya .

Targets were selected in two redshift windows ($z=0.028-0.109$ and
$0.134-0.190$) that allows \lya\ to be captured with a combination of long pass
filters in SBC that don't transmit the geocoronal \lya\ line, offering a
significantly improved sensitivity compared to the available $z\sim0$ \lya\
filter (F122M) in SBC.

In order to provide a useful comparison sample, we strived to cover a range of
far UV (FUV) luminosities that is similar to the range observed for high
($z>2$) redshift \lya\ emitters and Lyman Break Galaxies.  To provide maximum
spatial resolution, preference was given to sources towards the lower end of
the redshift range, but since luminous galaxies are intrinsically rarer, higher
redshift sources were included in order to populate the high luminosity bins.
When galaxies with available and relevant archival HST UV/visual imaging, or
existing/planned COS spectroscopy, satisfied our selection criteria, these were
give priority in order to lower the orbit request and increase the scientific
value of the dataset. Twelve of the targets fall in the low-$z$ window and two
in the high-$z$ one. Seven targets have archival COS spectroscopy, and the
remainder were approved in Cycle 19 meaning that all LARS galaxies have both
HST \lya\ imaging and spectroscopy. In addition several projects to enhance the
LARS dataset with other relevant data (optical IFU spectroscopy of the ionized
gas kinematics, \hi\ single dish and interferometric observations, far IR and
CO observations, deep ground based optical and near IR imaging) are underway
and will be the subject of future LARS papers. 

In the current paper we also presented \lya\ imaging and COS spectroscopy
results for \mrk\ (LARS \#1), the first galaxy in the LARS sample. This galaxy
is metal-poor and  has a very extended asymmetric \lya\ halo with strong
emission in the north-eastern parts of the galaxy, and absorption in the
south-western components. The remote emission regions have very high \lya\
equivalent widths and \lya/\ha\ ratios suggesting that those \lya\ photons have
been resonantly scattered from the more centrally loctated star forming regions. The very
highest UV surface brightness regions show \lya\ in absorption.  While the far
UV continuum radial surface brightness profile is well fitted by a Sersic $n=4$
profile the \lya\ profile is purely exponential, and the \lya\ escape fraction
is seen to be monotonically increasing.  The integrated \lya\ luminosity is
log$(L_{Ly\alpha})=42.1$ \egs\ and the equivalent width is 45 \AA\ meaning that
it would be detectable in high-$z$ \lya\ imaging surveys. The integrated \lya\
escape fraction is 12\%, an unusually high value for low redshift galaxies. The
COS spectroscopy displays a \lya\ emission line from the centre with a p-Cyg
profile and blue shifted ISM absorption lines, indicating outflow of neutral
gas.  A radiative transfer model with a homogeneous expanding shell is able to
reproduce the line profile with parameters that are in overall agreement with
those that we derive from the \lya\ imaging and SDSS spectroscopy.
 
 As we have argued, a better physical understanding of what regulates the
ability (or not) of a galaxy to emit the \lya\ photons that are produced in its
star-forming regions, is required in order to be able to reliably use \lya\ as
a probe of cosmology and the evolution of the galaxy population.  The LARS
project provides an important step along this line and  is, as the name implies, intended
to produce a local reference sample for aiding interpretation of high redshift
\lya\ studies where information is often scarcer.  A number of papers have been/or are in  the
process of being written: A letter (LARS paper 0) presented the ubiquitous
extended \lya\ haloes seem in the majority of the sample is already published
\citep{Hayes2013}. Paper\,II \citep{Hayes2014} presents further information on
the large scale \lya\ properties, paper\,III (Pardy et al. 2014, accepted)
integrated \hi\ masses,  paper\,IV (Rivera-Thorsen et al. 2014 in prep)  will
focus on the ISM kinematics as inferred from COS spectroscopy, paper\,V (Duval
et al 2014b, in prep) a detailed study of LARS\#5, paper VI (Orlitova et al.
2014, in prep) radiative  transfer modelling, paper\,VII (Guaita et al. 2014,
in prep)  a morphological analysis including the predicted appearance of LARS
galaxies at high redshift. Subsequent papers will adress IFU data from the 
Potsdam Multi Aperture Spectrometer \citep[PMAS,][]{Roth2005}, \hi\
interferometry from JVLA and GMRT, CO observations, optical/IR imaging, 
integrated SED fits, clumpy ISM RT models, etc,  and a community data release. 

In HST cycle 21 an extension of LARS, eLARS, was approved. Compared to LARS it
is twice as big, comprising 28 targets, and with the \wha\ cut reduced to
40\AA\ in order to provide more galaxies that fall along the star formation
main sequence at low and high-$z$. COS spectroscopy for the full sample was
furthermore approved in cycle 22.

\acknowledgments
This research has made use of the NASA/IPAC Extragalactic Database (NED) which
is operated by the Jet Propulsion Laboratory, California Institute of
Technology, under contract with the National Aeronautics and Space
Administration.  This work was supported by the Swedish Research Council (VR)
and the Swedish National Space Board (SNSB).  G\"O is a Royal Swedish Academy
of Science research fellow (supported by a grant from the Knut \& Alice
Wallenbergs foundation).  HA and DK  acknowledge support from the Centre
National d'Etudes Spatiales (CNES). HA is supported by the European Research
Council (ERC) advanced grant ÒLight on the DarkÓ (LIDA). PL acknowledges
support from the ERC-StG grant EGGS-278202.  IO acknowledges the Sciex
fellowship of RectorsÕ Conference of Swiss Universities, grant 14-20666P of
Czech Science Foundation, and the long-term institutional development grant
RVO:67985815.  MMR acknowledges support from BMBF under grant no. 03Z2AN11. HOF
is funded by a postdoctoral UNAM grant.  PL acknowledges support from the
ERC-StG grant EGGS-278202. DK has been financially supported by the CNES.

{\it Facilities:} \facility{HST (ACS,WFC3,COS)}, SDSS.

%
%

\appendix

\section{A\,1. ~NGC\,6090 -- Proof of concept}
\begin{figure*}[t!]
\centering
\includegraphics[angle=0,scale=0.71]{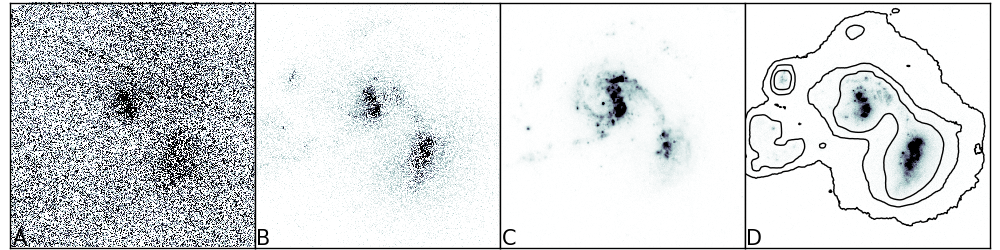}
\caption{NGC6090: Comparison of results obtained for NGC6090 ($z=0.029$) with
F122M vs F125LP--F140LP.  From left to right: {\bf A} \lya\ image produced from
F122M observations (\"Ostlin et. al. 2009);  {\bf B} \lya\ image produced by
using archival F125LP and F140LP data (PI McCandliss, program 11110) and  the
method described in Hayes et. al. 2009; {\bf C} \ha\ image; and finally  {\bf
D} F140LP image with F140LP contours overlaid (image in C and D are from
\"Ostlin et al 2009). While the integrated photometry of the two \lya\ images
are fully consistent, the improvement in image quality is striking, amounting
to an increase in  the S/N of a factor of 7. Also for this galaxy is \lya\
much more diffuse that \ha\ or the UV continuum.} 
\label{fig:ngc6090}
\end{figure*}

For the starburst galaxy NGC\,6090, at $z=0.029$, there are HST images taken
both with F122M (from GO program 9470, PI Kunth), and F125LP (program 11110 by
PI McCandliss) as well as F140LP (both programs).  In addition there are near
UV and optical broadband and \ha\ images from program 10575 (PI \"Ostlin) and
the results from programs 9470+10575 are presented in
\citet{2009AJ....138..923O}.  This galaxy hence offers the opportunity to
verify the new method using the long-pass filter combination, and compare it to
our previous approach using F122M. In Fig. \ref{fig:ngc6090} we perform this
comparison.  The left panel shows the resulting \lya\ image when F122M is used
to capture the line, and the second panel from the left shows result when the
F125LP--F140LP combination is used. The following panels show the \ha\ image
and the F140LP continuum images. A comparison of the results show that the use
of F125LP increases the S/N with a factor of $\sim7$ despite the exposure time
being three times smaller, which can be attributed to the factor 5 improvement
in throughput combined with the background being a factor of 20 lower during
the part of the orbit in which HST is in earth's shadow.

\section{A\,2. Re-calibrating the Bandpasses of the Solar Blind Channel }

We compared the fluxes measured by synthetic photometry in the COS/G130M and
SDSS spectra, to those measured by standard aperture photometry in matched
apertures of the HST images, using  {\tt PySynphot}.  In doing so we found
remarkable agreement between HST/WFC3 optical photometry and spectral
photometry from the SDSS -- discrepancies no larger than a few per cent were
discovered, even in the narrowband filters. However when comparing the ACS/SBC
aperture-matched flux densities with those obtained by spectral photometry in
the COS/G130M spectrum, substantial differences were noticed.  Specifically,
the fluxes measured in the SBC imaging were systematically brighter, which
persisted after accounting for vignetting in the Primary Science aperture of
COS. Furthermore, a noticeable wavelength dependency was noticed in the
discrepancy, which increases towards the blue, and its magnitude amounts to
11\% in F150LP, 13\% in F140LP, and 22\% in F125LP. According to the instrument
handbooks for COS and SBC, we can expect absolute accuracies to be better than
10\% (COS) and 3\% (SBC); our discrepancy at F125LP is double this combined
difference, and while formally consistent in F140LP and F150LP, the offset is
suspicious and a cause for concern. 

Note that the vignetting correction depends upon the distribution of light
within the PSA, and the correction for the continuum may be different from 
\lya, which can be more extended. \lya\ is transmitted only by F125LP, 
where the discrepancy is seen to be largest, so we also computed a vignetting 
correction for a flat surface that fills the aperture and applied it to the 
\lya\ region of the COS spectrum. However the F125LP bandpass is so broad that 
this correction can account for no more than 1\% absolute, or 4\% of the 
relative offset seen in F125LP. I.e. COS vignetting is only a minor contributor 
to the photometric discrepancy. 

Assuming that the cause of the discrepancy lies with one instrument alone, it
can be explained if either COS (or SBC) becomes systematically less (more) 
sensitive than expected with decreasing wavelength. Specifically with respect
to SBC, this could be attributed to either the response of the MAMA detector or
the throughput of the filters. Unfortunately even with the 
extensive sets of calibration data that have been obtained, not one single 
object from the COS calibration data-sets has been observed with SBC imaging in
the same filters that we have used. However, a rather complete SBC calibration
campaign has been carried out, targeting the globular cluster NGC\,6681 in all 
filters, and also with both PR110L and PR130L low resolution siltless prism 
spectroscopy. Thus for a substantial number of stars, the possibility exists to
test the response of SBC against itself in imaging and spectroscopic modes. 

The philosophy behind this experiment is as follows. The SBC LP bandpasses 
are defined on the red side by the sensitivity of the MAMA detector, which 
decreases with increasing wavelength. On the blue side they are described by 
a sharp cut on; thus all filters in principle share identical red-wings,
which, modulo the prism response function, applies also in the two spectroscopic
modes. The SBC has a well-known red leak, but the same 
principle applies here: each mode should be subject to the same leak, and its
contribution to the integrated countrate should be the same in every case. Were
the photometric discrepancy that we notice attributable to the red leak, it 
would not become more pronounced for the bluer filters.

\begin{figure*}[h!] 
\centering
\includegraphics[angle=00,scale=0.8]{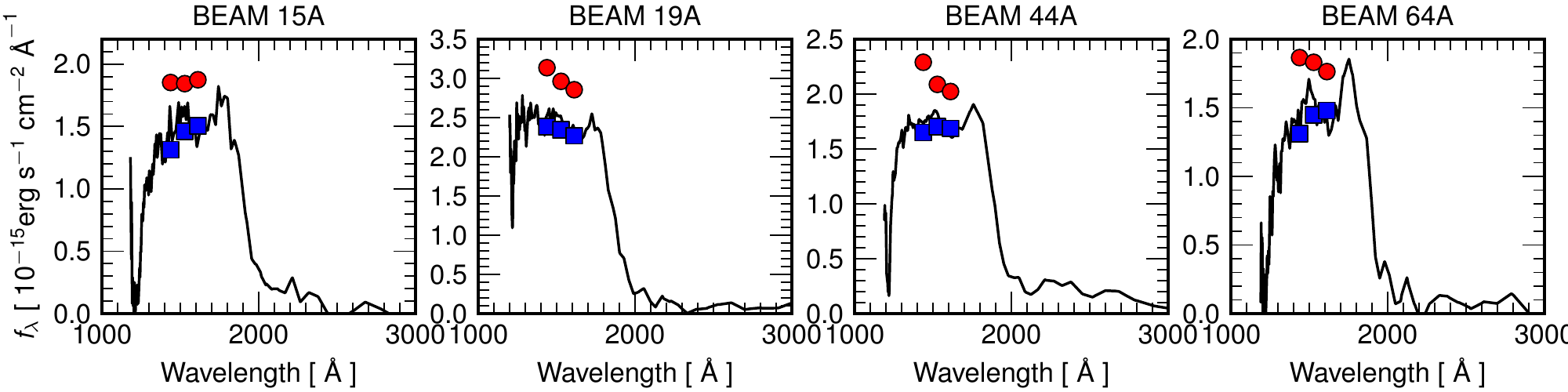}
\caption{SBC/PR110L spectra for four representative stars in NGC\,6681. Black
lines show the spectra, while synthetic photometry performed on these spectra
are shown by blue squares. Direct aperture photometry on the images is shown
by red circles. Filters from short to long wavelength are 
F125LP, F140LP, and F150LP. There is an undoubtable systematic trend 
for fluxes measured in the images to exceed those measured in the spectra, 
and furthermore, the magnitude of the this effect appears stronger in the 
bluer filters.} \label{fig:calsbc1}
\end{figure*}

We obtained the SBC data for NGC\,6681 from the archive, opting to use 2009 
calibration epoch; we obtained images in F125LP (data-sets ja8v02c7q and 
ja8v02c8q), F140LP (ja8v03imq and ja8v03inq), and F150LP (ja8v02c5q and 
ja8v02c6q), and prism spectral images in PR110L (ja8v02c3q and ja8v02c4q) which
extends all the way across the F125LP bandpass.  Using aXe V2.3 software 
\citep{Kummel2006} and SExtractor \citep{Bertin1996}, we performed a spectral 
extraction of 80 objects. We first identified and extracted the sources in each
individual direct image. The object positions are then used for the trace 
calculation and the wavelength calibration of the corresponding prism image.  We
used the prism reference files from cycle 13 and adopted a constant extraction 
direction, local background subtraction, and aperture size of 13 pixels. We 
then re-grided and combine the two spectra extracted for each star. We examined
the stamps of the spectral extractions and remove any spectra that are not 
well-centered in the aperture, and any objects for which the projection of a 
nearby star falls near enough to contaminate the main spectrum.  We further restrict
our analysis only to objects that appear relatively blue in the calibrated low 
resolution spectrum (approximately $\beta<1$); this is to insure 
that the majority of the counts in the images comes from the spectral region 
covered by the bandpass and not the red leak. We are then left with 20 stars.
Using  {\tt PySynphot} we convolve these spectra with the F125LP, F140LP, and 
F150LP bandpasses, and measure the total flux that would hypothetically be 
seen in imaging mode. 

We then perform aperture photometry of the same stars, using custom scripts. 
The spectral extractions described above use 13 pixel apertures on the 
cross-dispersion direction, but are essentially infinite in the dispersion 
direction, at least for photometric purposes. We therefore do our photometry
in non-standard rectangular slit-like apertures, 13 native pixels high, and 
1 arcsec (33 pixels) in length. This is to get a reasonable picture of total 
flux that will be throughput in spectroscopic mode. Figure~\ref{fig:calsbc1} 
shows the fluxes obtained through spectroscopic and photometric measurements 
of a few representative stars from the sample.

\begin{figure*}[h!]
\centering
\includegraphics[angle=00,scale=1]{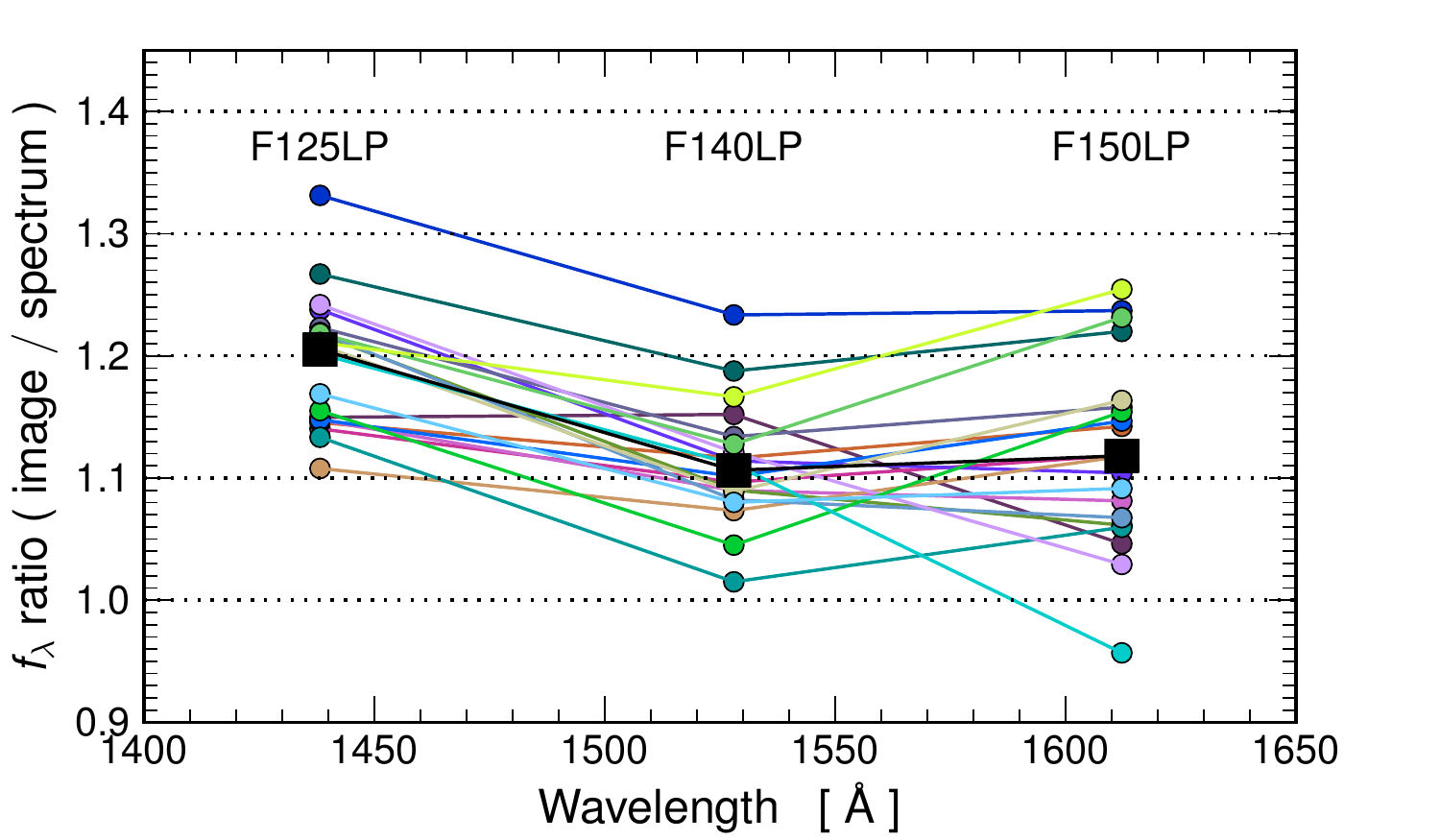}
\caption{The relative corrections factors, defined as flux(image)/flux(spectrum) 
for the 20 stars in NGC\,6681 (coloured circles). The large black squares show
the luminosity-weighted average ratio. 
} \label{fig:calsbc2}
\end{figure*}

Figure~\ref{fig:calsbc1} shows a clear systematic: fluxes measured directly by 
aperture photometry in the images exceed those measured by synthetic 
spectrophotometry in the same filters. The effect appears more pronounced in the bluer
filters, and decreases towards the red. For whatever reason, this is exactly
the same phenomenon that we saw when comparing the SBC photometry to that of 
COS. The offsets for each
filter are shown in Figure~\ref{fig:calsbc2}. Without applying any outlier rejection, 
we compute the luminosity-weighted average ratio to be for each of the three filters:
F125LP: $1.193  \pm 0.053$, 
F140LP: $1.113   \pm  0.047$, 
F150LP: $1.126    \pm 0.074$.  When calibrating our science data
we rescale the fluxes measured in the images down by these factors.

We note finally that our approach does not completely debug the photometric 
issue, but does reconcile SBC and COS spectrophotometry to within the relative 
calibration levels of the COS. More dedicated work on the cross-calibration of
these two instruments would be needed to fully resolve it.

\bibliographystyle{aa} 
\bibliography{lars}

\begin{thebibliography}{103}
\expandafter\ifx\csname natexlab\endcsname\relax\def\natexlab#1{#1}\fi

\bibitem[{{Adams} {et~al.}(2011){Adams}, {Blanc}, {Hill}, {Gebhardt}, {Drory},
  {Hao}, {Bender}, {Byun}, {Ciardullo}, {Cornell}, {Finkelstein}, {Fry},
  {Gawiser}, {Gronwall}, {Hopp}, {Jeong}, {Kelz}, {Kelzenberg}, {Komatsu},
  {MacQueen}, {Murphy}, {Odoms}, {Roth}, {Schneider}, {Tufts}, \&
  {Wilkinson}}]{2011ApJS..192....5A}
{Adams}, J.~J., {Blanc}, G.~A., {Hill}, G.~J., {et~al.} 2011, \apjs, 192, 5

\bibitem[{{Aller}(1984)}]{1984ASSL..112.....A}
{Aller}, L.~H., ed. 1984, Astrophysics and Space Science Library, Vol. 112,
  {Physics of thermal gaseous nebulae}

\bibitem[{{Anders} \& {Fritze-v.~Alvensleben}(2003)}]{2003A&A...401.1063A}
{Anders}, P. \& {Fritze-v.~Alvensleben}, U. 2003, \aap, 401, 1063

\bibitem[{{Atek} {et~al.}(2008){Atek}, {Kunth}, {Hayes}, {{\"O}stlin}, \&
  {Mas-Hesse}}]{Atek08}
{Atek}, H., {Kunth}, D., {Hayes}, M., {{\"O}stlin}, G., \& {Mas-Hesse}, J.~M.
  2008, \aap, 488, 491

\bibitem[{{Atek} {et~al.}(2009){Atek}, {Kunth}, {Schaerer}, {Hayes},
  {Deharveng}, {{\"O}stlin}, \& {Mas-Hesse}}]{2009A&A...506L...1A}
{Atek}, H., {Kunth}, D., {Schaerer}, D., {et~al.} 2009, \aap, 506, L1

\bibitem[{{Atek} {et~al.}(2014){Atek}, {Kunth}, {Schaerer}, {Mas-Hesse},
  {Hayes}, {{\"O}stlin}, \& {Kneib}}]{Atek2014}
{Atek}, H., {Kunth}, D., {Schaerer}, D., {et~al.} 2014, \aap, 561, A89

\bibitem[{{Baldwin} {et~al.}(1981){Baldwin}, {Phillips}, \&
  {Terlevich}}]{1981PASP...93....5B}
{Baldwin}, J.~A., {Phillips}, M.~M., \& {Terlevich}, R. 1981, \pasp, 93, 5

\bibitem[{{Bergvall} \& {{\"O}stlin}(2002)}]{2002A&A...390..891B}
{Bergvall}, N. \& {{\"O}stlin}, G. 2002, \aap, 390, 891

\bibitem[{{Bertin} \& {Arnouts}(1996)}]{Bertin1996}
{Bertin}, E. \& {Arnouts}, S. 1996, \aaps, 117, 393

\bibitem[{{Bigiel} \& {Blitz}(2012)}]{2012ApJ...756..183B}
{Bigiel}, F. \& {Blitz}, L. 2012, \apj, 756, 183

\bibitem[{{Bohlin} {et~al.}(1978){Bohlin}, {Savage}, \& {Drake}}]{Bohlin78}
{Bohlin}, R.~C., {Savage}, B.~D., \& {Drake}, J.~F. 1978, \apj, 224, 132

\bibitem[{{Bouwens} {et~al.}(2014){Bouwens}, {Illingworth}, {Oesch}, {Trenti},
  {Labbe'}, {Bradley}, {Carollo}, {van Dokkum}, {Gonzalez}, {Holwerda},
  {Franx}, {Spitler}, {Smit}, \& {Magee}}]{bouwens2014}
{Bouwens}, R.~J., {Illingworth}, G.~D., {Oesch}, P.~A., {et~al.} 2014, ArXiv
  e-prints

\bibitem[{{Brinchmann} {et~al.}(2004){Brinchmann}, {Charlot}, {White},
  {Tremonti}, {Kauffmann}, {Heckman}, \& {Brinkmann}}]{2004MNRAS.351.1151B}
{Brinchmann}, J., {Charlot}, S., {White}, S.~D.~M., {et~al.} 2004, \mnras, 351,
  1151

\bibitem[{{Calzetti} {et~al.}(1994){Calzetti}, {Kinney}, \&
  {Storchi-Bergmann}}]{1994ApJ...429..582C}
{Calzetti}, D., {Kinney}, A.~L., \& {Storchi-Bergmann}, T. 1994, \apj, 429, 582

\bibitem[{{Cantalupo} \& {Porciani}(2011)}]{2011MNRAS.411.1678C}
{Cantalupo}, S. \& {Porciani}, C. 2011, \mnras, 411, 1678

\bibitem[{{Cappellari} \& {Copin}(2003)}]{Cappellari2003}
{Cappellari}, M. \& {Copin}, Y. 2003, \mnras, 342, 345

\bibitem[{{Cardelli} {et~al.}(1989){Cardelli}, {Clayton}, \&
  {Mathis}}]{1989ApJ...345..245C}
{Cardelli}, J.~A., {Clayton}, G.~C., \& {Mathis}, J.~S. 1989, \apj, 345, 245

\bibitem[{{Chapman} {et~al.}(2005){Chapman}, {Blain}, {Smail}, \&
  {Ivison}}]{2005ApJ...622..772C}
{Chapman}, S.~C., {Blain}, A.~W., {Smail}, I., \& {Ivison}, R.~J. 2005, \apj,
  622, 772

\bibitem[{{Charlot} \& {Fall}(1993)}]{1993ApJ...415..580C}
{Charlot}, S. \& {Fall}, S.~M. 1993, \apj, 415, 580

\bibitem[{{Cowie} {et~al.}(2010){Cowie}, {Barger}, \&
  {Hu}}]{2010ApJ...711..928C}
{Cowie}, L.~L., {Barger}, A.~J., \& {Hu}, E.~M. 2010, \apj, 711, 928

\bibitem[{{Cowie} \& {Hu}(1998)}]{1998AJ....115.1319C}
{Cowie}, L.~L. \& {Hu}, E.~M. 1998, \aj, 115, 1319

\bibitem[{{Curtis-Lake} {et~al.}(2012){Curtis-Lake}, {McLure}, {Pearce},
  {Dunlop}, {Cirasuolo}, {Stark}, {Almaini}, {Bradshaw}, {Chuter}, {Foucaud},
  \& {Hartley}}]{Curtis-Lake2012}
{Curtis-Lake}, E., {McLure}, R.~J., {Pearce}, H.~J., {et~al.} 2012, \mnras,
  422, 1425

\bibitem[{{Deharveng} {et~al.}(2008){Deharveng}, {Small}, {Barlow},
  {P{\'e}roux}, {Milliard}, {Friedman}, {Martin}, {Morrissey}, {Schiminovich},
  {Forster}, {Seibert}, {Wyder}, {Bianchi}, {Donas}, {Heckman}, {Lee},
  {Madore}, {Neff}, {Rich}, {Szalay}, {Welsh}, \& {Yi}}]{Deharveng2008}
{Deharveng}, J.-M., {Small}, T., {Barlow}, T.~A., {et~al.} 2008, \apj, 680,
  1072

\bibitem[{{Dessauges-Zavadsky} {et~al.}(2010){Dessauges-Zavadsky}, {D'Odorico},
  {Schaerer}, {Modigliani}, {Tapken}, \& {Vernet}}]{Dessauges10}
{Dessauges-Zavadsky}, M., {D'Odorico}, S., {Schaerer}, D., {et~al.} 2010, \aap,
  510, A26

\bibitem[{{Diehl} \& {Statler}(2006)}]{Diehl2006}
{Diehl}, S. \& {Statler}, T.~S. 2006, \mnras, 368, 497

\bibitem[{{Dijkstra} {et~al.}(2006){Dijkstra}, {Haiman}, \&
  {Spaans}}]{2006ApJ...649...14D}
{Dijkstra}, M., {Haiman}, Z., \& {Spaans}, M. 2006, \apj, 649, 14

\bibitem[{{Dijkstra} \& {Wyithe}(2010)}]{2010MNRAS.408..352D}
{Dijkstra}, M. \& {Wyithe}, J.~S.~B. 2010, \mnras, 408, 352

\bibitem[{{Duval} {et~al.}(2014){Duval}, {Schaerer}, {{\"O}stlin}, \&
  {Laursen}}]{Duval2014}
{Duval}, F., {Schaerer}, D., {{\"O}stlin}, G., \& {Laursen}, P. 2014, \aap,
  562, A52

\bibitem[{{Elbaz} {et~al.}(2011){Elbaz}, {Dickinson}, {Hwang},
  {D{\'{\i}}az-Santos}, {Magdis}, {Magnelli}, {Le Borgne}, {Galliano},
  {Pannella}, {Chanial}, {Armus}, {Charmandaris}, {Daddi}, {Aussel}, {Popesso},
  {Kartaltepe}, {Altieri}, {Valtchanov}, {Coia}, {Dannerbauer}, {Dasyra},
  {Leiton}, {Mazzarella}, {Alexander}, {Buat}, {Burgarella}, {Chary}, {Gilli},
  {Ivison}, {Juneau}, {Le Floc'h}, {Lutz}, {Morrison}, {Mullaney}, {Murphy},
  {Pope}, {Scott}, {Brodwin}, {Calzetti}, {Cesarsky}, {Charlot}, {Dole},
  {Eisenhardt}, {Ferguson}, {F{\"o}rster Schreiber}, {Frayer}, {Giavalisco},
  {Huynh}, {Koekemoer}, {Papovich}, {Reddy}, {Surace}, {Teplitz}, {Yun}, \&
  {Wilson}}]{2011A&A...533A.119E}
{Elbaz}, D., {Dickinson}, M., {Hwang}, H.~S., {et~al.} 2011, \aap, 533, A119

\bibitem[{{Fan} {et~al.}(2006){Fan}, {Carilli}, \&
  {Keating}}]{2006ARA&A..44..415F}
{Fan}, X., {Carilli}, C.~L., \& {Keating}, B. 2006, \araa, 44, 415

\bibitem[{{Gawiser} {et~al.}(2006){Gawiser}, {van Dokkum}, {Gronwall},
  {Ciardullo}, {Blanc}, {Castander}, {Feldmeier}, {Francke}, {Franx},
  {Haberzettl}, {Herrera}, {Hickey}, {Infante}, {Lira}, {Maza}, {Quadri},
  {Richardson}, {Schawinski}, {Schirmer}, {Taylor}, {Treister}, {Urry}, \&
  {Virani}}]{2006ApJ...642L..13G}
{Gawiser}, E., {van Dokkum}, P.~G., {Gronwall}, C., {et~al.} 2006, \apjl, 642,
  L13

\bibitem[{{Giavalisco} {et~al.}(1996){Giavalisco}, {Koratkar}, \&
  {Calzetti}}]{1996ApJ...466..831G}
{Giavalisco}, M., {Koratkar}, A., \& {Calzetti}, D. 1996, \apj, 466, 831

\bibitem[{{Gronwall} {et~al.}(2007){Gronwall}, {Ciardullo}, {Hickey},
  {Gawiser}, {Feldmeier}, {van Dokkum}, {Urry}, {Herrera}, {Lehmer}, {Infante},
  {Orsi}, {Marchesini}, {Blanc}, {Francke}, {Lira}, \&
  {Treister}}]{2007ApJ...667...79G}
{Gronwall}, C., {Ciardullo}, R., {Hickey}, T., {et~al.} 2007, \apj, 667, 79

\bibitem[{{Haiman}(2002)}]{2002ApJ...576L...1H}
{Haiman}, Z. 2002, \apjl, 576, L1

\bibitem[{{Hansen} \& {Oh}(2006)}]{2006MNRAS.367..979H}
{Hansen}, M. \& {Oh}, S.~P. 2006, \mnras, 367, 979

\bibitem[{{Hashimoto} {et~al.}(2013){Hashimoto}, {Ouchi}, {Shimasaku}, {Ono},
  {Nakajima}, {Rauch}, {Lee}, \& {Okamura}}]{Hashimoto2013}
{Hashimoto}, T., {Ouchi}, M., {Shimasaku}, K., {et~al.} 2013, \apj, 765, 70

\bibitem[{{Hayes} {et~al.}(2007){Hayes}, {{\"O}stlin}, {Atek}, {Kunth},
  {Mas-Hesse}, {Leitherer}, {Jim{\'e}nez-Bail{\'o}n}, \&
  {Adamo}}]{2007MNRAS.382.1465H}
{Hayes}, M., {{\"O}stlin}, G., {Atek}, H., {et~al.} 2007, \mnras, 382, 1465

\bibitem[{{Hayes} {et~al.}(2014){Hayes}, {{\"O}stlin}, {Duval}, {Sandberg},
  {Guaita}, {Melinder}, {Adamo}, {Schaerer}, {Verhamme}, {Orlitov{\'a}},
  {Mas-Hesse}, {Cannon}, {Atek}, {Kunth}, {Laursen}, {Ot{\'{\i}}-Floranes},
  {Pardy}, {Rivera-Thorsen}, \& {Herenz}}]{Hayes2014}
{Hayes}, M., {{\"O}stlin}, G., {Duval}, F., {et~al.} 2014, \apj, 782, 6 (paper
  II)

\bibitem[{{Hayes} {et~al.}(2009){Hayes}, {{\"O}stlin}, {Mas-Hesse}, \&
  {Kunth}}]{2009AJ....138..911H}
{Hayes}, M., {{\"O}stlin}, G., {Mas-Hesse}, J.~M., \& {Kunth}, D. 2009, \aj,
  138, 911

\bibitem[{{Hayes} {et~al.}(2005){Hayes}, {{\"O}stlin}, {Mas-Hesse}, {Kunth},
  {Leitherer}, \& {Petrosian}}]{2005A&A...438...71H}
{Hayes}, M., {{\"O}stlin}, G., {Mas-Hesse}, J.~M., {et~al.} 2005, \aap, 438, 71

\bibitem[{{Hayes} {et~al.}(2010){Hayes}, {{\"O}stlin}, {Schaerer}, {Mas-Hesse},
  {Leitherer}, {Atek}, {Kunth}, {Verhamme}, {de Barros}, \&
  {Melinder}}]{Hayes10}
{Hayes}, M., {{\"O}stlin}, G., {Schaerer}, D., {et~al.} 2010, \nat, 464, 562

\bibitem[{{Hayes} {et~al.}(2013){Hayes}, {{\"O}stlin}, {Schaerer}, {Verhamme},
  {Mas-Hesse}, {Adamo}, {Atek}, {Cannon}, {Duval}, {Guaita}, {Herenz}, {Kunth},
  {Laursen}, {Melinder}, {Orlitov{\'a}}, {Ot{\'{\i}}-Floranes}, \&
  {Sandberg}}]{Hayes2013}
{Hayes}, M., {{\"O}stlin}, G., {Schaerer}, D., {et~al.} 2013, \apjl, 765, L27

\bibitem[{{Hayes} {et~al.}(2011){Hayes}, {Schaerer}, {{\"O}stlin}, {Mas-Hesse},
  {Atek}, \& {Kunth}}]{2011ApJ...730....8H}
{Hayes}, M., {Schaerer}, D., {{\"O}stlin}, G., {et~al.} 2011, \apj, 730, 8

\bibitem[{{Heckman} {et~al.}(2005){Heckman}, {Hoopes}, {Seibert}, {Martin},
  {Salim}, {Rich}, {Kauffmann}, {Charlot}, {Barlow}, {Bianchi}, {Byun},
  {Donas}, {Forster}, {Friedman}, {Jelinsky}, {Lee}, {Madore}, {Malina},
  {Milliard}, {Morrissey}, {Neff}, {Schiminovich}, {Siegmund}, {Small},
  {Szalay}, {Welsh}, \& {Wyder}}]{Heckman2005}
{Heckman}, T.~M., {Hoopes}, C.~G., {Seibert}, M., {et~al.} 2005, \apjl, 619,
  L35

\bibitem[{{Hoopes} {et~al.}(2007){Hoopes}, {Heckman}, {Salim}, {Seibert},
  {Tremonti}, {Schiminovich}, {Rich}, {Martin}, {Charlot}, {Kauffmann},
  {Forster}, {Friedman}, {Morrissey}, {Neff}, {Small}, {Wyder}, {Bianchi},
  {Donas}, {Lee}, {Madore}, {Milliard}, {Szalay}, {Welsh}, \&
  {Yi}}]{Hoopes2007}
{Hoopes}, C.~G., {Heckman}, T.~M., {Salim}, S., {et~al.} 2007, \apjs, 173, 441

\bibitem[{{Hu} \& {McMahon}(1996)}]{1996Natur.382..231H}
{Hu}, E.~M. \& {McMahon}, R.~G. 1996, \nat, 382, 231

\bibitem[{{Hummer} \& {Storey}(1987)}]{1987MNRAS.224..801H}
{Hummer}, D.~G. \& {Storey}, P.~J. 1987, \mnras, 224, 801

\bibitem[{{Kauffmann} {et~al.}(2003){Kauffmann}, {Heckman}, {White}, {Charlot},
  {Tremonti}, {Brinchmann}, {Bruzual}, {Peng}, {Seibert}, {Bernardi},
  {Blanton}, {Brinkmann}, {Castander}, {Cs{\'a}bai}, {Fukugita}, {Ivezic},
  {Munn}, {Nichol}, {Padmanabhan}, {Thakar}, {Weinberg}, \&
  {York}}]{2003MNRAS.341...33K}
{Kauffmann}, G., {Heckman}, T.~M., {White}, S.~D.~M., {et~al.} 2003, \mnras,
  341, 33

\bibitem[{{Kennicutt}(1998)}]{Kennicutt1998}
{Kennicutt}, Jr., R.~C. 1998, \araa, 36, 189

\bibitem[{{K{\"u}mmel} {et~al.}(2006){K{\"u}mmel}, {Larsen}, \&
  {Walsh}}]{Kummel2006}
{K{\"u}mmel}, M., {Larsen}, S.~S., \& {Walsh}, J.~R. 2006, in The 2005 HST
  Calibration Workshop: Hubble After the Transition to Two-Gyro Mode, ed. A.~M.
  {Koekemoer}, P.~{Goudfrooij}, \& L.~L. {Dressel}, 85

\bibitem[{{Kunth} {et~al.}(2003){Kunth}, {Leitherer}, {Mas-Hesse},
  {{\"O}stlin}, \& {Petrosian}}]{2003ApJ...597..263K}
{Kunth}, D., {Leitherer}, C., {Mas-Hesse}, J.~M., {{\"O}stlin}, G., \&
  {Petrosian}, A. 2003, \apj, 597, 263

\bibitem[{{Kunth} {et~al.}(1994){Kunth}, {Lequeux}, {Sargent}, \&
  {Viallefond}}]{1994A&A...282..709K}
{Kunth}, D., {Lequeux}, J., {Sargent}, W.~L.~W., \& {Viallefond}, F. 1994,
  \aap, 282, 709

\bibitem[{{Kunth} {et~al.}(1998){Kunth}, {Mas-Hesse}, {Terlevich}, {Terlevich},
  {Lequeux}, \& {Fall}}]{1998A&A...334...11K}
{Kunth}, D., {Mas-Hesse}, J.~M., {Terlevich}, E., {et~al.} 1998, \aap, 334, 11

\bibitem[{{Laursen} {et~al.}(2013){Laursen}, {Duval}, \&
  {{\"O}stlin}}]{Laursen13}
{Laursen}, P., {Duval}, F., \& {{\"O}stlin}, G. 2013, \apj, 766, 124

\bibitem[{{Laursen} {et~al.}(2009){Laursen}, {Sommer-Larsen}, \&
  {Andersen}}]{Laursen09}
{Laursen}, P., {Sommer-Larsen}, J., \& {Andersen}, A.~C. 2009, \apj, 704, 1640

\bibitem[{{Lehnert} {et~al.}(2010){Lehnert}, {Nesvadba}, {Cuby}, {Swinbank},
  {Morris}, {Cl{\'e}ment}, {Evans}, {Bremer}, \& {Basa}}]{2010Natur.467..940L}
{Lehnert}, M.~D., {Nesvadba}, N.~P.~H., {Cuby}, J.-G., {et~al.} 2010, \nat,
  467, 940

\bibitem[{{Leitherer} {et~al.}(2013){Leitherer}, {Chandar}, {Tremonti},
  {Wofford}, \& {Schaerer}}]{Leitherer13}
{Leitherer}, C., {Chandar}, R., {Tremonti}, C.~A., {Wofford}, A., \&
  {Schaerer}, D. 2013, \apj, 772, 120

\bibitem[{{Leitherer} {et~al.}(1999){Leitherer}, {Schaerer}, {Goldader},
  {Gonz{\'a}lez Delgado}, {Robert}, {Kune}, {de Mello}, {Devost}, \&
  {Heckman}}]{1999ApJS..123....3L}
{Leitherer}, C., {Schaerer}, D., {Goldader}, J.~D., {et~al.} 1999, \apjs, 123,
  3

\bibitem[{{Lequeux} {et~al.}(1995){Lequeux}, {Kunth}, {Mas-Hesse}, \&
  {Sargent}}]{1995A&A...301...18L}
{Lequeux}, J., {Kunth}, D., {Mas-Hesse}, J.~M., \& {Sargent}, W.~L.~W. 1995,
  \aap, 301, 18

\bibitem[{{Lidman} {et~al.}(2012){Lidman}, {Hayes}, {Jones}, {Schaerer},
  {Westra}, {Tapken}, {Meisenheimer}, \& {Verhamme}}]{Lidman12}
{Lidman}, C., {Hayes}, M., {Jones}, D.~H., {et~al.} 2012, \mnras, 420, 1946

\bibitem[{{Malhotra} \& {Rhoads}(2002)}]{2002ApJ...565L..71M}
{Malhotra}, S. \& {Rhoads}, J.~E. 2002, \apjl, 565, L71

\bibitem[{{Malhotra} \& {Rhoads}(2004)}]{2004ApJ...617L...5M}
{Malhotra}, S. \& {Rhoads}, J.~E. 2004, \apjl, 617, L5

\bibitem[{{Mallery} {et~al.}(2012){Mallery}, {Mobasher}, {Capak}, {Kakazu},
  {Masters}, {Ilbert}, {Hemmati}, {Scarlata}, {Salvato}, {McCracken},
  {LeFevre}, \& {Scoville}}]{Mallery2012}
{Mallery}, R.~P., {Mobasher}, B., {Capak}, P., {et~al.} 2012, \apj, 760, 128

\bibitem[{{Martins} {et~al.}(2005){Martins}, {Schaerer}, \&
  {Hillier}}]{2005A&A...436.1049M}
{Martins}, F., {Schaerer}, D., \& {Hillier}, D.~J. 2005, \aap, 436, 1049

\bibitem[{{Mas-Hesse} \& {Kunth}(1999)}]{Mas-Hesse1999}
{Mas-Hesse}, J.~M. \& {Kunth}, D. 1999, \aap, 349, 765

\bibitem[{{Mas-Hesse} {et~al.}(2003){Mas-Hesse}, {Kunth}, {Tenorio-Tagle},
  {Leitherer}, {Terlevich}, \& {Terlevich}}]{2003ApJ...598..858M}
{Mas-Hesse}, J.~M., {Kunth}, D., {Tenorio-Tagle}, G., {et~al.} 2003, \apj, 598,
  858

\bibitem[{{Meurer} {et~al.}(1999){Meurer}, {Heckman}, \&
  {Calzetti}}]{1999ApJ...521...64M}
{Meurer}, G.~R., {Heckman}, T.~M., \& {Calzetti}, D. 1999, \apj, 521, 64

\bibitem[{{Neufeld}(1991)}]{1991ApJ...370L..85N}
{Neufeld}, D.~A. 1991, \apjl, 370, L85

\bibitem[{{Nilsson} {et~al.}(2007){Nilsson}, {M{\o}ller}, {M{\"o}ller},
  {Fynbo}, {Micha{\l}owski}, {Watson}, {Ledoux}, {Rosati}, {Pedersen}, \&
  {Grove}}]{2007A&A...471...71N}
{Nilsson}, K.~K., {M{\o}ller}, P., {M{\"o}ller}, O., {et~al.} 2007, \aap, 471,
  71

\bibitem[{{Nilsson} {et~al.}(2009){Nilsson}, {Tapken}, {M{\o}ller},
  {Freudling}, {Fynbo}, {Meisenheimer}, {Laursen}, \&
  {{\"O}stlin}}]{2009A&A...498...13N}
{Nilsson}, K.~K., {Tapken}, C., {M{\o}ller}, P., {et~al.} 2009, \aap, 498, 13

\bibitem[{{{\"O}stlin} {et~al.}(2009){{\"O}stlin}, {Hayes}, {Kunth},
  {Mas-Hesse}, {Leitherer}, {Petrosian}, \& {Atek}}]{2009AJ....138..923O}
{{\"O}stlin}, G., {Hayes}, M., {Kunth}, D., {et~al.} 2009, \aj, 138, 923

\bibitem[{{Ouchi} {et~al.}(2008){Ouchi}, {Shimasaku}, {Akiyama}, {Simpson},
  {Saito}, {Ueda}, {Furusawa}, {Sekiguchi}, {Yamada}, {Kodama}, {Kashikawa},
  {Okamura}, {Iye}, {Takata}, {Yoshida}, \& {Yoshida}}]{2008ApJS..176..301O}
{Ouchi}, M., {Shimasaku}, K., {Akiyama}, M., {et~al.} 2008, \apjs, 176, 301

\bibitem[{{Overzier} {et~al.}(2008){Overzier}, {Heckman}, {Kauffmann},
  {Seibert}, {Rich}, {Basu-Zych}, {Lotz}, {Aloisi}, {Charlot}, {Hoopes},
  {Martin}, {Schiminovich}, \& {Madore}}]{2008ApJ...677...37O}
{Overzier}, R.~A., {Heckman}, T.~M., {Kauffmann}, G., {et~al.} 2008, \apj, 677,
  37

\bibitem[{{Prevot} {et~al.}(1984){Prevot}, {Lequeux}, {Prevot}, {Maurice}, \&
  {Rocca-Volmerange}}]{Prevot84}
{Prevot}, M.~L., {Lequeux}, J., {Prevot}, L., {Maurice}, E., \&
  {Rocca-Volmerange}, B. 1984, \aap, 132, 389

\bibitem[{{Reddy} {et~al.}(2008){Reddy}, {Steidel}, {Pettini}, {Adelberger},
  {Shapley}, {Erb}, \& {Dickinson}}]{Reddy2008}
{Reddy}, N.~A., {Steidel}, C.~C., {Pettini}, M., {et~al.} 2008, \apjs, 175, 48

\bibitem[{{Roth} {et~al.}(2005){Roth}, {Kelz}, {Fechner}, {Hahn}, {Bauer},
  {Becker}, {B{\"o}hm}, {Christensen}, {Dionies}, {Paschke}, {Popow}, {Wolter},
  {Schmoll}, {Laux}, \& {Altmann}}]{Roth2005}
{Roth}, M.~M., {Kelz}, A., {Fechner}, T., {et~al.} 2005, \pasp, 117, 620

\bibitem[{{Salim} {et~al.}(2007){Salim}, {Rich}, {Charlot}, {Brinchmann},
  {Johnson}, {Schiminovich}, {Seibert}, {Mallery}, {Heckman}, {Forster},
  {Friedman}, {Martin}, {Morrissey}, {Neff}, {Small}, {Wyder}, {Bianchi},
  {Donas}, {Lee}, {Madore}, {Milliard}, {Szalay}, {Welsh}, \&
  {Yi}}]{2007ApJS..173..267S}
{Salim}, S., {Rich}, R.~M., {Charlot}, S., {et~al.} 2007, \apjs, 173, 267

\bibitem[{{Salzer} {et~al.}(2001){Salzer}, {Gronwall}, {Lipovetsky}, {Kniazev},
  {Moody}, {Boroson}, {Thuan}, {Izotov}, {Herrero}, \&
  {Frattare}}]{2001AJ....121...66S}
{Salzer}, J.~J., {Gronwall}, C., {Lipovetsky}, V.~A., {et~al.} 2001, \aj, 121,
  66

\bibitem[{{Santos}(2004)}]{2004MNRAS.349.1137S}
{Santos}, M.~R. 2004, \mnras, 349, 1137

\bibitem[{{Scarlata} {et~al.}(2009){Scarlata}, {Colbert}, {Teplitz}, {Panagia},
  {Hayes}, {Siana}, {Rau}, {Francis}, {Caon}, {Pizzella}, \&
  {Bridge}}]{2009ApJ...704L..98S}
{Scarlata}, C., {Colbert}, J., {Teplitz}, H.~I., {et~al.} 2009, \apjl, 704, L98

\bibitem[{{Schaerer} \& {de Barros}(2009)}]{2009A&A...502..423S}
{Schaerer}, D. \& {de Barros}, S. 2009, \aap, 502, 423

\bibitem[{{Schaerer} \& {de Barros}(2010)}]{2010A&A...515A..73S}
{Schaerer}, D. \& {de Barros}, S. 2010, \aap, 515, A73+

\bibitem[{{Schaerer} {et~al.}(2011){Schaerer}, {Hayes}, {Verhamme}, \&
  {Teyssier}}]{Schaerer11}
{Schaerer}, D., {Hayes}, M., {Verhamme}, A., \& {Teyssier}, R. 2011, \aap, 531,
  A12

\bibitem[{{Schaerer} \& {Verhamme}(2008)}]{Schaerer08}
{Schaerer}, D. \& {Verhamme}, A. 2008, \aap, 480, 369

\bibitem[{{Shapley} {et~al.}(2003{\natexlab{a}}){Shapley}, {Steidel},
  {Pettini}, \& {Adelberger}}]{2003ApJ...588...65S}
{Shapley}, A.~E., {Steidel}, C.~C., {Pettini}, M., \& {Adelberger}, K.~L.
  2003{\natexlab{a}}, \apj, 588, 65

\bibitem[{{Shapley} {et~al.}(2003{\natexlab{b}}){Shapley}, {Steidel},
  {Pettini}, \& {Adelberger}}]{Shapley03}
{Shapley}, A.~E., {Steidel}, C.~C., {Pettini}, M., \& {Adelberger}, K.~L.
  2003{\natexlab{b}}, \apj, 588, 65

\bibitem[{{Shimasaku} {et~al.}(2006){Shimasaku}, {Kashikawa}, {Doi}, {Ly},
  {Malkan}, {Matsuda}, {Ouchi}, {Hayashino}, {Iye}, {Motohara}, {Murayama},
  {Nagao}, {Ohta}, {Okamura}, {Sasaki}, {Shioya}, \&
  {Taniguchi}}]{2006PASJ...58..313S}
{Shimasaku}, K., {Kashikawa}, N., {Doi}, M., {et~al.} 2006, \pasj, 58, 313

\bibitem[{{Stark} {et~al.}(2010){Stark}, {Ellis}, {Chiu}, {Ouchi}, \&
  {Bunker}}]{2010MNRAS.408.1628S}
{Stark}, D.~P., {Ellis}, R.~S., {Chiu}, K., {Ouchi}, M., \& {Bunker}, A. 2010,
  \mnras, 408, 1628

\bibitem[{{Steidel} {et~al.}(1996){Steidel}, {Giavalisco}, {Dickinson}, \&
  {Adelberger}}]{Steidel1996}
{Steidel}, C.~C., {Giavalisco}, M., {Dickinson}, M., \& {Adelberger}, K.~L.
  1996, \aj, 112, 352

\bibitem[{{Tapken} {et~al.}(2007){Tapken}, {Appenzeller}, {Noll}, {Richling},
  {Heidt}, {Meink{\"o}hn}, \& {Mehlert}}]{2007A&A...467...63T}
{Tapken}, C., {Appenzeller}, I., {Noll}, S., {et~al.} 2007, \aap, 467, 63

\bibitem[{{Tasitsiomi}(2006)}]{2006ApJ...645..792T}
{Tasitsiomi}, A. 2006, \apj, 645, 792

\bibitem[{{Tenorio-Tagle} {et~al.}(1999){Tenorio-Tagle}, {Silich}, {Kunth},
  {Terlevich}, \& {Terlevich}}]{1999MNRAS.309..332T}
{Tenorio-Tagle}, G., {Silich}, S.~A., {Kunth}, D., {Terlevich}, E., \&
  {Terlevich}, R. 1999, \mnras, 309, 332

\bibitem[{{van Breukelen} {et~al.}(2005){van Breukelen}, {Jarvis}, \&
  {Venemans}}]{2005MNRAS.359..895V}
{van Breukelen}, C., {Jarvis}, M.~J., \& {Venemans}, B.~P. 2005, \mnras, 359,
  895

\bibitem[{{Vanzella} {et~al.}(2009){Vanzella}, {Giavalisco}, {Dickinson},
  {Cristiani}, {Nonino}, {Kuntschner}, {Popesso}, {Rosati}, {Renzini}, {Stern},
  {Cesarsky}, {Ferguson}, \& {Fosbury}}]{Vanzella2009}
{Vanzella}, E., {Giavalisco}, M., {Dickinson}, M., {et~al.} 2009, \apj, 695,
  1163

\bibitem[{{Vanzella} {et~al.}(2010){Vanzella}, {Grazian}, {Hayes},
  {Pentericci}, {Schaerer}, {Dickinson}, {Cristiani}, {Giavalisco}, {Verhamme},
  {Nonino}, \& {Rosati}}]{Vanzella10}
{Vanzella}, E., {Grazian}, A., {Hayes}, M., {et~al.} 2010, \aap, 513, A20

\bibitem[{{V{\'a}zquez} \& {Leitherer}(2005)}]{2005ApJ...621..695V}
{V{\'a}zquez}, G.~A. \& {Leitherer}, C. 2005, \apj, 621, 695

\bibitem[{{Verhamme} {et~al.}(2012){Verhamme}, {Dubois}, {Blaizot}, {Garel},
  {Bacon}, {Devriendt}, {Guiderdoni}, \& {Slyz}}]{2012A&A...546A.111V}
{Verhamme}, A., {Dubois}, Y., {Blaizot}, J., {et~al.} 2012, \aap, 546, A111

\bibitem[{{Verhamme} {et~al.}(2008{\natexlab{a}}){Verhamme}, {Schaerer},
  {Atek}, \& {Tapken}}]{2008A&A...491...89V}
{Verhamme}, A., {Schaerer}, D., {Atek}, H., \& {Tapken}, C. 2008{\natexlab{a}},
  \aap, 491, 89

\bibitem[{{Verhamme} {et~al.}(2008{\natexlab{b}}){Verhamme}, {Schaerer},
  {Atek}, \& {Tapken}}]{Verhamme08}
{Verhamme}, A., {Schaerer}, D., {Atek}, H., \& {Tapken}, C. 2008{\natexlab{b}},
  \aap, 491, 89

\bibitem[{{Verhamme} {et~al.}(2006){Verhamme}, {Schaerer}, \&
  {Maselli}}]{Verhamme06}
{Verhamme}, A., {Schaerer}, D., \& {Maselli}, A. 2006, \aap, 460, 397

\bibitem[{{Wofford} {et~al.}(2013){Wofford}, {Leitherer}, \&
  {Salzer}}]{Wofford2013}
{Wofford}, A., {Leitherer}, C., \& {Salzer}, J. 2013, \apj, 765, 118

\bibitem[{{Yajima} {et~al.}(2012){Yajima}, {Li}, {Zhu}, \&
  {Abel}}]{2012MNRAS.424..884Y}
{Yajima}, H., {Li}, Y., {Zhu}, Q., \& {Abel}, T. 2012, \mnras, 424, 884

\bibitem[{{Yin} {et~al.}(2007){Yin}, {Liang}, {Hammer}, {Brinchmann}, {Zhang},
  {Deng}, \& {Flores}}]{2007A&A...462..535Y}
{Yin}, S.~Y., {Liang}, Y.~C., {Hammer}, F., {et~al.} 2007, \aap, 462, 535

\end{thebibliography}

\clearpage

\end{document}